\documentclass{lmcs} 
\pdfoutput=1
\usepackage[utf8]{inputenc}

\usepackage{lastpage}
\lmcsdoi{20}{3}{2}
\lmcsheading{}{\pageref{LastPage}}{}{}%
{Dec.~01,~2022}{Jul.~02,~2024}{}


\usepackage{hyperref}
\usepackage{float}
\usepackage{microtype}
\usepackage{wrapfig}
\usepackage{tikz}
\usetikzlibrary{decorations.pathreplacing,calligraphy}

\newcommand{\Aa}{\mathcal{A}}
\newcommand{\Bb}{\mathcal{B}}
\newcommand{\Gg}{\mathcal{G}}
\newcommand{\Ll}{\mathcal{L}}
\newcommand{\incl}{\subseteq}

\newcommand{\RR}{\mathbb{R}}
\newcommand{\Rpos}{\RR_{\ge 0}}
\newcommand{\Rneg}{\RR_{\le 0}}

\newcommand{\Nat}{\mathbb{N}}

\newcommand{\ezg}{\ensuremath{\mathsf{EZG}}}

\newcommand{\xra}[1]{\xrightarrow{#1}}

\newcommand\sem[1]{{[\![ #1 ]\!]}}

\newcommand{\prophecy}[1]{\overrightarrow{#1}}
\newcommand{\history}[1]{\overleftarrow{#1}}
\newcommand{\A}{\mathcal{A}}
\newcommand{\G}{\mathcal{G}}
\newcommand{\asplit}{\mathsf{split}}
\newcommand{\pre}[2]{\mathsf{pre}(#1,#2)}
\newcommand{\leqlt}{\mathrel{\triangleleft}}
\newcommand{\ua}{{\uparrow}}
\newcommand{\da}{{\downarrow}}
\newcommand{\GG}{\mathbb{G}}
\newcommand{\graph}[1]{\GG(#1)}
\newcommand{\graphv}[2]{\GG_{#1}(#2)}
\newcommand{\V}{\mathbb{V}}

\newcommand{\elapse}[1]{\overrightarrow{#1}}

\begin{document}

\title[Simulations for Event-Clock Automata]{Simulations for Event-Clock Automata}
\titlecomment{A preliminary version of this paper appeared in~\cite{AGGS-CONCUR22}.}
\thanks{This work was supported by DST/CEFIPRA/INRIA Project EQuaVE.
S. Akshay was partially supported by DST/SERB Matrics Grant MTR/2018/000744.
Paul Gastin was partially supported by ANR project Ticktac (ANR-18-CE40-0015).
The bulk of this work was carried out while R.~Govind was a postdoctoral fellow at Indian Institute of Technology Bombay.}	

\author[S.~Akshay]{S. Akshay\lmcsorcid{0000-0002-2471-5997}}[a]
\author[P.~Gastin]{Paul Gastin\lmcsorcid{0000-0002-1313-7722}}[b,d]
\author[R.~Govind]{R. Govind\lmcsorcid{0000-0002-1634-5893}}[a,e]
\author[B.~Srivathsan]{B. Srivathsan\lmcsorcid{0000-0003-2666-0691}}[c,d]

\address{Department of Computer Science and Engineering, Indian Institute of Technology Bombay, Mumbai, India}	
\email{akshayss@cse.iitb.ac.in}  

\address{Université Paris-Saclay, ENS Paris-Saclay, CNRS, LMF, 91190,
Gif-sur-Yvette, France}	
\email{paul.gastin@ens-paris-saclay.fr}  

\address{Chennai Mathematical Institute, India}	
\email{sri@cmi.ac.in}  

\address{CNRS, ReLaX, IRL 2000, Siruseri, India}	

\address{Uppsala University, Sweden}	
\email{govind.rajanbabu@it.uu.se}  

\begin{abstract}
  Event-clock automata (ECA) are a well-known \emph{semantic} subclass of timed automata (TA) which enjoy admirable theoretical properties, e.g., determinizability, and are practically useful to  capture timed specifications.
  However, unlike for timed automata, there exist no implementations   for checking non-emptiness of event-clock automata. As ECAs contain special \emph{prophecy clocks} that guess and maintain the time to the next occurrence of specific events, they cannot be seen as a syntactic subclass of TA. Therefore, implementations for TA cannot be directly used for ECAs, and moreover the translation of an ECA to a semantically equivalent TA is expensive. Another reason for the lack of ECA implementations is the difficulty in adapting zone-based algorithms, critical in the timed automata setting, to the event-clock automata setting. This difficulty was studied by Geeraerts et al. in 2011, where the authors proposed a zone enumeration procedure that uses \emph{zone extrapolations} for finiteness.

  In this article, we propose a different zone-based algorithm to solve the
  reachability problem for event-clock automata, \emph{using simulations} for finiteness. A surprising consequence of our result is that for event-predicting automata, the subclass of event-clock automata that only use prophecy clocks, we obtain finiteness even without any simulations.  For general event-clock automata, our new algorithm exploits the $\Gg$-simulation framework, which is the coarsest known simulation relation in timed automata literature, and has been recently used for advances in other extensions of timed automata. 
\end{abstract}

\maketitle

\section{Introduction}\label{sec:intro}

Timed automata (TA)~\cite{Alur:TCS:1994} are a well-established model
for real-time systems and form the basis for employing model-checking
techniques.  The most popular property that has been considered in
these systems is (control state) reachability. Reachability in timed
automata is a well-studied problem and was shown to be decidable (and
PSPACE-complete) using the so-called region
construction~\cite{Alur:TCS:1994}. This construction was primarily of
theoretical interest, as the number of regions, which are collections
of reachable configurations, explodes both in theory and in
practice. On the other hand, timed automata have been implemented in
several tools: UPPAAL~\cite{Larsen:1997:UPPAAL,BehrmannDLHPYH06},
KRONOS~\cite{KRONOS}, PAT~\cite{PAT}, RED~\cite{RED},
TChecker~\cite{TChecker}, Theta~\cite{Theta}, LTS-Min~\cite{LTSmin},
Symrob~\cite{RoussanalySM19},
MCTA~\cite{DBLP:conf/cav/KupferschmidWNP08}, etc. Most of these tools
have a common underlying algorithm which is an explicit
enumeration of reachable configurations stored as
\emph{zones}~\cite{Bengtsson:LCPN:2003}.
Since the late 90s, a substantial effort has been invested in
improving zone enumeration techniques, the common challenge being how
to get a sound and complete enumeration while exploring as few zones
as possible.

The more general model-checking problem asks whether the system represented by a TA $\Aa$ satisfies the specification given by a TA $\Bb$. This problem reduces to checking language inclusion $\Ll(\Aa) \incl
\Ll(\Bb)$. There are two challenges here: first, the inclusion problem
is undecidable in its full generality, and second, having clocks,
though excellent for timed implementations, is often less than ideal
for modeling timed specifications. This has led to the introduction of
event-clocks and the corresponding model of \emph{event-clock
  automata (ECA)}~\cite{AlurFH99}. Event-clock automata make use of
special clocks that track the time since the last occurrence of an
event (history clocks) or the time until the next occurrence of an
event (prophecy clocks). On one hand this makes writing timed
specifications more natural. Indeed, the role of prophecy clocks is in
the same spirit as future modalities in temporal logics. This has led
to several extensions of temporal logics with
event-clocks~\cite{DSouzaT04,AkshayBG13,RaskinS99},
which are often used as specification languages and can be converted
into ECA. On the other hand, ECAs can be determinized and hence
complemented. 
So, model-checking event-clock specifications over TA models can be reduced to
the emptiness problem on the product of the TA with an ECA (for the
complement of the specification). This product contains usual
clocks, history clocks and prophecy clocks.  The usual clocks can be treated in the same
way as history clocks for the zone analysis.  Therefore, if we solve ECA reachability
(with history and prophecy clocks) using zones, we can incorporate usual clocks into the procedure seamlessly.

Note that ECAs are a semantic (and not syntactic) subclass of TAs., i.e., while the class of languages recognized by ECA are a subclass of the class of languages recognized by TA, syntactically ECAs are not a subclass of TAs. Thus, even though there are several efficient zone-based algorithms based on time-abstract simulations for timed automata, these cannot be directly applied to event-clock automata. In the work that first introduced ECA~\cite{AlurFH99}, a translation from an event-clock automaton to an equivalent timed automaton was also proposed.
One could, in principle, first translate a given ECA into a timed automaton using the translation proposed in \cite{AlurFH99}, 
and then run the state-of-the-art reachability algorithm of \cite{GastinMS18,Gastin0S19, Gastin0S20} on this timed automaton.
However, the translation from ECA to TA is not efficient: in the worst
case, this translation incurs a blowup that is linear in the number of clocks and exponential in the number of clocks~\cite{GeeraertsRS14}.

Thus, in this paper, we focus on the core problem of building efficient, zone-based algorithms for reachability in ECA. This problem
turns out to be significantly different compared to zone-based reachability algorithms in usual TA, precisely due to prophecy
clocks. Our goal is to align the zone-based reachability algorithms for ECAs with recent approaches for TAs that have shown significant
gains.

As mentioned earlier, the core of an efficient TA reachability algorithm is an enumeration of zones, where the central challenge is
that na\"ive enumeration does not terminate. One approach to guarantee termination is to make use of an \emph{extrapolation} operation on
zones: each new zone that is enumerated is extrapolated to a bigger zone. Any freshly enumerated zone that is contained in an existing
zone is discarded. More recently, a new \emph{simulation} approach to
zone enumeration has been designed, where enumerated zones are left
unchanged (i.e., no extrapolation). Instead,
with each fresh zone it is checked whether the fresh zone is simulated by an already seen zone. If yes, the fresh zone is
discarded. Otherwise, it is kept for further exploration. Different simulations have been considered: the
$LU$-simulation~\cite{HerbreteauSW12} which is based on maximum
constants appearing in lower and
upper bounded constraints in the TA, or the $\Gg$-simulation~\cite{Gastin0S20}, which is based on a
carefully-chosen set of constraints from the TA. Coarser simulations lead to fewer zones being enumerated.
The $\Gg$-simulation is currently the coarsest-known simulation that can be efficiently applied in the simulation approach. The simulation
based approach offers several gains over the extrapolation approach:
(1) since concrete zones are maintained, one could use dynamic simulation parameters and dynamic simulations, starting from a coarse
simulation and refining whenever necessary~\cite{DBLP:conf/cav/HerbreteauSW13}, (2) the simulation
approach has been extended to richer models like timed automata with
diagonal constraints~\cite{Gastin0S19,GastinMS18}, updatable timed
automata~\cite{Gastin0S20}, weighted timed
automata~\cite{DBLP:conf/cav/BouyerCM16} and pushdown timed
automata~\cite{AkshayGP21}. In these richer models, extrapolation has
either been shown to be impossible~\cite{Bouyer04} or is unknown.

Surprisingly, for ECA, an arguably more basic and well-known model, it
turns out that there are no existing simulation-based
approaches. However, an extrapolation approach using maximal constants
has been studied for ECA in~\cite{GeeraertsRS11,GeeraertsRS14}. In
this work, the authors start by showing that prophecy clocks exhibit
fundamental differences as compared to usual clocks. To begin with, it
was shown that there is no finite time-abstract bisimulation for ECA
in general. This is in stark contrast to TA where the region
equivalence forms a finite time-abstract bisimulation. The correctness
of extrapolation is strongly dependent on the region equivalence. Therefore, in order to get an algorithm, the authors define a weak semantics for ECA and a
corresponding notion of weak regions which is a finite time-abstract
bisimulation for the weak semantics and show that the weak semantics
is sound for reachability. Building on this, they define an
extrapolation operation for the zone enumeration.

\paragraph*{Our contributions.} Given the advantages of using
simulations with respect to extrapolations in the TA setting described
above, we extend the $\Gg$-simulation approach to ECA.  Here are the
technical contributions leading to the result.
\begin{itemize}
\item We start with a slightly modified presentation of zones in ECA
  and provide a clean algebra for manipulating weights in the graph
  representation for such event-zones. This simplifies the reasoning
  and allows us to adapt many ideas for simulation developed in the TA
  setting directly to the ECA setting.
  
\item The $\Gg$-simulation is parameterized by a set of constraints at
  each state of the automaton. We adapt the constraint computation and
  the definition of the simulation to the context of ECA, the main
  challenge being the handling of prophecy clocks.

\item We give a simulation test between two zones that runs in time
  quadratic in the number of clocks. This is an extension of the
  similar test that exists for timed automata, but now it incorporates
  new conditions that arise due to prophecy clocks.
  
\item Finally, we show that the reachability algorithm using the
  $\Gg$-simulation terminates for ECA: for every sequence
  $Z_0, Z_1, \dots$ of event-zones that are \emph{reachable} during a
  zone enumeration of an ECA, there exist $i < j$ such that $Z_j$ is
  simulated by $Z_i$. This is a notable difference to the existing
  methods in TA, where finiteness is guaranteed for all zones, not
  only the reachable zones. In the ECA case, this is not true: we can
  construct an infinite sequence of event-zones which are incomparable
  with respect to the new $\Gg$-simulation.  However, we show that
  finiteness does hold when restricting to reachable zones, and this
  is sufficient to prove termination of the zone enumeration
  algorithm. Our argument involves identifying some crucial invariants
  in reachable zones, specially, involving the prophecy clocks.
\end{itemize}

The fundamental differences in the behaviour of prophecy clocks as
compared to usual clocks constitute the major challenge in developing
efficient procedures for the analysis of ECAs.  In our work, we have
developed methods to incorporate prophecy clocks alongside the usual
clocks.  We prove a surprising property: in all \emph{reachable}
event-zones, the constraints involving \emph{prophecy} clocks come
from a finite set.  A direct consequence of this observation is that
the event-zone graph of an ECA containing only prophecy clocks (known
as Event-Predicting Automata (EPA)) is always finite. This
  observation is similar in spirit to an early work on
  automata with timers~\cite{Dill89}, whose symbolic analysis was
  shown to terminate without additional constructions. 

A preliminary version of this paper appeared in the
  conference proceedings~\cite{AGGS-CONCUR22}. The current version includes
  cleaner proofs and additional intuitions. Most importantly, we
  identify a mistake in the conference version and fix it
  here. Event-clocks can sometimes take \emph{undefined values} -- for
  instance, before seeing the first $a$, the history clock recording
  the time since the previous $a$ is undefined, and similarly after
  seeing the last $a$, the prophecy clock predicting the next
  occurrence to $a$ is undefined. One of the key contributions of this
  work is the novel mechanism to represent these undefined values
  using $+\infty$ and $-\infty$ and seamlessly integrating the
  manipulation of these quantities along with the finite (real)
  values. This required the definition of a new algebra to handle
  weights in the graph representation of event-zones. In the
  conference version, we had wrongly claimed that an event-zone is
  empty iff its graph representation has a negative cycle. We explain
  in Section~\ref{sec:DBM} that this is not true as it is stated, due
  to the subtle interplay between finite and infinity weights. We fix
  this by introducing a \emph{standard form} for the graph
  representation, and a \emph{normal form} that is obtained from graphs in
  standard form. These
  changes have a significant impact on many of the proofs in the later
  sections and lead to a clearer presentation.  As a result we are
  able to avoid certain cases, while in other places we use our new
  lemmas to get more succinct proofs.

  As a follow-up to~\cite{AGGS-CONCUR22}, an extended model of \emph{Generalized
    Timed Automata} has been proposed, implemented in the open source
  tool TChecker, and is publicly available~\cite{AkshayGGJS23}. This
  model subsumes event-clock automata, and also contains diagonal
  constraints. Reachability is undecidable for this model, and a
  decidable fragment has been identified. The decidable fragment strongly subsumes
  event-clock automata, and requires further sophisticated analysis of
  the objects that we develop in this document.

\paragraph*{Organization of the paper.}
Section~\ref{sec:prelims} recalls ECA and describes a slightly
modified presentation of the ECA
semantics. Section~\ref{sec:event-zones} introduces event-zones, event
zone graph and the simulation-based reachability
framework. Section~\ref{sec:DBM} introduces the new algebra for
representing event-zones and describes some operations needed to build
the zone graph.  Section~\ref{sec:simulation} introduces the
$\Gg$-simulation for event-clock automata and gives the simulation
test. Section \ref{sec:termination} proves finiteness of the
simulation when restricted to reachable zones.

\section{Event-Clock Automata and Valuations}\label{sec:prelims}
Let $X$ be a finite set of real-valued variables called \emph{clocks}.
Let $\overline{\mathbb{R}}=\mathbb{R}\cup \{-\infty,+\infty\}$ denote the set of
all real numbers along with $-\infty$ and $+\infty$.   
The usual $<$ order on reals is extended to deal with $\{-\infty, +\infty\}$ as: $-\infty < c$, $c < +\infty$ for all $c \in \mathbb{R}$ and $-\infty < +\infty$.
Similarly, $\overline{\mathbb{Z}}=\mathbb{Z}\cup \{-\infty,+\infty\}$ denotes the set of all integers
along with $-\infty$ and $+\infty$. 
Let $\Rpos$ (resp. $\Rneg$) be the set of non-negative (resp. non-positive) reals.
Let $\Phi(X)$ denote a set of clock constraints generated by the following
grammar: $\varphi ::= x \leqlt c \mid c \leqlt x \mid \varphi \land \varphi$
where $x \in X$,
$c \in \overline{\mathbb{Z}}$ and
${\leqlt} \in \{<, \le\}$.  The base constraints of the form
$x \leqlt c$ and $c \leqlt x$ will be called \emph{atomic
  constraints}.  Constraints $x< -\infty$ and $+\infty<x$ are
equivalent to \emph{false} and constraints $-\infty\leq x$ and
$x\leq +\infty$ are equivalent to \emph{true}.

Given a finite alphabet $\Sigma$, we define a set 
$X_H = \{\history{a} \mid a \in \Sigma\}$ of \emph{history clocks} 
and a set $X_P = \{\prophecy{a} \mid a \in \Sigma\}$ of \emph{prophecy clocks}. 
Together, history and prophecy clocks are called \emph{event-clocks}.  In this paper, all
clocks will be event-clocks, thus we set $X = X_H \cup X_P$.

\begin{wrapfigure}{r}{7cm}
  \centering
	\begin{tikzpicture}[state/.style={draw, thick, circle, inner sep=2pt}]
  
	  \begin{scope}[->, >=stealth, thick]
  		\draw (2,7) to (6,7);
	  	\draw (6,7) to (2,7);
	  \end{scope}
	
	  \node at (1.3,7)  {$-\infty$ };
	  \node at (6.7,7)  {$+\infty$ };
	  \node at (3,6.5)  { $\prophecy{a}$ };
	  \node at (5,6.5)  { $\history{a}$ };
    \node at (4.2,6.5)  { $0$ };

    \node at (3,7) [circle,fill,inner sep=1pt]{};
	  \node at (5,7) [circle,fill,inner sep=1pt]{};
	  \node at (4,7) [circle,fill,inner sep=1pt]{};
	\end{tikzpicture}
	\caption{Valuation of event-clocks.}
  \label{fig:event-clocks}
\end{wrapfigure}
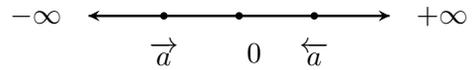

\begin{defi}[Valuation]
  A valuation of event-clocks is a function
  $v\colon X \mapsto \overline{\RR}$ which
  maps history clocks to $\Rpos\cup\{+\infty\}$ and prophecy clocks to
  $\Rneg \cup \{-\infty\}$.  
  We say a history clock $\history{a}$, for some $a\in \Sigma$ is \emph{undefined} (resp.\
  \emph{defined}) when $v(\history{a}) = +\infty$ (resp.\ $v(\history{a}) < +\infty$) and a
  prophecy clock $\prophecy{a}$ is undefined (resp.\ defined) when $v(\prophecy{a}) = -\infty$
  (resp.\ $-\infty< v(\prophecy{a})$).
  We denote by $\V(X)$ or simply by $\V$ the set of valuations over $X$.
\end{defi}

\begin{figure}
  \centering
  \begin{tikzpicture}[state/.style={draw, thick, circle, inner sep=2pt}]

    \begin{scope}[->, >=stealth, thick]
      \draw (0,4) to (8,4);
      \draw (8,4) to (0,4);
  
      \draw (0,7) to (8,7);
      \draw (8,7) to (0,7);
    \end{scope}
  
    \node at (1,4) [circle,fill,inner sep=1pt]{};
    \node at (7,4) [circle,fill,inner sep=1pt]{};
  
    \node at (1,8)  {\small \textcolor{black}{Timestamp of} };
    \node at (1,7.6)  {\small \textcolor{black}{previous $a$} };
    \node at (7,8)  {\small \textcolor{black}{Timestamp of } };
    \node at (7,7.6)  {\small \textcolor{black}{next $a$} };
    
    \node at (-1,7)  {$-\infty$ };
    \node at (9,7)  {$+\infty$ };
    \node at (1,6.5)  { $t'$ };
    \node at (7,6.5)  { $t''$ };
    \node at (1,7) [circle,fill,inner sep=1pt]{};
    \node at (7,7) [circle,fill,inner sep=1pt]{};
    \node at (4,7) [circle,fill,inner sep=1pt]{};
  
    \node at (4,4) [circle,fill,inner sep=1pt]{};
    \node at (4.2,6.5)  { $t_{present}$ };
  
    \draw [thick, decorate,
    decoration = {calligraphic brace}] (4,6.25) --  (1,6.25);
    \draw [thick, decorate,
    decoration = {calligraphic brace}] (7,6.25) --  (4,6.25);
  
    \node at (2.5,5.8)   { \textcolor{black}{$v(\history{a})$} };
    \node at (2.5,5.3)  { $t_{present} - t'$ };
    \node at (5.5,5.8)  { \textcolor{black}{$v(\prophecy{a})$} };
    \node at (5.5,5.3)  { $t_{present} - t''$ };

    \node at (-1,4)  {$-\infty$ };
    \node at (9,4)  {$+\infty$ };
    \node at (1,3.5)  { $t'$ };
    \node at (7,3.5)  { $t''$ };
    \node [red] at (5.7,3.5)  { $t_{present}$ };
    \node at (4,4) [circle,fill,inner sep=1pt]{};
    \node at (4.2,3.5)  { \textcolor{gray}{$t_{present}$}};
    \node at (5.5,4) [circle,fill,inner sep=1pt]{};
  
      \begin{scope}[->, >=stealth, thick]
        \draw [thick,red] (4,4.25) to (5.5,4.25);
      \end{scope}
      \node at (4.75,4.5) {$\delta$};
  
      \node [red] at (5.5,4) [circle,fill,inner sep=1pt]{};

      \draw [thick, decorate,
      decoration = {calligraphic brace}] (5.5,3) --  (1,3);
      \draw [thick, decorate,
      decoration = {calligraphic brace}] (7,3) --  (5.5,3);
  
      \node at (5.7,3.5)  { \textcolor{red}{$t_{present}$} };
      \node at (3.25,2.5)   { \small \textcolor{black}{$\textcolor{red}{v'(\history{a})} = v(\history{a}) + \delta$} };
      \node at (3.25,2)  { \textcolor{red}{$t_{present}$}$ - t'$ };
      \node at (7,2.5)  {\small \textcolor{black}{$\textcolor{red}{v'(\prophecy{a})} = v(\prophecy{a}) + \delta$} };
      \node at (6.25,2)  { \textcolor{red}{$t_{present}$}$ - t''$ };
  \end{tikzpicture}
    \caption{Representation of valuations in event-clock automata. Here, $v' = v+ \delta$.}
  \label{fig:eca-valuation}
\end{figure}
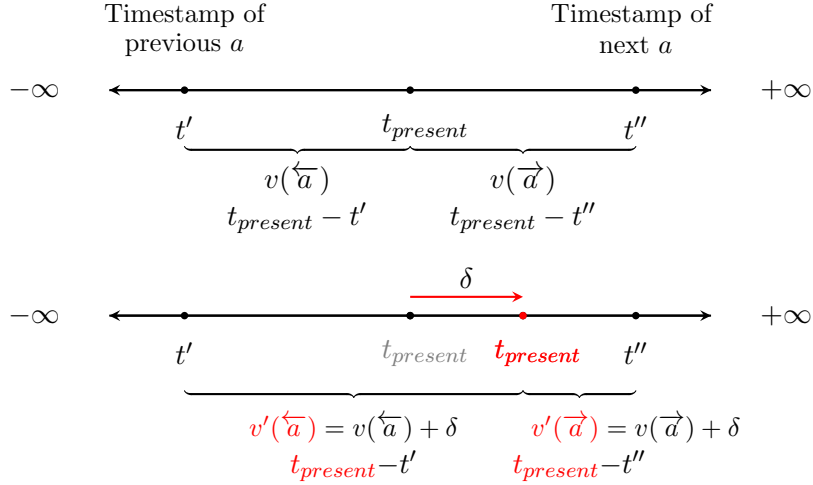

We remark that the history clock and the prophecy clock of an event
$a$ are symmetric notions,  
as illustrated in Figure~\ref{fig:eca-valuation}.
In the semantics that we introduce in this
paper, history clock $\history{a}$ stores the amount of time elapsed after
seeing the last $a$, measuring how far ahead in the
future we are w.r.t.\ the last occurrence of $a$. Before we see an $a$
for the first time, $\history{a}$ is set to $+\infty$.  The prophecy clock
$\prophecy{a}$ stores the negative of the amount of time that needs to be
elapsed before seeing the next $a$.  In other words, $-\prophecy{a}$ tells
us how far behind in the past we are w.r.t.\ the next occurrence of
$a$. If no more $a$'s are going to be seen, then the prophecy clock 
of $a$ is set to $-\infty$, i.e., $\prophecy{a} = -\infty$.

Notice that for history (resp.\ prophecy) clocks, \emph{useful}
constraints use non-negative (resp.\ non-positive) constants. Also,
$\history{a}<0$ and $0<\prophecy{a}$ are equivalent to \emph{false} whereas
$0 \le \history{a}$, $\history{a}\leq+\infty$, $\prophecy{a} \le 0$ and 
$-\infty\leq\prophecy{a}$ are equivalent to \emph{true}. A
constraint $c\leqlt\history{a}$ does not imply that the history clock
$\history{a}$ is defined, whereas a constraint $\history{a}\leqlt c$ with
$(\leqlt,c)\neq(\leq,+\infty)$ does.  The same applies to prophecy
clocks where a constraint $c\leqlt\prophecy{a}$ with
$(c,\leqlt)\neq(-\infty,\leq)$ implies that $\prophecy{a}$ is defined,
whereas $\prophecy{a}\leqlt c$ does not;
in fact, $\prophecy{a}\leq-\infty$ states that $\prophecy{a}$ is undefined.

\begin{rem}
  In the earlier works on ECA~\cite{AlurFH99,GeeraertsRS14}, prophecy
  clocks assumed non-negative values and decreased along with time.  This allowed to
  write guards on prophecy clocks with non-negative constants, e.g., $\prophecy{a} \le 5$ means
  that the next $a$ occurs in at most $5$ time units.  In our convention, this would be
  written as $-5 \le \prophecy{a}$.  Secondly, an undefined clock (history or prophecy) was
  assigned a special symbol $\bot$ in earlier works.  We have changed this to use
  $-\infty$ and $+\infty$ for undefined prophecy and history clocks respectively.
  We adopt these new conventions as they allow to treat both history clocks and prophecy
  clocks in a symmetric fashion, and a clean integration of undefined values
  when we describe zones and simulations.
\end{rem}

We say that a valuation $v$ satisfies a constraint $\varphi$, denoted as $v \models
\varphi$, if $\varphi$ evaluates to \emph{true}, when each variable $x$ in $\varphi$ is
replaced by its value $v(x)$.

We write $[\history{a}]v$ to denote the valuation $v'$ obtained from $v$
by resetting the history clock $\history{a}$ to $0$, keeping the value of
other clocks unchanged.  We denote by $[\prophecy{a}]v$ the \emph{set} of
valuations $v'$ obtained from $v$ by setting the prophecy clock
$\prophecy{a}$ non-deterministically to some value in $[-\infty,0]$,
keeping the value of other clocks unchanged. We see
  $[\prophecy{a}]v$ as the result of an operation $[\prophecy{a}]$ on
  valuation $v$, which we call the \emph{release} of $a$ from $v$. The
  idea is that $v$ maintains an exact value for the next occurrence of
  $a$. On releasing $a$ from $v$, this value is forgotten and a fresh
  guess is made for the next $a$. This fresh guess has no constraints
  and can take any value between $-\infty$ and $0$. As an example,
  consider a valuation $v$ over two events $a$ and $b$, with
  $v(\prophecy{a}) = 0, v(\prophecy{b}) 
  = -2$, and some arbitrary values for the history clocks
  $v(\history{a})$ and $v(\history{b})$. This valuation $v$
  predicts that the next $a$ will occur immediately, whereas $b$ will
  occur in two time units from now. By releasing $a$ from $v$, we get a set
  $[\prophecy{a}]v$ with valuations containing the different possible
  orders between 
  the next occurrences of $a$ and $b$: for instance, pick $v_1, v_2 \in
  [\prophecy{a}]v$ with $v_1(\prophecy{a}) = -3$, $v_2(\prophecy{a}) =
  -0.5$. In both the valuations, the value of $\prophecy{b}$ retains
  the value $-2$ from $v$. Therefore, valuation $v_1$ predicts that
  $a$ will occur after $b$, 
  whereas $v_2$ predicts that $a$ will occur before $b$. There is also
a valuation $v_3 \in [\prophecy{a}]v$ with $v_3(\prophecy{a}) =
-\infty$. This valuation says that $a$ will never occur from hereon.

The next operation on valuations is that of a \emph{time elapse}.
We denote by $v + \delta$ the valuation obtained by increasing the value of all clocks from the valuation $v$ by $\delta \in \Rpos$. 
The time elapse operation is illustrated in Figure~\ref{fig:eca-valuation}.
Not every time elapse may be possible from a valuation, since prophecy clocks need to
stay at most $0$. For example, if there are two events $a,b$, then
a valuation with $v(\prophecy{a}) = -3$ and $v(\prophecy{b}) = -2$ can elapse at
most $2$ time units.

\begin{defi}[Event-clock automata~\cite{AlurFH99}]\label{defn:eca}
  An \emph{event-clock automaton (ECA) } $\A$ is given by a tuple
  $(Q, \Sigma, X, T, q_0, F)$, where $Q$ is a finite set of states,
  $\Sigma$ is a finite alphabet of actions, $X = X_H \cup X_P$ is the set of event clocks for $\Sigma$, $q_0 \in Q$ is the initial state,
  $F \subseteq Q$ is the set of accepting states and
  $T \subseteq Q \times \Sigma \times \Phi(X) \times Q$ is a
  finite set of transitions.

  \noindent  The semantics of an ECA $\A=(Q,\Sigma,X,T,q_0,F)$ is given by a
  transition system $S_{\A}$ whose states are \emph{configurations} $(q,v)$ of $\A$, where $q \in Q$ and $v$ is a valuation.
  A configuration $(q,v)$ is initial if $q=q_{0}$, $v(x)= +\infty$ for all $x\in X_H$.
  A configuration $(q,v)$ is accepting if $q\in F$, and $v(x)=-\infty$ for all
  $x\in X_P$. 
  Transitions of $S_{\A}$ are of two forms:
  \begin{itemize}
    \item \textbf{Delay transition}:
      $(q,v) \xra{\delta} (q, v + \delta)$, if
      $(v + \delta)(x) \leq 0$ for all $x \in X_P$.
      
    \item \textbf{Action transition}: $(q,v) \xra{t} (q',[\history{a}]v')$
      if $t=(q,a,g,q')$ is a transition in $\A$, $v(\prophecy{a})=0$,
      $v'\in[\prophecy{a}]v$ and $v'\models g$.
      
      A transition with action $a$ can be taken when the value of the
      prophecy clock $\prophecy{a}$ is $0$, then a new value in
      $[-\infty,0]$ for $\prophecy{a}$ is non-deterministically guessed so
      that the resulting valuation $v'$ satisfies the guard $g$,
      and finally, the history clock $\history{a}$ is reset to $0$.
    \end{itemize}
\end{defi}

\begin{rem}
  In the semantics for ECA defined above, we say that a configuration is accepting only if $v(x)=-\infty$ for all prophecy clocks. Analogously, for initial configurations, we ask that all the history clocks are set to $+\infty$.
\end{rem}

  Here is an example to illustrate the transition semantics. A
  transition $t = (q, a, g, q')$ with guard $g := (\history{a} \le 5) \land (-3 \le
  \prophecy{a}) \land (\prophecy{b} \le -\infty)$ can be taken if the
  previous $a$ was seen within $5$ units, the next $a$ will be
  seen within $3$ units and there will be no $b$ anymore.
  A valuation $v$ can take this transition provided $v(\history{a}) \le 5$,
  $v(\prophecy{b}) = -\infty$ and $v(\prophecy{a}) = 0$. At this point, the guard
  $-3 \le \prophecy{a}$ is not used. This constraint plays a role in the value
  to which $\prophecy{a}$ is released to: the only possible values of
  $\prophecy{a}$ after taking the transition lie in the interval
  $[-3, 0]$. More precisely, if $(q, v) \xra{t} (q', [\history{a}]v')$,
  then $v'(\prophecy{a}) \in [-3, 0]$. Notice that in
  $[\history{a}]v'$, the value of the history clock $\history{a}$
 is set to $0$ to record the information that $a$ was just
 seen. Figure~\ref{fig:eca-zone-graph-ex-1} gives an example
 of an ECA. Intuitively, this ECA recognizes sequences of the form
 $b^+a$ where there exists a $b$ (not necessarily the last one) at time $1$ before the final letter $a$.

An ECA is called an event-recording automaton (ERA) if it only contains history clocks and
event-predicting automaton (EPA) if it only contains prophecy clocks.  A \emph{run} of an
event-clock automaton is a finite sequence of transitions from an initial configuration of
$S_{\A}$.  A run is said to be \emph{accepting} if its last configuration is
accepting.  We are interested in the \emph{reachability problem} of an event-clock automaton. 
Formally, 

\begin{defi}[Reachability problem for ECA]\label{defn:reach-problem}
  The reachability problem for an event-clock automaton $\A$ is to
  decide whether $\A$ has an accepting run.
\end{defi}

Different solutions based on regions and zones have been proposed in
\cite{AlurFH99,GeeraertsRS11,GeeraertsRS14}.  For ERA, the standard region and zone-based algorithms for timed automata work directly.  However, for EPA (and ECA), this is not the
case.  In fact,~\cite{GeeraertsRS11} show that the standard region abstraction is not
possible, as there exists no finite bisimulation due to the behavior
of prophecy clocks. We provide an intuitive reasoning for this
  fact. Pick some $\alpha \ge 0$. From a valuation $v_\alpha$ with
  $v_\alpha(\prophecy{a}) = -\alpha$ and $v_{\alpha}(\prophecy{b})=0$, 
  we can see $\lfloor \alpha \rfloor$ (the greatest integer smaller than or equal to
  $\alpha$) occurrences of $b$, if we assume
  that the time between consecutive $b$'s is $1$ (see Figure 3 of
  \cite{GeeraertsRS14} or Figure~\ref{fig:eca-zone-graph-ex-2} for a
  detailed example).  This creates a
  difference between $v_\alpha$, and say $v_{\alpha + 1}$, since there
  is an extra occurrence of $b$ possible from $v_{\alpha + 1}$. Therefore, there is
  no constant $M \ge 0$ such that we can put all valuations $v_\alpha$
  with $\alpha < -M$ into one ``equivalence class''. In other words,
  there is no $M$ for which we can forget the actual value of
  $\prophecy{a}$ as long as it is less than $-M$. This is in sharp contrast
  with history clocks (and classical timed automata) where once a
  clock goes beyond a maximum constant $M$, its actual value can be
  forgotten. In some sense, time elapse moves history clocks above
  some $M$ only further away from $M$, 
  whereas for prophecy clocks time elapse brings clocks which are $<
  -M$ closer to
  $-M$ and at some point they cross $-M$ and are no longer $<
  -M$. This creates a fundamental issue causing the impossibility of finite
  bisimulations. 
Also, the standard definition of zones used for timed automata is not sufficient to handle
valuations with undefined clocks.  The papers \cite{GeeraertsRS11,GeeraertsRS14} make use
of special symbols $\bot$ and $?$ for this purpose.  In this work, we use a different
formulation of zones by making use of $+\infty$ and $-\infty$.
Instead of using $x=\bot$ (resp.\ $x\neq\bot$) to state that a clock is undefined
(resp.\ defined) as in~\cite{GeeraertsRS11,GeeraertsRS14}, we write $+\infty\leq x$
or $x\leq-\infty$ (resp.\ $x<+\infty$ or $-\infty<x$) depending on whether $x$ is a
history clock or a prophecy clock.  This distinction between being undefined for history
and prophecy clocks plays an important role.

\section{Event-zones and simulation-based reachability}\label{sec:event-zones}

The most widely used approach for checking reachability in a timed automaton is based on
reachability in a graph called the \emph{zone graph} of a timed automaton~\cite{Daws}.
Roughly, \emph{zones}~\cite{Bengtsson:LCPN:2003} are sets of valuations that can be
represented efficiently using constraints between differences of clocks.  In this section,
we introduce an analogous notion for event-clock automata.  We consider \emph{event
zones}, which are special sets of valuations of event-clock automata.
  
\begin{defi}[Weights]\label{def:weights}
  Let $\mathcal{C} = \{(\leqlt, c) \mid c \in \overline{\mathbb{R}} \text{ and } 
  {\leqlt} \in \{\leq, <\}\}$, called the set of weights.
\end{defi}

We will use constraints of the form $y-x\leqlt c$ with $(\leqlt,c)\in\mathcal{C}$ and
$x,y\in X\cup\{0\}$ (event-clocks are extended with the special constant clock $0$,
meaning that $v(0)=0$ for all valuations $v\in\V$).  Such constraints, between two clocks
are sometimes called diagonal constraints.  The introduction of the special constant clock
allows us to treat constraints with just a single clock (sometimes referred to as
non-diagonal constraints) as special cases.  Indeed, for all weights
$(\leqlt,c)\in\mathcal{C}$, the constraint $x\leqlt c$ is equivalent to $x-0\leqlt c$ and
the constraint $c\leqlt x$ is equivalent to $0-x\leqlt -c$.

To evaluate such constraints, we extend addition on real numbers with the convention
that $(+\infty)+\alpha = \alpha+(+\infty) = +\infty$ for all $\alpha\in\overline{\RR}$ and
$(-\infty)+\beta = \beta+(-\infty) = -\infty$, as long as $\beta\neq+\infty$.  
We also extend the unary minus operation from real numbers to
$\overline{\mathbb{R}}$ by setting $-(+\infty)=-\infty$ and $-(-\infty)=+\infty$.  Abusing
notation, we write $\beta-\alpha$ for $\beta+(-\alpha)$. 
Notice that with this definition of extended addition, the minus operation does not
distribute over addition.\footnote{
  Notice that $-(a+b)=(-a)+(-b)$ when $a$ or $b$ is finite or when $a=b$. But,
  when $a=+\infty$ and $b=-\infty$ then $-(a+b)=-\infty$ whereas $(-a)+(-b)=+\infty$.}
We now highlight a few more important features of this definition.

\begin{rem}\label{rem:extended-addition}
  This extended addition has the following properties that are easy to check:
  \begin{enumerate}
    \item $(\overline{\mathbb{R}},+,0)$ is a monoid with $0$ as neutral element.  In
    particular, the extended addition is associative.

    \item $(\overline{\mathbb{R}},+,0)$ is not a group, since $-\infty$ and $+\infty$ have
    no opposite values.  Note that, $\alpha+(-\alpha)=0$ when $\alpha\in\mathbb{R}$ is
    finite but $\alpha+(-\alpha)=+\infty$ when $\alpha\in\{-\infty,+\infty\}$.  As a
    consequence, in an equation $\alpha+\beta=\alpha+\gamma$, we can cancel $\alpha$ and
    deduce $\beta=\gamma$ when $\alpha$ is finite, but not when $\alpha$ is infinite.
    
    \item The order $\leq$ is monotone on $\overline{\mathbb{R}}$: $b\leq c$ implies
    $a+b\leq a+c$, but the converse implication only holds when $a$ is finite.

    \item The strict order $<$ is only monotone with respect to finite values: when $a$ is
    finite, $b<c$ iff $a+b<a+c$.

    \item For all $a,b\in\overline{\mathbb{R}}$ and $(\leqlt,c)\in\mathcal{C}$, we have
    $a\leqlt b$ iff $-b\leqlt -a$.  Further, $a-b\leqlt c$ implies $a\leqlt b+c$. 
    The converse of the latter statement holds when $b$ is finite, but may be false when $b$ is infinite.\footnote{
    For instance, if $a<+\infty=b$ then $a<b+(-\infty)$, but $a-b=-\infty\not<-\infty$.
    
    If $a<+\infty$ and $b=-\infty$ then $a<b+\infty$, but $a-b=+\infty\not<+\infty$.
    
    If $a=b\in\{-\infty,+\infty\}$ and $c$ is finite then $a\leq b+c$, but
    $a-b=+\infty\not\leq c$.}
  \end{enumerate}  
\end{rem}

\begin{defi}\label{def:diagonal-semantics}
  Let $x,y\in X\cup\{0\}$ be event-clocks (including 0) and $(\leqlt,c)\in\mathcal{C}$ be
  a weight.  For valuations $v\in\V$, define $v\models y-x \leqlt c$ as $v(y)-v(x)\leqlt c$.
\end{defi}

\begin{rem}
  From Definition~\ref{def:diagonal-semantics}, we easily check that the constraint
  $y-x\leqlt c$ is equivalent to \emph{true} (resp.\ \emph{false}) when
  $(\leqlt,c)=(\leq,+\infty)$ (resp.\ $(\leqlt,c)=(<,-\infty)$): for all valuations
  $v\in\V$, $v\models y-x\leq +\infty$ and $v\not\models y-x< -\infty$.
  Constraints that are equivalent to \emph{true} or \emph{false} will be called trivial,
  whereas all others are non-trivial constraints.

  If $(\leqlt,c)\neq(\leq,+\infty)$ then $v\models y-x \leqlt c$ never holds when
  $v(x)=-\infty$.  
  
  Also, if $v(x)=v(y)\in\{-\infty,+\infty\}$ then $v\models y-x \leqlt c$ only holds for
  $(\leqlt,c)=(\leq,+\infty)$.
  
  Consider now a non-trivial constraint $y-x \leqlt c$ with weight
  $(\leqlt,c)\in\mathcal{C}\setminus\{(<,-\infty),(\leq,+\infty)\}$.  
  We have $v\models y-x\leqlt c$ iff  $v(y)<+\infty=v(x)$ or ($v(x)$ is finite and 
  $v(y) \leqlt v(x) + c$).

  Let us consider special cases of Definition~\ref{def:diagonal-semantics}.  
  \begin{itemize}
    \item $v\models y-x \leq -\infty$ iff $v(y)<+\infty=v(x)$ or $v(y)=-\infty<v(x)$.
  
    \item $v\models y-x < +\infty$ iff $v(x)\neq-\infty$ and $v(y)\neq+\infty$.
    \qed
  \end{itemize}
\end{rem}
  
\begin{defi}[\textbf{Event-zones}]\label{defn:event-zones}
  An event-zone is a set of valuations satisfying a conjunction of constraints of the form
  $y-x \leqlt c$, where $x,y \in X\cup\{0\}$,
  $c\in\overline{\mathbb{Z}}=\mathbb{Z}\cup\{-\infty,+\infty\}$ and
  ${\leqlt}\in\{\leq,<\}$.
\end{defi}

Let $W$ be an arbitrary set of valuations (not necessarily an event-zone) and $q$ be a state. For transition
$t:= (q, a, g, q_1)$, we write $(q, W)$ $\xra{t}$ $(q_1, W_t)$ if
$W_t = \{ v_1 \mid \text{$v_1$ is a valuation and } (q, v) \xra{t} \xra{\delta} (q_1, v_1) \text{ for
some } v \in W \text{ and } \delta \in \mathbb{R}_{\geq0} \}$. 
Notice that
we have action $\xra{t}$ followed by delay $\xra{\delta}$ in this
definition, a common convention adopted in the timed automata
literature. This results in the set $W_t$ being closed under
time-successors.  We will show in the 
next section that starting from an event-zone $Z$, the successors
are also event-zones: $(q, Z) \xra{t} (q_1, Z_t)$ implies $Z_t$ is an
event-zone too. We use this feature to define an event-zone
graph. 

\begin{defi}[\textbf{Event-zone graph}]
  Given an ECA $\A$, its {\em event-zone graph, denoted \ezg($\A$)}, is defined as follows:
  Nodes are of the form $(q, Z)$ where $q$ is a state and $Z$ is an event-zone.  The
  initial node is $(q_0, Z_0)$ where $q_0$ is the initial state and $Z_0$ is given by
  $\bigwedge_{a \in \Sigma} \big((+\infty\leq\history{a}) \wedge (\prophecy{a} \leq 0)\big)$. 
  This is the set of all initial valuations, 
  which is already closed under time elapse\footnote{
    By definition, the time elapse operation only permits delays that keep prophecy clocks non-positive.}. 
  For every node $(q, Z)$ and every
  transition $t := (q, a, g, q_1)$ there is a transition
  $(q, Z) \xra{t} (q_1, Z_t)$ in the event-zone graph.
  A node $(q,Z)$ is accepting if $q\in F$ and $Z\cap Z_{f}$ is non-empty where the final 
  zone $Z_{f}$ is defined by $\bigwedge_{a\in\Sigma} \prophecy{a}\leq-\infty$.
\end{defi}

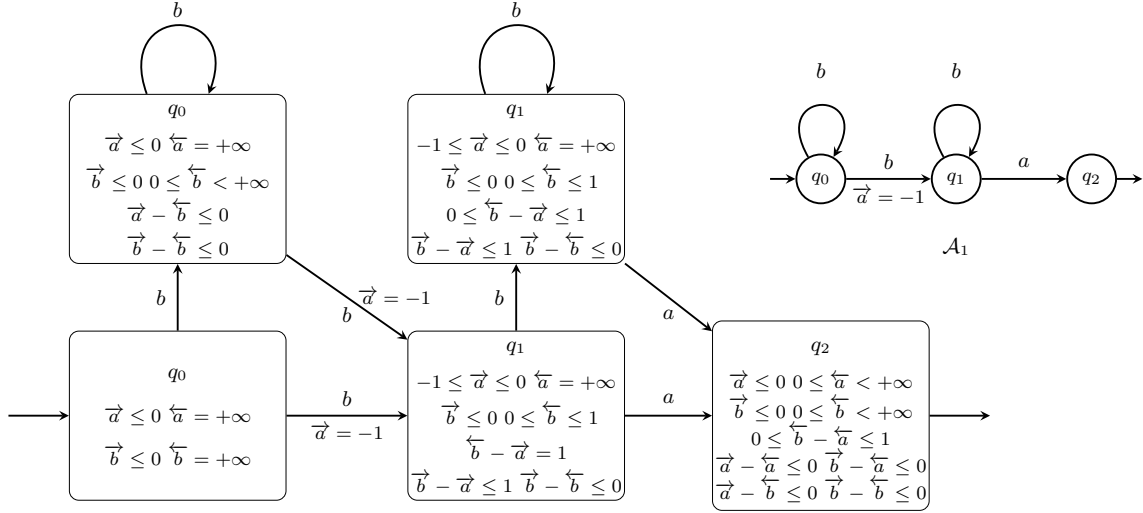
\begin{figure}[!h]
  \centering	
	\scalebox{0.9}{
    \begin{tikzpicture}
      [state/.style={draw, thick, circle, inner sep=4pt}]
      \begin{scope}[xshift=3.5cm, yshift=0.5cm]
      \begin{scope}[every node/.style={state}]
        \node (q0) at (4, 3) {\scriptsize $q_0$}; 
        \node (q1) at (6, 3) {\scriptsize $q_1$}; 
        \node (q2) at (8, 3) {\scriptsize $q_2$}; 
      \end{scope}	
      
      \node at (6, 2) {\footnotesize $\Aa_1$}; 

      \begin{scope}[->, >=stealth, thick]
        \draw (3.25, 3) to (q0); 
        \draw (q0) to [out=120,in=60,looseness=8] (q0);
        \draw (q1) to [out=120,in=60,looseness=8] (q1);  
        \draw (q0) to (q1); 
        \draw (q1) to (q2); 
        \draw (q2) to (8.75,3); 
      \end{scope}
      
      \node at (4, 4.6) {\footnotesize $b$}; 
      \node at (6, 4.6) {\footnotesize $b$}; 
      \node at (5, 3.25) {\footnotesize $b$}; 
      \node at (5, 2.75) {\scriptsize $\prophecy{a}=-1$}; 

      \node at (7, 3.25) {\footnotesize $a$}; 

    \end{scope}

    \begin{scope}[xshift=-2cm, yshift=0cm]

      \begin{scope}[xshift=0cm, yshift=0cm, zone/.style={draw,rectangle,minimum width=3.2cm, minimum height=2.5cm, rounded corners}]
        \node [zone] (z0) at (0, 0) {}; 
        \node [zone] (z1) at (0, 3.5) {};
        \node [zone] (z2) at (5, 0) {}; 
        \node [zone] (z3) at (5, 3.5) {}; 
        
      \end{scope}
      
      \begin{scope}[xshift=0cm, yshift=0cm, zone/.style={draw,rectangle,minimum width=3.2cm, minimum height=2.8cm, rounded corners}]
        \node [zone] (z4) at (9.5, 0) {};
      \end{scope}

      \begin{scope}[->, >=stealth, thick]
        \draw (-2.5, 0) to (z0); 
        \draw (z0) to (z1); 
      \draw (z0) to (z2);
      \draw (z1) to (z2);
      \draw (z2) to (z3);
      \draw (z2) to (z4);
      \draw (z3) to (z4);
      \draw (z1) to [out=110,in=70,looseness=4] (z1);
      \draw (z3) to [out=110,in=70,looseness=4] (z3);
      \draw (z4) to (12,0);
      \end{scope}
      \node at (0, 0.6) {\footnotesize $q_0$}; 
      \node at (0, 0) {\scriptsize $\prophecy{a} \leq 0~~~~\history{a} = +\infty$}; 
      \node at (0, -0.6) {\scriptsize $\prophecy{b} \leq 0~~~~\history{b} = +\infty$}; 

      \node at (0, 4.5) {\footnotesize $q_0$}; 
      \node at (0, 4) {\scriptsize $\prophecy{a} \leq 0~~~~\history{a} = +\infty$}; 
      \node at (0, 3.5) {\scriptsize $\prophecy{b} \leq 0~~~~0 \leq \history{b} < +\infty$}; 
      \node at (0, 3) {\scriptsize $\prophecy{a} - \history{b} \leq 0$}; 
      \node at (0, 2.5) {\scriptsize $\prophecy{b} - \history{b} \leq 0$};

      \node at (5, 1) {\footnotesize $q_1$}; 
      \node at (5, 0.5) {\scriptsize $-1 \leq \prophecy{a} \leq 0~~~~\history{a} = +\infty$}; 
      \node at (5, 0) {\scriptsize $~~\prophecy{b} \leq 0~~~~~0 \leq \history{b} \leq 1$}; 
      \node at (5, -0.5) {\scriptsize $\history{b} - \prophecy{a} = 1$}; 
      \node at (5, -1) {\scriptsize $\prophecy{b} - \prophecy{a} \leq 1~~~\prophecy{b} - \history{b} \leq 0$};

      \node at (5, 4.5) {\footnotesize $q_1$}; 
      \node at (5, 4) {\scriptsize $-1 \leq \prophecy{a} \leq 0~~~~\history{a} = +\infty$}; 
      \node at (5, 3.5) {\scriptsize $~~\prophecy{b} \leq 0~~~~~0 \leq \history{b} \leq 1$}; 
      \node at (5, 3) {\scriptsize $0 \leq \history{b} - \prophecy{a} \leq 1$}; 
      \node at (5, 2.5) {\scriptsize $\prophecy{b} - \prophecy{a} \leq 1~~~\prophecy{b} - \history{b} \leq 0$}; 

      \node at (9.5, 1) {\footnotesize $q_2$}; 
      \node at (9.5, 0.5) {\scriptsize $\prophecy{a} \leq 0~~~~~~0 \leq \history{a} < +\infty$}; 
      \node at (9.5, 0.1) {\scriptsize $\prophecy{b} \leq 0~~~~~~0 \leq \history{b} < +\infty$}; 
      \node at (9.5, -0.3) {\scriptsize $0 \leq \history{b} - \history{a} \leq 1$}; 
      \node at (9.5, -0.7) {\scriptsize $\prophecy{a} - \history{a} \leq 0~~\prophecy{b} - \history{a} \leq 0$}; 
      \node at (9.5, -1.1) {\scriptsize $\prophecy{a} - \history{b} \leq 0~~\prophecy{b} - \history{b} \leq 0$}; 

      \node at (-0.25, 1.75) {\footnotesize $b$}; 
      
      \node at (0, 6) {\footnotesize $b$}; 

      \node at (2.5, 0.25) {\footnotesize $b$}; 
      \node at (2.5, -0.25) {\scriptsize $\prophecy{a}=-1$}; 

      \node at (2.5, 1.5) {\footnotesize $b$}; 
      \node at (3.2, 1.75) {\scriptsize $\prophecy{a}=-1$}; 

      \node at (4.75, 1.75) {\footnotesize $b$}; 
      \node at (5, 6) {\footnotesize $b$}; 

      \node at (7.25, 1.5) {\footnotesize $a$}; 

      \node at (7.25, 0.25) {\footnotesize $a$}; 
    \end{scope}
  \end{tikzpicture} 
  } \caption[Example of event-zone graph]{The event-clock automaton $\Aa_1$ recognizes the
  language $\{b^n a \mid n \ge 1 \}$ such that there exists some $b$ which occurs exactly
  one time unit before $a$.  Its event-zone graph is on the left.}
  \label{fig:eca-zone-graph-ex-1}
\end{figure}

An example of ECA with its event-zone graph is given in Figure~\ref{fig:eca-zone-graph-ex-1}. 
We use some shorthands when writing some constraints.  For instance, we write
$\history{b}-\prophecy{a}=1$ for the conjunction of $\history{b}-\prophecy{a}\leq 1$ and
$\prophecy{a}-\history{b}\leq -1$.\footnote{
  Notice that for $a,b,c\in\overline{\mathbb{R}}$ and ${\leqlt}\in\{\leq,<\}$, if $a-b\leqlt
  c$ then $-c\leqlt b-a$.  The converse is true when $a,b,c$ are finite, but not in general.
  For instance, when $a=b\in\{-\infty,+\infty\}$ and the weight
  $(\leqlt,c)\in\mathcal{C}\setminus\{(<,-\infty),(\leq,+\infty)\}$ is non-trivial, then
  $-c\leqlt b-a=+\infty$ but $a-b=+\infty\not\leqlt c$.}
Another example is $0\leq\history{b}-\prophecy{a}\leq 1$ which stands for the conjunction of
$\history{b}-\prophecy{a}\leq 1$ and $\prophecy{a}-\history{b}\leq 0$.

Similar to the case of timed automata, the event-zone graph can be used to decide
reachability.  The next lemma follows by a straightforward adaptation of the corresponding
proof~\cite{Daws} from timed automata.

\begin{prop}
  The event-zone graph of an ECA is sound and complete for reachability.
\end{prop}

However, as in the case of zone graphs for timed automata, the event-zone graph for an ECA
is also not guaranteed to be finite.  In fact, in Geeraerts et
al.~\cite{GeeraertsRS11,GeeraertsRS14}, it was shown that there are ECAs for which there is
no finite time-abstract bisimulation relation on valuations, e.g., the ECA $\A_2$ in
Figure~\ref{fig:eca-zone-graph-ex-2}.

In the rest of this section, we will define what a simulation is and see how it can be
used to get a finite truncation of the event-zone graph for ECAs, which is still sound and
complete for reachability.

\begin{defi}[\textbf{Simulation}]
  A simulation relation on the semantics of an ECA is a reflexive,
  transitive relation $(q, v) \preceq (q, v')$ relating configurations
  with the same control state and (1) for every
  $(q, v) \xra{\delta} (q, v+\delta)$, we have
  $(q, v') \xra{\delta} (q, v'+\delta)$ and
  $(q, v+\delta) \preceq (q, v'+ \delta)$, (2) for every transition
  $t$, if $(q, v) \xra{t} (q_1, v_1)$ for some valuation $v_{1}$, then
  $(q, v') \xra{t} (q_1, v'_1)$ for some valuation $v'_{1}$ 
  with $(q_1, v_1) \preceq (q_1, v'_1)$.

  For two event-zones $Z, Z'$, we say $(q, Z) \preceq (q, Z')$ if for
  every $v \in Z$ there exists $v' \in Z'$ such that
  $(q, v) \preceq (q, v')$. 
    Adopting the terminology from ~\cite{GastinMS18,Gastin0S19,Gastin0S20}, we say that the simulation $\preceq$ is 
  {\em finite}
  if for every sequence $(q_{1}, Z_1), (q_{2}, Z_2), \dots$ of \emph{reachable} nodes (i.e., those having a path from initial node),  there exists $j > i$ such that $(q_{j}, Z_j) \preceq (q_{i}, Z_i)$. That is, the converse or transpose of $\preceq$ is a well-quasi order over reachable nodes, and thus the set of downward closed sets is finite.
\end{defi}

The reachability algorithm enumerates the nodes of the event-zone
graph and uses $\preceq$ to truncate nodes that are smaller with
respect to the simulation.

\begin{defi}[\textbf{Reachability algorithm}] \label{def:reach-algo}
  Let $\Aa$ be an ECA and
  $\preceq$ a finite simulation for $\Aa$. Add the initial node of the
  event-zone graph $(q_0, Z_0)$ to a Waiting list. Repeat the following
  until Waiting list is empty:
  \begin{itemize}
    \item Pop a node $(q, Z)$ from the Waiting list and add it to the
    Passed list.
    \item For every $(q, Z) \xra{t} (q_1, Z_1)$: if there exists a
    $(q_1, Z'_1)$ in the Passed or Waiting lists such that
    $(q_1, Z_1) \preceq (q_1, Z'_1)$, discard $(q_1, Z_1)$; else add
    $(q_1, Z_1)$ to the Waiting list.
  \end{itemize}
  If some accepting node is reached,
  the algorithm terminates and returns a Yes.
  Else, it continues until there are no further nodes to be explored and returns a No answer.
\end{defi}

The correctness of the reachability algorithm follows
from the fact that $\preceq$ is a simulation relation. 
Moreover, termination is guaranteed when the simulation used is finite.

\begin{thm}
  An ECA has an accepting run iff the reachability algorithm returns Yes.
\end{thm}

We have now presented the framework for the simulation approach in its entirety. However, to make it functional, we will need the following.
\begin{enumerate}
  \item An efficient representation for event-zones and algorithms to
  compute successors.
  \item A concrete simulation relation $\preceq$ for ECA with an
  efficient simulation test $(q, Z) \preceq (q, Z')$.
  \item A proof that $\preceq$ is finite, to
  guarantee termination of the reachability algorithm.  
\end{enumerate}
In the rest of the paper, we show how these can be achieved.  To start with, for standard
timed automata, zones are represented using Difference-Bound-Matrices
(DBMs)~\cite{Dill89}.  For such a representation to work on event-zones, we will need to
incorporate the fact that valuations can now take $+\infty$ and $-\infty$.  In
Section~\ref{sec:DBM}, we propose a way to merge $+\infty$ and $-\infty$ seamlessly into
the DBM technology.  In the subsequent Section~\ref{sec:simulation}, we define a
simulation for ECA based on $\Gg$-simulation, develop some technical machinery and present
an efficient simulation test.  Finally, in Section~\ref{sec:termination}, we deal with the
main problem of showing finiteness.  For this, we prove some non-trivial
invariants on the event-zones that are reachable in ECA and use them to show a surprising
property regarding prophecy clocks.  More precisely, we show that constraints involving 
prophecy clocks in reachable event-zones come from a finite set depending on the maximal 
constant of the ECA only.


\section{Computing with event-zones and distance graphs}\label{sec:DBM}

We now show that event-zones can be represented using
Difference-Bound-Matrices (DBMs) and the operations required for the reachability
algorithm can be implemented using DBMs. Each entry in a DBM encodes a
constraint of the form $x - y \leqlt c$. For timed automata analysis,
the entries could be $(<,+\infty)$ or $(\leqlt, c)$ with $c \in \mathbb{R}$ 
and ${\leqlt} \in \{ <, \leq \}$.  For ECA, we 
need to deal with valuations $+\infty$ or $-\infty$.  For this purpose, we 
use more general weights as introduced in Definition~\ref{def:weights}
and we extend the algebra of weights to the new entries in a natural
way. However, we need to rework the basic results on DBMs in our
  extended setting, since as we have seen in
  Remark~\ref{rem:extended-addition}, some seemingly obvious properties
  that hold for finite weights do not carry over to the extended
  weights. It turns out that even with the restricted versions of the
  properties that continue to hold, as
  listed in Remark~\ref{rem:extended-addition}, we can achieve the
  same results in the extended DBMs. This is the subject of this section.

\subsection{Extending the algebra on weights}

\begin{defi}[Order and sum of weights]\label{def:order-sum-weights}
  Let $(\leqlt,c), (\leqlt',c')\in\mathcal{C}$ be weights.
  
  \noindent\textbf{Order.} Define
  $(\leqlt, c) < (\leqlt', c')$ when either (1) $c < c'$, or (2)
  $c = c'$ and $\leqlt$ is $<$ while $\leqlt'$ is $\leq$.  This is a
  total order with
  $(<,-\infty) < (\le, -\infty) < (\leqlt, c) < (<,+\infty) < (\le,
  +\infty)$ for all $c \in \mathbb{R}$.
 
  \smallskip\noindent\textbf{Sum.}  We define the \emph{commutative}
  sum operation as follows.
  \begin{align*}
    (<,-\infty)+\alpha &= (<,-\infty) 
    &&\text{if } \alpha\in\mathcal{C}
    \\
    (\leq,+\infty)+\alpha &= (\leq,+\infty) 
    &&\text{if } \alpha\in\mathcal{C}\setminus\{(<,-\infty)\}
    \\
    (\leq,-\infty)+\alpha &= (\leq,-\infty) 
    &&\text{if } \alpha\in\mathcal{C}\setminus\{(<,-\infty),(\leq,+\infty)\}
    \\
    (<,+\infty)+\alpha &= (<,+\infty)
    &&\text{if } \alpha\in\mathcal{C}\setminus\{(<,-\infty),(\leq,-\infty),(\leq,+\infty)\}
    \\
    (\leqlt,c) + (\leqlt',c') &= (\leqlt'',c+c')
    &&\text{if } c,c'\in\mathbb{R} \text{ and } 
       {\leqlt''}={\leq} \text{ if } {\leqlt} = {\leqlt}' = {\leq}
       \text{ and } {\leqlt''}={<} \text{ otherwise.}
  \end{align*}
  Notice that the sum of weights is an associative operation and
  $\alpha+(\leq,0)=\alpha$ for all $\alpha\in\mathcal{C}$.
\end{defi}

\begin{rem}
Note that our set of weights $\mathcal{C}$ has a sequence of partially  absorbant elements with decreasing strength - $(<,-\infty)$ is an absorbant for
  $\mathcal{C}$, $(\leq,+\infty)$ is an absorbant for $\mathcal{C} \setminus
  \{(<,-\infty)\}$, $(\leq,-\infty)$ is an absorbant for $\mathcal{C} \setminus
  \{(<,-\infty),(\leq,+\infty)\}$, and $(<,+\infty)$ is an absorbant for $\mathcal{C}
  \setminus \{(<,-\infty), (\leq,+\infty), (\leq,-\infty)\}$.  This implies that all the sums
  are well-defined.
\end{rem}

The intuition behind the above definition of order is that when
$(\leqlt,c)<(\leqlt',c')$, the set of valuations that satisfies a
constraint $y - x \leqlt c$ is contained in the solution set of
$y - x \leqlt' c'$.  For the sum, the following lemma gives the idea
behind our choice of definition.

\begin{lem}\label{lem:sum-weights}
  Let $x, y, z\in X\cup\{0\}$ be event-clocks,
  $(\leqlt_1, c_1), (\leqlt_2, c_2)\in\mathcal{C}$ be weights and
  $(\leqlt, c) = (\leqlt_1, c_1) + (\leqlt_2, c_2)$.  For all
  valuations $v\in\V$, if $v\models y-x \leqlt_1 c_1$ and
  $v\models z-y \leqlt_2 c_2$, then $v\models z-x \leqlt c$.
\end{lem}

\begin{proof}
  When $(\leqlt,c)\in\{(<,-\infty),(\le,+\infty)\}$, this is clear.  We
  assume for the rest of the proof that
  $(\leqlt,c)\in\mathcal{C}\setminus\{(<,-\infty),(\le,+\infty)\}$ is
  non-trivial. Consider a valuation $v\in\V$ such that
  $v\models y-x \leqlt_1 c_1$ and $v\models z-y \leqlt_2 c_2$.
  
  From $v(y)-v(x)\leqlt_1 c_1$ and $(\leqlt_1,c_1)\neq(\leq,+\infty)$
  we deduce that $v(y)\neq+\infty$ and $v(x)\neq-\infty$.  Similarly,
  from $v(z)-v(y)\leqlt_2 c_2$ and $(\leqlt_2,c_2)\neq(\leq,+\infty)$
  we deduce that $v(z)\neq+\infty$ and $v(y)\neq-\infty$. Hence,
  $v(y)$ is finite.
  If $v(x)=+\infty$, or $v(x)<+\infty$ and $v(z)=-\infty$ then
  $v(z)-v(x)=-\infty$, hence $v\models z-x \leqlt c$.
  
  The remaining case is when $v(x),v(y),v(z)$ are all finite.  This
  rules out $(\leqlt,c)=(\leq,-\infty)$.  If $(\leqlt,c)=(<,+\infty)$
  then $v\models z-x \leqlt c$.  Otherwise, $c=c_{1}+c_{2}$ is finite
  and $v(z)-v(x)=v(z)-v(y)+v(y)-v(x)\leqlt c_{2}+c_{1}$. This
  completes the proof.
\end{proof}

The following technical lemma will be useful in many proofs.

\begin{lem}\label{lem:weight-properties}\hfill
  \begin{enumerate}
  \item Let $(\leqlt,c)$ be a weight and
    $\alpha\in\overline{\mathbb{R}}$. Then,
    \begin{itemize}
    \item $\alpha\leqlt c$ iff $(\leq,\alpha)\leq(\leqlt,c)$ iff
      $(\leq,0)\leq(\leq,-\alpha)+(\leqlt,c)$,
      
    \item $\alpha\not\leqlt c$ iff $(\leqlt,c)<(\leq,\alpha)$ iff
      $(\leq,-\alpha)+(\leqlt,c)<(\leq,0)$ iff
      $(\leq,-\alpha)+(\leqlt,c)\leq(<,0)$.
    \end{itemize}
    
  \item Let $(\leqlt,c),(\leqlt',c'),(\leqlt'',c'')$ be weights with
    $(\leq,0)\leq(\leqlt,c)+(\leqlt',c')$.
    
    Then, there exists $\alpha\in\overline{\mathbb{R}}$ such that
    $\alpha\leqlt c$ and $-\alpha\leqlt' c'$.
    
    If in addition we have $(\leqlt'',c'')<(\leqlt,c)$ then there
    exists such an $\alpha$ with $\alpha\not\leqlt'' c''$.
  \end{enumerate}
\end{lem}

\begin{proof}\hfill
  \begin{enumerate}
  \item The first item is easy to prove, checking separately the cases
    where $\alpha$ or $c$ is infinite. The second item follows from
    the first one using the fact that $\leq$ is a total order on
    weights.
  
  \item Assume first that ${\leqlt}={\leq}$.  Take $\alpha=c$.  Using
    1, we easily check that $\alpha$ satisfies the desired properties.
    
    Next, assume that ${\leqlt}={<}$. From
    $(\leq,0)\leq(\leqlt,c)+(\leqlt',c')$ we deduce that
    $c\neq-\infty\neq c'$. Next, we see that $-c'<c$ (consider
    separately the cases $c=+\infty$ or $c'=+\infty$). Take $\alpha$
    such that $-c'<\alpha<c$.  We obtain $\alpha\leqlt c$ and
    $-\alpha<c'$. The latter inequality implies $-\alpha\leqlt' c'$.
    
    If in addition we have $(\leqlt'',c'')<(<,c)$, then $c''<c$ and we
    take $\alpha$ such that $\max(-c',c'')<\alpha<c$.  
    As above, we have $\alpha\leqlt c$ and $-\alpha\leqlt' c'$.  
    We also get $\alpha\not\leqlt'' c''$.
    \qedhere%
  \end{enumerate}
\end{proof}

Equipped with the weights and the arithmetic over it, we will work
with a graph representation of zones (called distance graphs), instead
of matrices (i.e., DBMs), since this makes the analysis more
convenient.  We wish to highlight that our definition of weights,
order and sum have been chosen to ensure that this notion of distance
graphs remains identical to the one for usual TA. As a consequence, we
are able to adapt many of the well-known properties about distance
graphs for ECA.

\subsection{Distance graphs over the extended algebra}

\begin{defi}[Distance graphs]\label{def:distance-graph}
  A distance graph $\GG$ is a weighted directed graph without self-loops, with vertex set
  being $X\cup\{0\}=X_P \cup X_H \cup \{0\}$, edges being labeled with weights from
  $\mathcal{C}\setminus\{(<,-\infty)\}$.\footnote{
    If we allowed an edge with weight $\GG_{xy}=(<,-\infty)$ then we would get
    $\sem{\GG}=\emptyset$ since the constraint $y-x<-\infty$ is equivalent to false.}
  We define $\sem{\GG} := \{ v\in\V \mid v \models y - x \leqlt c \text{ for all edges } 
  x \xra{\leqlt\,c} y \text{ in } \GG \}$.
  The weight of edge $x\to y$ is denoted $\GG_{xy}$ and we set 
  $\GG_{xy}=(\leq,+\infty)$ if there is no edge $x\to y$.
  
  We say that $\GG$ is in standard form if it satisfies the following conditions;
  \begin{enumerate}
    \item $\GG_{0x}\leq(\leq,0)$ for all $x\in X_{P}$ and
    $\GG_{x0}\leq(\leq,0)$ for all $x\in X_{H}$.

    \item For all $x,y\in X$, if $\GG_{xy}\neq(\leq,+\infty)$ 
    then $\GG_{x0}\neq(\leq,+\infty)$ and $\GG_{0y}\neq(\leq,+\infty)$.
  \end{enumerate}
  
  We extend the order on weights to distance graphs pointwise:
  Let $\GG$, $\GG'$ be distance graphs, we write $\GG\leq\GG'$ when 
  $\GG_{xy}\leq\GG'_{xy}$ for all edges $x\to y$. Notice that this implies 
  $\sem{\GG}\subseteq\sem{\GG'}$.
\end{defi}

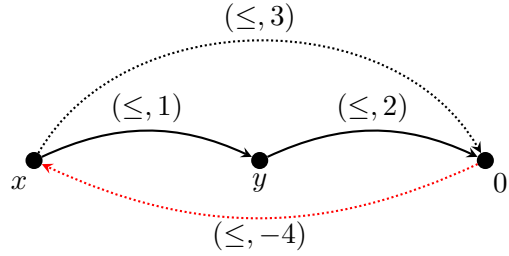
\begin{figure}
	\centering
    \begin{tikzpicture}
        \begin{scope}[state/.style={draw, thick, circle, inner sep=2pt}]
          \node [state, fill=black] (p) at (2, 6) {}; 
          \node [state, fill=black] (q) at (5, 6) {}; 
          \node [state, fill=black] (r) at (8, 6) {}; 
        \end{scope}
        
        \begin{scope}[->, >=stealth, thick]
          \draw [bend left=25] (p) to (q); 
          \draw [bend left=25] (q) to (r); 
        \end{scope}
  
        \begin{scope}[->, >=stealth, densely dotted, thick]
          \draw [color=red, bend left=25] (r) to (p); 
          \draw [bend left=60] (p) to (r); 
        \end{scope}
  
        \node at (1.8,5.7)  {$x$};
        \node at (5,5.7)  {$y$};
        \node at (8.2,5.7)  {$0$};
    
        \node at (3.5,6.7)  {$(\leq, 1)$};
        \node at (6.5,6.7)  {$(\leq, 2)$};
        \node at (5,7.9)  {$(\leq, 3)$};
        \node at (5,5)  {$(\leq,-4)$};
    \end{tikzpicture}  
    \caption[]{
    An example of a distance graph for timed automata with clocks $x, y$.  The dotted edge in black denotes that the edge is obtained by canonicalization, and the dotted edge in red depicts an edge introduced by a guard.}
    \label{fig:dist-graph-ta}
\end{figure}

Here is an example to illustrate the need for a standard form.
Suppose that valuations are \emph{finite} and that we have constraints: $y - x \le 1$ and
$-y \le 2$.  The distance graph representing these constraints is depicted with black
edges in Figure~\ref{fig:dist-graph-ta}.  From these constraints, we can infer $-x\le 3$
just by adding the inequalities - this corresponds to the edge depicted by dotted lines
from $x$ to $0$.  If there was another constraint $x\le -4$ (denoted by the red dotted
edge from $0$ to $x$), we will get unsatisfiability,
witnessed by a negative cycle $0 \xra{\le -4} x \xra{\le 3} 0$, as depicted by the dotted
edges in Figure~\ref{fig:dist-graph-ta}.  Basically, adding the weights of $x \to y$ and
$y \to 0$, we get the strongest possible constraint about $x \to 0$.

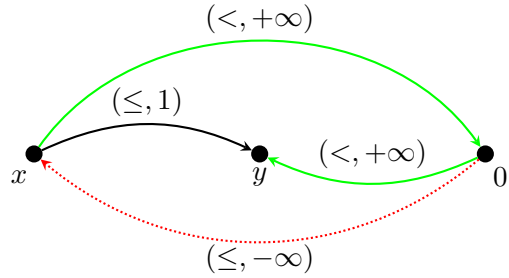
\begin{figure}
	\centering
    \begin{tikzpicture}
        \begin{scope}[state/.style={draw, thick, circle, inner sep=2pt}]
          \node [state, fill=black] (p) at (2, 6) {}; 
          \node [state, fill=black] (q) at (5, 6) {}; 
          \node [state, fill=black] (r) at (8, 6) {}; 
        \end{scope}
        
        \begin{scope}[->, >=stealth, thick]
          \draw [bend left=25] (p) to (q); 
          \draw [color=green, bend left=55] (p) to (r); 
          \draw [color=green, bend left=25] (r) to (q); 
        \end{scope}

        \begin{scope}[->, >=stealth, densely dotted, thick]
          \draw [color=red, bend left=40] (r) to (p); 
        \end{scope}
  
        \node at (1.8,5.7)  {$x$};
        \node at (5,5.7)  {$y$};
        \node at (8.2,5.7)  {$0$};
    
        \node at (3.5,6.7)  {$(\leq, 1)$};
        \node at (5,4.6)  {$(\leq,-\infty)$};
        \node at (6.5,6)  {$(<, +\infty)$};
        \node at (5,7.8)  {$(<, +\infty)$};
    \end{tikzpicture}
    \caption[]{Distance graph with standardization using the extended algebra.  Note that the green edges are the edges induced by standardization.}
    \label{fig:dist-graph-gta}
\end{figure}

This holds no more in the extended algebra, due to the fundamental difference while adding weight $(\le, +\infty)$.
Consider the constraint $y-x\le 1$ 
corresponding to the edge $x\xra{\leq 1}y$ 
in the distance graph.
Recall that missing edges correspond to the constraint $(\leq,+\infty)$ which is
equivalent to \emph{true}, hence we have implicit edges $x\xra{\leq\infty}0$ and
$0\xra{\leq\infty}y$.  These are \emph{not} the strongest possible constraints on $x$ and
$y$ induced by $y-x\le 1$.  Indeed, any valuation $v$ satisfying $y-x\le 1$
should have $v(y)\neq\infty$ and $v(x)\neq-\infty$.
Now, if we also had a constraint $x-0\le -\infty$, corresponding to edge
$0\xra{\leq-\infty}x$, we have no negative cycle in the corresponding distance graph, as
can be seen by considering the part of the graph without green edges in
Figure~\ref{fig:dist-graph-gta}.  But, as explained above, the set of constraints $y-x\le 1$ and $x-0\le-\infty$ is not satisfiable.

In order to get a correspondence between negative cycles and empty solution
sets, we propose the standard form. The standard form 
equips the graph with the additional information that when $y - x$ is
bounded by a finite value, that is, edge $x \to y$ does not have weight
$(\le, +\infty)$, the constraints $0 - x$ and $y-0$ are at most $(<,+\infty)$ - the
edges induced by standardization are depicted in green in Figure~\ref{fig:dist-graph-gta}.
With this information, we get negative cycles whenever there is a contradiction.

Now, each distance graph $\GG$ can be transformed into an equivalent distance 
graph $\GG'$ which is in standard form. By equivalent, we mean $\sem{\GG}=\sem{\GG'}$.
First, we set $\GG'_{0x}=\min(\GG_{0x},(\leq,0))$ for $x\in X_{P}$ and
$\GG'_{x0}=\min(\GG_{x0},(\leq,0))$ for $x\in X_{H}$.
Moreover, if $x\in X_{P}$ then we set $\GG'_{x0}=\min(\GG_{x0},(<,+\infty))$ if
$\GG_{xy}\neq(\leq,+\infty)$ for some $y\neq x$, otherwise we keep $\GG'_{x0}=\GG_{x0}$.
Similarly, if $y\in X_{H}$ then we set $\GG'_{0y}=\min(\GG_{0y},(<,+\infty))$ if
$\GG_{xy}\neq(\leq,+\infty)$ for some $x\neq y$, otherwise we keep $\GG'_{0y}=\GG_{0y}$.
Finally, for $x,y\in X$ with $x\neq y$ we set $\GG'_{xy}=\GG_{xy}$.
The graph $\GG'$ constructed above is called the standardization of $\GG$.

\begin{lem}\label{lem:standard-form}
  The standardization $\GG'$ of a distance graph $\GG$ is in standard form and
  $\sem{\GG'}=\sem{\GG}$.
\end{lem}

\begin{proof}
  The graph $\GG'$ is in standard form by construction. Clearly, $\GG'\leq\GG$, hence
  $\sem{\GG'}\subseteq\sem{\GG}$. Now, let $v\in\sem{\GG}$. We show that $v$ satisfies 
  all constraints in $\GG'$. Let $x\in X_{P}$ be a prophecy clock. We have $v(x)\leq 0$. 
  Hence $v$ satisfies the constraint $\GG'_{0x}=\min(\GG_{0x},(\leq,0))$.
  Assume now that $\GG_{xy}=(\leqlt,c)<(\leq,+\infty)$ for some $y\neq x$.  Since
  $v\in\sem{\GG}$, we have $v(y)-v(x)\leqlt c$, which implies $v(x)\neq-\infty$.  
  Hence $v$ satisfies the constraint $\GG'_{x0}=\min(\GG_{x0},(<,+\infty))$.
  We proceed similarly for the new constraints when $x\in X_{H}$ is a
  history clock.
\end{proof}

\begin{defi}
  The weight of a path in a distance graph $\GG$ is the sum of the weights of its edges.  A
  cycle in $\GG$ is said to be negative if its weight is strictly less than $(\le, 0)$.
  
  A distance graph $\GG$ is in \emph{normal form} if it is in standard form, has no
  negative cycles, and the weight of each edge $x \to y$ is not greater than the weight of
  any path from $x$ to $y$.
\end{defi}

A standard distance graph with no negative cycles can be normalized.
Let $\GG$ be a standard distance graph with no negative cycles.  
Let $\GG'$ be the graph where the weight of an edge $x\to y$ is set to the
minimum of the weights of the paths from $x$ to $y$ in $\GG$.\footnote{
  When $\GG$ has no negative cycles, the weight of a non-simple path $x\cdots z\cdots
  z\cdots y$ is at least the weight of the subpath $x\cdots z\cdots y$. Hence, we may 
  restrict to simple paths when taking the minimum.} 
The graph $\GG'$ constructed above is called the \emph{normalization} of $\GG$.

\begin{lem}[Normalization]\label{lem:normalization}
  Let $\GG$ be a standard distance graph with no negative cycles.
  The normalization $\GG'$ of $\GG$ is in normal form and $\sem{\GG}=\sem{\GG'}$.
\end{lem}

\begin{proof}
  First, we check that $\GG'$ is in standard form.
  Let $x\in X_{P}$.  We have $\GG'_{0x}\leq\GG_{0x}\leq(\leq,0)$.  Now assume that
  $\GG'_{x0}=(\leq,+\infty)$.  Then $\GG'_{x0}\leq\GG_{x0}=(\leq,+\infty)$ and
  $\GG_{xy}=(\leq,+\infty)$ for all $y\neq x$.  Since all paths in $\GG$ from
  $x$ to some $y\neq x$ start with some edge $x\to z$ of weight $(\leq,+\infty)$, we obtain
  $\GG'_{xy}=(\leq,+\infty)$ for all $y\neq x$.
  The condition for history clocks is proved similarly.
  
  It is easy to see that $\GG'$ is in normal form.
  We have $\GG'\leq\GG$ by construction, hence $\sem{\GG'}\subseteq\sem{\GG}$. 
  Conversely, let $v\in\sem{\GG}$ and let $x_1 \to x_2 \to \cdots \to x_n$ be an arbitrary
  path in $\GG$ with weight $(\leqlt,c)$.
  As $v \in \sem{\GG}$, we have $v\models x_{i+1}-x_i \leqlt_i c_i$
  with $\GG_{x_{i}x_{i+1}}=(\leqlt_i, c_i)$.
  By Lemma~\ref{lem:sum-weights}, we get $v\models x_{n}-x_{1}\leqlt c$.
  We deduce that $v\in\sem{\GG'}$. 
\end{proof}

Given two distance graphs $\GG'$, $\GG''$, we define $\GG=\min(\GG',\GG'')$ as the
distance graph obtained by setting $\GG_{xy}=\min(\GG'_{xy},\GG''_{xy})$ for all $x\neq
y$.  Notice that $\sem{\GG}=\sem{\GG'}\cap\sem{\GG''}$.  Moreover, if both $\GG'$ and
$\GG''$ are in standard form, then so is $\min(\GG',\GG'')$.  On the other hand, even when
$\GG'$ and $\GG''$ are in normal form, $\min(\GG',\GG'')$ need not be in normal
form.

\medskip
We will make use of important properties, which have been shown when weights come from
$\mathcal{C} \setminus \{(<,-\infty),(\le,-\infty),(\le,+\infty)\}$, and continue to hold
even with the new weights.
First, we show that a classical and crucial property of distance graphs for timed
automata extends to the \emph{standard} distance graphs over the extended algebra that we
have defined above for event-clock automata.

\begin{lem}
  \label{lem:empty-iff-negative-cycle}
  Let $\GG$ be a \emph{standard} distance graph. Then, $\sem{\GG}\neq\emptyset$ iff
  $\GG$ has no negative cycles.
\end{lem}

\begin{proof}
  Suppose $\sem{\GG}$ is non-empty. Let $v \in \sem{\GG}$. 
  Let $x_1 \to x_2 \to \cdots \to x_n \to x_{n+1}=x_1$ be an arbitrary cycle in $\GG$
  and let $(\leqlt,c)$ be the weight of the cycle. 
  As $v \in \sem{\GG}$, we have $v\models x_{i+1}-x_i \leqlt_i c_i$
  with $(\leqlt_i, c_i)$ the weight of $x_i \to x_{i+1}$ in $\GG$. 
  By Lemma~\ref{lem:sum-weights}, we deduce that $v\models x_{n+1}-x_{1}\leqlt c$.
  Since $x_{n+1}=x_{1}$, we get $v(x_{n+1})-v(x_{1})\in\{0,+\infty\}$.
  Hence, the cycle is not negative.  
  
  Conversely, let $\GG$ be a standard distance graph with no negative cycles.  We will
  tighten the distance graph, progressively fixing the value of each clock, while
  preserving a standard graph with no negative cycles.  
  By Lemma~\ref{lem:normalization}, we assume w.l.o.g. that $\GG$ is in normal form.  
  
  Let $x\in X$ be a clock.  Considering the cycle $0\to x\to 0$, we get
  $(\leq,0)\leq\GG_{0x}+\GG_{x0}$.  By Lemma~\ref{lem:weight-properties}, we find
  $\alpha\in\overline{\mathbb{R}}$ with $(\leq,\alpha)\leq\GG_{0x}$ and
  $(\leq,-\alpha)\leq\GG_{x0}$.
  Notice that $0\leq\alpha$ if $x$ is a history clock and $\alpha\leq0$ if $x$ is a
  prophecy clock.
  
  Define the graph $\GG'$ obtained from $\GG$ by setting
  $\GG'_{0x}=(\leq,\alpha)$, $\GG'_{x0}=(\leq,-\alpha)$ and keeping the other weights
  unchanged.  We can easily check that $\GG'$ is in standard form.
  We claim that $\GG'$ has no negative cycles.  Cycles using only weights from $\GG$ are
  non-negative since $\GG$ has no negative cycles.  Consider now a cycle $0\to x \cdots \to0$.  If $x\to \cdots \to 0$ consists of a single edge, then the weight of the cycle in $\GG'$ is
  $(\leq,\alpha)+(\leq,-\alpha)$ which is either $(\leq,0)$ or $(\leq,+\infty)$ and hence 
  non-negative.  Otherwise, the path $x \cdots 0$ only uses weights from $\GG$.  Let
  $(\leqlt,c)$ be its weight.  We have $\GG_{x0}\leq(\leqlt,c)$ since $\GG$ is in normal
  form.  Now, using $(\leq,-\alpha)\leq\GG_{x0}$ and Lemma~\ref{lem:weight-properties}, we
  get $(\leq,0)\leq(\leq,\alpha)+(\leqlt,c)$ showing that the cycle is not negative.  We
  proceed similarly for cycles of the form $0 \to \cdots x\to 0$.

  The graph $\GG'$, denoted $\GG[x=\alpha]$, is obtained from $\GG$ by fixing the
  value of $x$ to $\alpha$.  By construction we have $\GG'\leq\GG$, hence
  $\sem{\GG'}\subseteq\sem{\GG}$.  Moreover, $v(x)=\alpha$ for all
  $v\in\sem{\GG'}$.
  
  Since $\GG'$ is standard and has no negative cycles, we can repeat the process, fixing
  the value of each clock in $X$ one after the other.  At the end of the process, we
  obtain a standard distance graph $\GG''$ with no negative cycles and where the value of
  each clock $x\in X$ has been fixed to some $\alpha_{x}\in\overline{\mathbb{R}}$.  We
  still have $\sem{\GG''}\subseteq\sem{\GG}$.  
  
  Let $v\in\V$ be defined by $v(x)=\alpha_{x}$ for all $x\in X$.  We show that
  $v\in\sem{\GG''}$.  Clearly, $v$ satisfies all the constraints
  $\GG''_{0x}=(\leq,\alpha_{x})$ and $\GG''_{x0}=(\leq,-\alpha_{x})$.  Now, consider an
  edge $x\to y$ with weight $\GG''_{xy}=(\leqlt,c)$.
  If $\alpha_{y}=+\infty$ then $y$ is a history clock and $\GG''_{0y}=(\leq,+\infty)$.
  Since $\GG''$ is standard, we deduce that $\GG''_{xy}=(\leq,+\infty)$.  Similarly, if
  $\alpha_{x}=-\infty$ then $x$ is a prophecy clock, $\GG''_{x0}=(\leq,+\infty)$ and
  $\GG''_{xy}=(\leq,+\infty)$. In both cases we get $v\models y-x\leqlt c$.
  We assume below that $\alpha_{y}\neq+\infty$ and $\alpha_{x}\neq-\infty$.
  The cycle $x\to y\to 0\to x$ is non-negative and has weight
  $(\leqlt,c)+(\leq,-\alpha_{y})+(\leq,\alpha_{x})=
  (\leqlt,c)+(\leq,\alpha_{x}-\alpha_{y})$.  We deduce that
  $(\leq,-(\alpha_{x}-\alpha_{y}))\leq(\leqlt,c)$.  Recall that $-(b-a)=a-b$ unless
  $a=b\in\{-\infty,+\infty\}$.  Since $\alpha_{y}\neq+\infty$ and $\alpha_{x}\neq-\infty$
  we obtain $(\leq,\alpha_{y}-\alpha_{x})\leq(\leqlt,c)$, which implies $v\models
  y-x\leqlt c$. This concludes the proof.
\end{proof}

We will now establish a stronger result, which will be used in several proofs, which will also show that a distance graph in normal form is \emph{canonical}, in a sense that we explain below.
The result (which generalizes the one from timed automata) intuitively says that if a
distance graph is in normal form, then for every pair of clocks $x,y\in X\cup\{0\}$,
there is a valuation $v$ in its semantics which assigns time-stamps to $x$ and $y$ such
that their difference (in the extended algebra) is a value arbitrarily close to the weight
of the edge $x\to y$.
Formally,

\begin{lem}
  \label{lem:fixing-values-distance-graph}
  Let $\GG$ be a distance graph in normal form, $x,y\in X\cup\{0\}$ be a pair of distinct clocks and $\alpha\in\overline{\mathbb{R}}$.  There is a valuation $v\in\sem{\GG}$ with
  $v(y)-v(x)=\alpha$ if and only if
  \begin{enumerate}
    \item $(\leq,\alpha)\leq\GG_{xy}$ and $(\leq,-\alpha)\leq\GG_{yx}$, and
  
    \item if $x,y\in X$ and $\alpha\in\mathbb{R}$ is finite then the weights 
    $\GG_{x0},\GG_{0x},\GG_{y0},\GG_{0y}$ are all different from $(\leq,-\infty)$, and
  
    \item if $x,y\in X$ and $\alpha=-\infty$ then $\GG_{0x}\neq(\leq,-\infty)\neq\GG_{y0}$.
  \end{enumerate}
\end{lem}

\begin{proof}
  We first show that the conditions are necessary. Let $v\in\sem{\GG}$ be a valuation with
  $v(y)-v(x)=\alpha$. 
  \begin{enumerate}
    \item  Let $\GG_{xy}=(\leqlt,c)$. We know that $v\models y-x\leqlt
      c$ and 
    we deduce that $\alpha\leqlt c$, which is equivalent to $(\leq,\alpha)\leq\GG_{xy}$.
    Let $\GG_{yx}=(\leqlt',c')$. We know that $v\models x-y\leqlt' c'$ and 
    we deduce that $v(x)-v(y)\leqlt' c'$. Recall that $-(b-a)=a-b$ unless
    $a=b\in\{-\infty,+\infty\}$, in which case $a-b=b-a=+\infty$.
    We deduce that $-\alpha\leq v(x)-v(y)\leqlt' c'$, which implies
    $(\leq,-\alpha)\leq\GG_{yx}$.
    
    \item  Assume that $\alpha\in\mathbb{R}$ is finite, then $v(x),v(y)\in\mathbb{R}$ 
    are both finite. Assume that $\GG_{x0}=(\leq,-\infty)$ (resp.\ 
    $\GG_{0x}=(\leq,-\infty)$). Then $v\in\sem{\GG}$ implies 
    $v(x)=+\infty$ (resp.\ $v(x)=-\infty$), a contradiction. Similarly, 
    $\GG_{y0},\GG_{0y}$ are not $(\leq,-\infty)$.
  
    \item Assume that $\alpha=-\infty$.  Then, $v(y)=-\infty\neq v(x)$ or
    $v(x)=+\infty\neq v(y)$.  Now, if $\GG_{0x}=(\leq,-\infty)$ then $v(x)=-\infty$, a
    contradiction.  Similarly, we must have $\GG_{y0}\neq(\leq,-\infty)$.
  \end{enumerate}

  Before proving the other direction, we give a short comparison
    with the timed automata setting. For distance graphs arising in
    classical timed automata, 
    Condition 1 is already sufficient to get the right-to-left
    direction: if we replace the edges $x \to y$ and $y \to
  x$ with $(\le, \alpha)$ and $(\le, -\alpha)$ respectively, we can
  deduce that the
  resulting graph has no negative cycles, and hence the semantics of
  this graph is non-empty. Any valuation $v$ in the
  semantics of the resulting graph has $v(y) - v(x) = \alpha$, which
  gives us the converse direction of the lemma.  On the other hand, in the current setting
with infinity weights, doing this same replacement might not result in
a graph in standard form. As we have seen earlier, for graphs not in
standard form, the absence of negative cycles
does not imply a non-empty solution set. This is why we need the second
and third conditions which indirectly allow to reason on a standard form for the
resulting graph, 
depending on the $\alpha$ that is chosen. Let us now prove this
direction.

  Assume that the three conditions are satisfied. We consider different cases.
  
  Assume first that $x=0$.  We fix the value of $y$ to $\alpha$ by
  considering the graph
  $\GG'=\GG[y=\alpha]$ defined in the proof of
  Lemma~\ref{lem:empty-iff-negative-cycle}.  We have seen that $\GG'$ is standard with no
  negative cycles.  By Lemma~\ref{lem:empty-iff-negative-cycle} we find
  $v\in\sem{\GG'}\subseteq\sem{\GG}$ and we have $v(y)=\alpha$.
  The case $y=0$ is handled similarly by considering $\GG'=\GG[x=-\alpha]$ which fixes the
  value of $x$ to $-\alpha$.
  For the rest of the proof, we assume that $x,y\in X$ and we
  distinguish three cases depending on $\alpha$.
  
  \medskip\noindent$\bullet$
  The first case is when $\alpha=+\infty$. Since $\GG$ is in normal form, we have 
  $(\leq,+\infty)\leq\GG_{xy}\leq\GG_{x0}+\GG_{0y}$. Hence, $\GG_{x0}=(\leq,+\infty)$ or 
  $\GG_{0y}=(\leq,+\infty)$. Assume that $\GG_{x0}=(\leq,+\infty)$. 
  We have $(\leq,-\infty)\leq\GG_{0x}$. 
  Hence we may fix the value of $x$ to $-\infty$ by taking $\GG'=\GG[x=-\infty]$.
  Applying Lemma~\ref{lem:empty-iff-negative-cycle} we find a valuation
  $v\in\sem{\GG'}\subseteq\sem{\GG}$. We have $v(x)=-\infty$. We deduce that 
  $v(y)-v(x)=+\infty=\alpha$. We proceed similarly when $\GG_{0y}=(\leq,+\infty)$, fixing 
  the value of $y$ to $+\infty$.
  
  \medskip\noindent$\bullet$
  The next case is when $\alpha\in\mathbb{R}$ is finite. To fix the value of $y-x$ to 
  $\alpha$ we define the graph $\GG'$ by setting
  $\GG'_{xy}=(\leq,\alpha)$, $\GG'_{yx}=(\leq,-\alpha)$,
  $\GG'_{x0}=\min(\GG_{x0},(<,+\infty))$, $\GG'_{0x}=\min(\GG_{0x},(<,+\infty))$,
  $\GG'_{y0}=\min(\GG_{y0},(<,+\infty))$, $\GG'_{0y}=\min(\GG_{0y},(<,+\infty))$, and the 
  weight of other edges $z\to z'$ are kept unchanged: $\GG'_{zz'}=\GG_{zz'}$.
  This graph $\GG'$ is denoted $\GG[y-x=\alpha]$.
  It is easy to see that $\GG'$ is in standard form. Since $\GG'\leq\GG$ we have 
  $\sem{\GG'}\subseteq\sem{\GG}$. Moreover, $v(y)-v(x)=\alpha$ for all $v\in\sem{\GG'}$.
  It remains to show that $\sem{\GG'}\neq\emptyset$, i.e., that $\GG'$ has no negative
  cycles.  Since $\GG$ is in normal form, we may restrict to cycles not using two
  consecutive weights from $\GG$.  In particular, cycles using only weights from $\GG$ are
  non negative.  It remains to consider the following cycles.
  \begin{itemize}
    \item  Cycle $x\to y\to x$ has weight $(\leq,\alpha)+(\leq,-\alpha)=(\leq,0)$ which 
    is non-negative.
  
    \item Cycle $0\to x\to 0$ with $\GG'_{0x}=(<,+\infty)$ or $\GG'_{x0}=(<,+\infty)$
    is not negative since we have $\GG_{x0}\neq(\leq,-\infty)\neq\GG_{0x}$.
    We argue similarly for cycle $0\to y\to 0$.
  
    \item  Cycle $0\to x\to y\to 0$ with $\GG'_{0x}=(<,+\infty)$ or $\GG'_{y0}=(<,+\infty)$
    is not negative since $\GG_{y0}\neq(\leq,-\infty)\neq\GG_{0x}$.  
    We argue similarly for cycle $0\to y\to x\to 0$.
  \end{itemize}
  Therefore, $\GG'$ has no negative cycles and we are done with the case 
  $\alpha\in\mathbb{R}$ finite.
  
  \medskip\noindent$\bullet$
  The last case is $\alpha=-\infty$. From $(\leq,-\alpha)\leq\GG_{yx}$ we get 
  $\GG_{yx}=(\leq,+\infty)$. Since $\GG$ is in normal form, 
  $\GG_{yx}\leq\GG_{y0}+\GG_{0x}$ and we deduce that $\GG_{y0}=(\leq,+\infty)$ or 
  $\GG_{0x}=(\leq,+\infty)$. 
  
  Assume that $\GG_{0x}=(\leq,+\infty)$.
  We define a graph $\GG'$ which fixes $x$ to $+\infty$ and forces $y$ to be different 
  from $+\infty$. We set $\GG'_{x0}=(\leq,-\infty)$, 
  $\GG'_{0y}=\min(\GG_{0y},(<,+\infty)$) and $\GG'_{zz'}=\GG_{zz'}$ otherwise.
  The graph $\GG'$ is denoted $\GG[x= +\infty,y<+\infty]$.
  It is easy to check that $\GG'$ is in standard form. Moreover, $\GG'\leq\GG$ and for 
  all $v\in\sem{\GG'}$ we have $v(x)=+\infty$ and $v(y)<+\infty$, so we get 
  $v(y)-v(x)=-\infty$ as desired.
  It remains to show that $\sem{\GG'}\neq\emptyset$, i.e., that $\GG'$ has no negative
  cycles.  Since $\GG$ is in normal form, we may restrict to cycles not using two
  consecutive weights from $\GG$.  In particular, cycles using only weights from $\GG$ are
  non negative.  It remains to consider the following cycles.
  \begin{itemize}
    \item Cycle $0\to x\to 0$ has weight $(\leq,+\infty)$ since $\GG_{0x}=(\leq,+\infty)$.

    \item Cycle $0\to y\to 0$ with $\GG'_{0y}=(<,+\infty)$ is not negative since
    $\GG_{y0}\neq(\leq,-\infty)$.
  
    \item  Cycle $0\to y\to x\to 0$ with $\GG'_{0y}=(<,+\infty)$ has weight 
    $(\leq,+\infty)$ since $\GG_{yx}=(\leq,+\infty)$.
  \end{itemize}
  Therefore, $\GG'$ has no negative cycles.
  
  We proceed similarly when $\GG_{y0}=(\leq,+\infty)$ by defining a graph 
  $\GG'=\GG[x\neq-\infty,y=-\infty]$ fixing $y$ to $-\infty$ and forcing $x$
  to be different from $-\infty$.
  This concludes the proof.
\end{proof}

To conclude, we show that a distance graph in \emph{normal form} is {\em canonical}, in
the sense that if two distance graphs in normal form have the same semantics, then they
must be identical.

\begin{lem}\label{lem:canonical}
  Let $\GG,\GG'$ be two distance graphs with $\GG$ in normal form. If 
  $\sem{\GG}\subseteq\sem{\GG'}$ then $\GG\leq\GG'$.
  In particular, if both $\GG,\GG'$ are in normal form and if $\sem{\GG}=\sem{\GG'}$ then 
  $\GG=\GG'$.
\end{lem}

\begin{proof}
  Assume that $\GG\not\leq\GG'$.  Let $x,y\in X\cup\{0\}$ such that $\GG'_{xy}<\GG_{xy}$.
  Since the cycle $x\to y\to x$ is not negative in $\GG$, by
  Lemma~\ref{lem:weight-properties}(2) we find $\alpha\in\overline{\mathbb{R}}$ such that
  $(\leq,\alpha)\leq\GG_{xy}$, $(\leq,-\alpha)\leq\GG_{yx}$, and
  $\GG'_{xy}<(\leq,\alpha)$.
  
  We will apply Lemma~\ref{lem:fixing-values-distance-graph}. We have chosen $\alpha$ so that 
  Condition 1 is satisfied. Notice that $\alpha\neq-\infty$ since 
  $\GG'_{xy}<(\leq,\alpha)$ (recall that weight $(<, -\infty)$ is not allowed in distance graphs). Assume that $x,y\in X$ and $\alpha$ finite. 
  From the choice of $\alpha$ in the proof of Lemma~\ref{lem:weight-properties},
  we see that if $\GG_{xy}=(\leq,+\infty)$ then $\alpha= +\infty$. 
  Hence, $\GG_{xy}\neq(\leq,+\infty)$. 
  Since $\GG$ is standard, 
  this implies $\GG_{x0}\neq(\leq,+\infty)\neq\GG_{0y}$. Since $\GG$ has no negative 
  cycles, we deduce that $\GG_{0x}\neq(\leq,-\infty)\neq\GG_{y0}$. Using 
  $(\leq,-\infty)\neq\GG_{xy}\leq\GG_{x0}+\GG_{0y}$ and 
  $\GG_{x0}\neq(\leq,+\infty)\neq\GG_{0y}$, we deduce that 
  $\GG_{x0}\neq(\leq,-\infty)\neq\GG_{0y}$. Therefore, Condition 2 is also satisfied.
  We apply Lemma~\ref{lem:fixing-values-distance-graph} and get a valuation $v\in\GG$ 
  with $v(y)-v(x)=\alpha$. From $\GG'_{xy}<(\leq,\alpha)$, we deduce that 
  $v\notin\sem{\GG'}$. Therefore, $\sem{\GG}\not\subseteq\sem{\GG'}$.
\end{proof}

This lemma finally allows us to define {\em the} canonical distance graph of a non-empty
event-zone $Z$, that can be computed from a distance graph defining the zone $Z$
(extending the result from usual zones in timed automata).

\begin{defi}\label{def:canonical}
  For a non-empty event-zone $Z$, we denote by $\graph{Z}$ the unique distance graph in
  normal form satisfying $\sem{\graph{Z}}=Z$.  $\graph{Z}$ is called the canonical
  distance graph of $Z$.  We denote by $Z_{xy}$ the weight of the edge $x\to y$ in
  $\graph{Z}$.
\end{defi}

\subsection{Successor computation}
To implement the computation of transitions $(q, Z) \xra{t} (q_1, Z_1)$ in an event-zone graph, we will make use of some operations on event-zones that we define below. Using distance graphs, we will show in Lemmas~\ref{lem:guard}, \ref{lem:reset}, \ref{lem:release} and~\ref{lem:time-elapse-prop} that these operations preserve event-zones, that is, starting from an event-zone and applying any of the operations leads to an event-zone again.
Thanks to the algebra over the new weights that we have defined, the arguments are very similar to the case of standard timed automata.

\begin{defi}[\textbf{Operations on event-zones}]
  Let $g$ be a guard and $Z$ an
  event-zone.
  \begin{itemize}
  \item Guard intersection:
    $Z \wedge g := \{v \mid v \in Z \text{ and } v \models g\}$

  \item Release: $[\prophecy{a}]Z= \bigcup_{v\in Z} [\prophecy{a}]v$

  \item Reset: $[\history{a}]Z=\{[\history{a}]v\mid v\in Z\}$

  \item Time elapse:
    $\elapse{Z} = \{v+\delta \mid v\in Z, \delta \in\Rpos \text{ s.t.\ }
    v+\delta \models \bigwedge_{a\in\Sigma} \prophecy{a} \leq 0 \}$
  \end{itemize}
  Let $M\in\mathbb{N}$.
  We say that an event-zone $Z$ is $M$-\emph{reachable} if it can be obtained starting from 
  the initial zone $Z_{0}$ and applying the above operations, where guards $g$ use 
  $M$-bounded finite constants, i.e., the constants allowed in a guard are from
  $\{-\infty,-M,\ldots,-1,0,1,\ldots,M,+\infty\}$.
  Recall that the initial zone $Z_0$ is given by
  $\bigwedge_{a \in \Sigma} \big((+\infty\leq\history{a}) \wedge (\prophecy{a} \leq 0)\big)$. 
\end{defi}
When $M$ is clear from context, we will sometimes abuse notation and just say reachable
zone instead of $M$-reachable zone.  A guard $g$ can be seen as yet another event-zone and
hence guard intersection is just an intersection operation between two event-zones.
By definition, for a transition $t:=(q, a, g, q')$ and a node $(q,Z)$ the
successor $(q, Z) \xra{t} (q',Z')$
can be computed in the following sequence:
\begin{align*}
  Z_1 := Z \cap (0 \le \prophecy{a}) \qquad Z_2 := [\prophecy{a}]Z_1 \qquad  Z_3 :=
  Z_2 \cap g \qquad Z_4 := [\history{a}]Z_3 \qquad Z' := \elapse{Z_4}
\end{align*}

\begin{figure}
  \centering	
	\scalebox{0.8}{
      \begin{tikzpicture}
      [state/.style={draw, thick, circle, inner sep=4pt}]
      \begin{scope}[xshift=-1cm, yshift=0cm]      
      \begin{scope}[xshift=0cm, yshift=0cm, zone/.style={draw,rectangle,minimum width=2.4cm, minimum height=2.4cm, rounded corners}]
        \node [zone] (z0) at (0, 0) {}; 
        \node [zone] (z1) at (3, 0) {};
        \node [zone] (z2) at (6, 0) {}; 
        \node [zone] (z3) at (9, 0) {};
        \node [zone] (z4) at (12, 0) {};
        \node [zone] (z5) at (15, 0) {};
      \end{scope}
  
      \begin{scope}[->, >=stealth, thick]
      \draw (z0) to (z1); 
      \draw (z1) to (z2);
      \draw (z2) to (z3);
      \draw (z3) to (z4);
      \draw (z4) to (z5);
      \end{scope}

      \node at (0, -1.6) {$Z$}; 
      \node at (3, -1.6) {$Z_1$}; 
      \node at (6, -1.6) {$Z_2$}; 
      \node at (9, -1.6) {$Z_3$}; 
      \node at (12, -1.6) {$Z_4$}; 
      \node at (15, -1.6) {$Z'$}; 

      \node at (0, 0.25) {\scriptsize $\prophecy{a} \leq 0~~\history{a} = +\infty$}; 
      \node at (0, -0.25) {\scriptsize $\prophecy{b} \leq 0~~\history{b} = +\infty$};

      \node at (3, 0.5) {\scriptsize $\prophecy{a} \leq 0~~\history{a} = +\infty$}; 
      \node at (3, 0) {\scriptsize $\prophecy{b} = 0~~\history{b} = +\infty$}; 
      \node at (3, -0.5) {\scriptsize $\prophecy{a} - \prophecy{b} \leq 0$}; 

      \node at (6, 0.25) {\scriptsize $\prophecy{a} \leq 0~~\history{a} = +\infty$}; 
      \node at (6, -0.25) {\scriptsize $\prophecy{b} \leq 0~~\history{b} = +\infty$};

      \node at (9, 0.5) {\scriptsize $\prophecy{a}=-1~~\history{a} = +\infty$}; 
      \node at (9, 0) {\scriptsize $\prophecy{b} \leq 0~~\history{b} = +\infty$}; 
      \node at (9, -0.5) {\scriptsize $\prophecy{b} - \prophecy{a} \leq 1$};

      \node at (12, 0.8) {\scriptsize $\prophecy{a}=-1~~\history{a} = +\infty$}; 
      \node at (12, 0.4) {\scriptsize $\prophecy{b} \leq 0~~\history{b} = 0$}; 
      \node at (12, 0) {\scriptsize $\prophecy{b} - \prophecy{a} \leq 1$}; 
      \node at (12, -0.4) {\scriptsize $\history{b} - \prophecy{a} = 1$}; 
      \node at (12, -0.8) {\scriptsize $\prophecy{b} - \history{b} \leq 0$}; 

      \node at (15, 1) {\scriptsize $-1 \leq \prophecy{a} \leq 0$}; 
      \node at (15, 0.6) {\scriptsize $0 \leq \history{b} \leq 1$}; 
      \node at (15, 0.2) {\scriptsize $\prophecy{b} \leq 0~~\history{a} = +\infty$}; 
      \node at (15, -0.2) {\scriptsize $\prophecy{b} - \prophecy{a} \leq 1$}; 
      \node at (15, -0.6) {\scriptsize $\history{b} - \prophecy{a} = 1$}; 
      \node at (15, -1) {\scriptsize $\prophecy{b} - \history{b} \leq 0$}; 
    \end{scope}

    \end{tikzpicture} 
  }
  \caption[]{Successor computation from event-zone $Z$ on an action $b$ with guard $\prophecy{a}=-1$}
  \label{fig:succ-computation}
\end{figure}
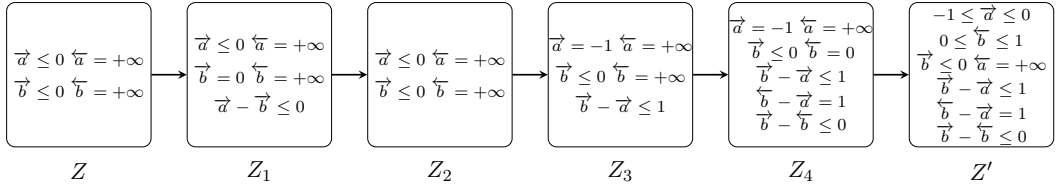

As an example, in Figure~\ref{fig:succ-computation}, suppose an action $b$ with guard
$(\prophecy{a}=-1)=(\prophecy{a}\leq -1) \wedge (-1 \leq \prophecy{a})$ is fired from Zone
$Z$ as depicted, applying the above sequence in order gives $Z_1 := Z
\cap (0 \le \prophecy{b})$, $Z_2 := [\prophecy{b}]Z_1$, $Z_3 := Z_2 \cap
(\prophecy{a} = - 1)$, $Z_4 := [\history{b}] Z_3$ resulting in
the successor zone $Z':= \elapse{Z_4}$, as depicted in the figure. 

We will now translate the operations from event-zones to distance graphs.

\begin{defi}[\textbf{Operations on distance graphs}]\label{defn:dist-graph-ops}
  Let $\GG$ be a distance graph in normal form. Let $g$ be a
  guard.
  \begin{itemize}
  \item Guard intersection: a distance graph $\GG_g$ is
    obtained from $\GG$ as follows,
    \begin{itemize}
    \item for each constraint $x \leqlt c$ in $g$, replace
      weight of edge $0 \to x$ with $\min(\GG_{0x},(\leqlt, c))$,
    \item for each constraint $d \leqlt y$ in $g$, replace
      weight of edge $y \to 0$ with $\min(\GG_{y0},(\leqlt, -d))$,
    \item normalize the resulting graph if it has no negative cycles.
    \end{itemize}
  \item Release: a distance graph $[\prophecy{a}]\GG$ is obtained
    from $\GG$ by
    \begin{itemize}
    \item removing all edges involving $\prophecy{a}$ and then
    \item adding the edges $0 \xra{(\leq,0)} \prophecy{a}$ and
      $\prophecy{a} \xra{(\leq,+\infty)} 0$, and then
    \item normalizing the resulting graph.
    \end{itemize}
  \item Reset: a distance graph $[\history{a}]\GG$ is obtained from
    $\GG$ by
    \begin{itemize}
    \item removing all edges involving $\history{a}$ and then
    \item adding the edges $0 \xra{(\leq,0)} \history{a}$ and
      $\history{a} \xra{(\leq, 0)} 0$, and then
    \item normalizing the resulting graph.
    \end{itemize}
  \item Time elapse: the distance graph $\elapse{\GG}$ is
    obtained by the following transformation:
    \begin{itemize}
    \item if $\GG_{0\history{x}}\neq(\leq,+\infty)$ then replace it with $(<,+\infty)$,
      
    \item if $\GG_{0\prophecy{x}}\neq(\leq,-\infty)$ then replace it with $(\leq,0)$,

    \item normalize the resulting graph.
    \end{itemize}
  \end{itemize}
\end{defi}

The theorem below says that the operations on event-zones translate easily to operations
on distance graphs and that the successor of an event-zone is an event-zone.  Note that,
except for the release operation $[\prophecy{a}]$, the rest of the operations are standard in
timed automata, but they do not use $(\le, +\infty), (\le, -\infty)$.  We show that we can
perform all these operations in the new algebra with quadratic complexity, as in timed
automata without diagonal constraints~\cite{DBLP:journals/ipl/ZhaoLZ05}.

\begin{thm}\label{thm:distance-graph-operations}
  Let $\GG$ be a distance graph in normal form,
  $a\in\Sigma$, and $g$ be a guard.  
  We can compute, in $\mathcal{O}(|X_P \cup X_H|^2)$ time, distance graphs 
  $\GG_g$, $[\prophecy a]\GG$, $[\history a]\GG$ and $\elapse{\GG}$ in normal form, such that
  $\sem{\GG} \land g = \sem{\GG_g}$,
  $[\prophecy a]\sem{\GG} = \sem{[\prophecy a]\GG}$,
  $[\history a]\sem{\GG} = \sem{[\history a]\GG}$, and
  $\elapse{\sem{\GG}} = \sem{\elapse{\GG}}$.
\end{thm}

This theorem follows from Lemmas~\ref{lem:guard},~\ref{lem:release},~\ref{lem:reset}
and~\ref{lem:time-elapse-prop} which handle respectively the operations on distance
graphs given in Definition~\ref{defn:dist-graph-ops}.
Other than normalization, it can be easily checked that these operations can be computed
in quadratic time.  We discuss the normalization procedure in the stated lemmas.

In general, the normal form of a distance graph 
  cannot always be
  computed in quadratic time. However, starting from an event zone in
  normal form, and applying the operations of
  Definition~\ref{defn:dist-graph-ops} gives us special event-zones
  whose normal forms can be computed in quadratic time. This is
  explained in Lemmas~\ref{lem:guard},~\ref{lem:release},~\ref{lem:reset}
  and~\ref{lem:time-elapse-prop}.

\begin{lem}\label{lem:guard}
  Let $\GG$ be a distance graph in normal form and let $g$ be a guard.  
  Then $\sem{\GG_g}=\sem{\GG}\wedge g$, 
  and $\GG_{g}$ can be computed in time $\mathcal{O}(|X|^{2})$.
\end{lem}
\begin{proof}
  According to the definition, we first construct an intermediate graph $\GG'$ by
  replacing weights of edges of the form $0 \to x$ and $y \to 0$ depending on the guards
  present.  It is easy to see that $\sem{\GG'}=\sem{\GG}\wedge g$.
  The normalization process does not change the solution set.
  
  Since $g$ has only non-diagonal guards (as $g$ is from an ECA, and our ECAs do not have
  diagonal guards) and $\GG$ is in normal form, we can check if
  there is a negative cycle in $\GG'$ in quadratic time (relevant cycles use one or two
  modified edges and are of the form $x\to 0\to x$ or $x \to 0 \to y \to x$).  If not, one
  first computes in quadratic time all shortest paths $x\to 0$ and $0\to x$.  The shortest
  path $x \to 0$ is obtained as $\GG''_{x0}=\min(\GG'_{x0},\GG_{xz}+\GG'_{z0})$ over all
  $z\to 0$ that comes from guard $g$.  Similarly the shortest path $0 \to x$ is
  $\GG''_{0x}=\min(\GG'_{0x},\GG'_{0z}+\GG_{zx})$ over all $0 \to z$ coming from guard
  $g$.  Finally, the shortest path $x\to y$ has weight
  $\min(\GG'_{xy},\GG''_{x0}+\GG''_{0y})$.
\end{proof}

\begin{lem}\label{lem:release}
  Let $\GG$ be a distance graph in normal form and $a\in\Sigma$.  
  Then, $\sem{[\prophecy{a}]\GG}=[\prophecy{a}]\sem{\GG}$,
  and $[\prophecy{a}]\GG$ can be computed in time $\mathcal{O}(|X|^{2})$.
  
  Moreover, the weights of edges in $[\prophecy{a}]\GG$ are given by
  \begin{itemize}
    \item $x \to y$ has weight $\GG_{xy}$, if $x,y \neq \prophecy{a}$,
    \item $\prophecy{a} \to x$ has weight $(\leq,+\infty)$ and
    $x \to \prophecy{a}$ has weight $\GG_{x0}$ if $x\neq\prophecy{a}$.
  \end{itemize}
\end{lem}

\begin{proof}
  The release of a prophecy clock $\prophecy{a}$ corresponds to removing all edges involving
  the node $\prophecy{a}$, and then adding the edges $0 \xra{(\leq,0)}
  \prophecy{a}$ and $\prophecy{a} \xra{(\leq,+\infty)} 0$.  Let $\GG'$ be the distance graph thus
  obtained from $\GG$.  It is easy to see $[\prophecy{a}]\sem{\GG}\subseteq\sem{\GG'}$.  For 
  the converse inclusion, we pick $v\in\sem{\GG'}$ and show that there exists some
  $u\in\sem{\GG}$ such that $u$ coincides with $v$ in all variables except $\prophecy{a}$.
  To see this, we construct a distance graph $\GG''$ by setting $\GG''_{0x}=(\leq,v(x))$
  and $\GG''_{x0}=(\leq,-v(x))$ for all $x\in X\setminus\{\prophecy{a}\}$; and
  $\GG''_{xy}=\GG_{xy}$ for all other edges $x\to y$.  We show below that (a)
  $\GG''\leq\GG$, (b) $\GG''$ is in standard form, and (c) $\GG''$ has no negative cycles.
  We deduce that $\emptyset\neq\sem{\GG''}\subseteq\sem{\GG}$ and $v\in[\prophecy{a}]u$ for all
  $u\in\sem{\GG''}$.
  \begin{enumerate}
    \item[(a)] Let $x\in X\setminus\{\prophecy{a}\}$. Since $v\in\sem{\GG'}$, we have 
    $\GG''_{0x}=(\leq,v(x))\leq\GG'_{0x}=\GG_{0x}$ and similarly
    $\GG''_{x0}=(\leq,-v(x))\leq\GG'_{x0}=\GG_{x0}$.
  
    \item[(b)] Let $x\in X_{H}$. We have $\GG''_{x0}=(\leq,-v(x))\leq(\leq,0)$.
    Assume that $\GG''_{0x}=(\leq,+\infty)$. From (a) we get $\GG_{0x}=(\leq,+\infty)$. 
    Since $\GG$ is standard, this implies $\GG''_{yx}=\GG_{yx}=(\leq,+\infty)$ for all 
    $y\neq x$. We proceed similarly when $x\in X_{P}\setminus\{\prophecy{a}\}$.
  
    \item[(c)] Since $\GG$ is in normal form, cycles of $\GG$ that are possibly negative are of 
    the form $0\to x\to 0$ or $0\to x\to y\to 0$ with $x,y\neq\prophecy{a}$. The weight of 
    $0\to x\to 0$ in $\GG''$ is $(\leq,0)$ or $(\leq,+\infty)$, hence not negative.  For 
    the other cycle, notice first that since $v\in\sem{\GG'}$ we have
    $(\leq,v(y)-v(x))\leq\GG'_{xy}=\GG_{xy}$, which is equivalent to
    $(\leq,0)\leq\GG_{xy}+(\leq,-(v(y)-v(x)))$. Recall that $-(b-a)\leq a-b$. Therefore,
    $(\leq,0)\leq\GG_{xy}+(\leq,v(x)-v(y))$, which is the weight of the cycle
    $0\to x\to y\to 0$.
  \end{enumerate}
  
  Then, $[\prophecy{a}]\GG$ is obtained by normalization of $\GG'$.  Observe that the weight of
  no edge was decreased: $\GG_{0\prophecy{a}}\leq(\leq,0)$ and
  $\GG_{\prophecy{a}0}\leq(\leq,+\infty)$.
  Therefore, this transformation does not lead to shorter paths: the edge $x \to y$ in
  $[\prophecy{a}]\GG$ has weight $\GG'_{xy}=\GG_{xy}$ if $x\neq\prophecy{a}\neq y$.
  Now, non-trivial paths in $\GG'$ starting from $\prophecy{a}$ start with weight
  $(\leq,+\infty)$, hence the weight of edge $\prophecy{a}\to x$ in $[\prophecy{a}]\GG$ is
  $(\leq,+\infty)$.
  Finally, the shortest path in $\GG'$ from $x$ to $\prophecy{a}$ is $x\to 0\to\prophecy{a}$, which 
  is of weight $\GG_{x0}+(\leq,0)=\GG_{x0}$.
  Hence, the weight of $x\to\prophecy{a}$ in $[\prophecy{a}]\GG$ is $\GG_{x0}$.
\end{proof}

\begin{lem}\label{lem:reset}
  Let $\GG$ be a distance graph in normal form and $a\in\Sigma$.  
  Then, $\sem{[\history{a}]\GG}=[\history{a}]\sem{\GG}$,
  and $[\history{a}]\GG$ can be computed in time $\mathcal{O}(|X|^{2})$.
  
  Moreover, the weights of edges in $[\history{a}]\GG$ are given by
  \begin{itemize}
    \item $x \to y$ has weight $\GG_{xy}$, if $x,y \neq \history{a}$,
    \item $x \to \history{a}$ has weight $\GG_{x0}$ and
    $\history{a} \to x$ has weight $\GG_{0x}$ if $x\neq\history{a}$ (including $x=0$ assuming 
    $\GG_{00}=(\leq,0)$).
  \end{itemize}
\end{lem}

\begin{proof}
  The reset of a history clock $\history{a}$ corresponds to removing all edges involving
  node $\history{a}$, and then setting its value to $0$ by adding the edges $0 \xra{(\leq,0)}
  \history{a}$ and $\history{a} \xra{(\leq, 0)} 0$.
  Let $\GG'$ be the distance graph thus obtained from $\GG$. Similar
  to timed automata it can be shown that $\sem{\GG'} = [\history{a}]\sem{\GG}$.   
  
  Then, $[\history{a}]\GG$ is obtained by normalization of $\GG'$.
  Normalization does not affect the weight of the edges $x\to y$ if $x,y\neq\history{a}$
  (as any path from $x$ to $y$ in $\GG'$ using the new edges would involve a cycle
  $\history{a}\to 0\to\history{a}$ or $0\to\history{a}\to 0$ of weight $(\leq,0)$).  
  Thus, the weight of edge $x\to y$ in $[\history{a}]\GG$ is equal to $\GG_{xy}$ if $x,y\neq\history{a}$.
  Now, paths in $\GG'$ from $x\neq\history{a}$ to $\history{a}$ ends with the edge $0\to\history{a}$
  of weight $(\leq,0)$.  Since $\GG$ is in normal form, we deduce that
  the weight of edge $x\to\history{a}$ in $[\history{a}]\GG$ is $\GG'_{x0}=\GG_{x0}$.
  Similarly, the weight of edge $\history{a}\to x$ in $[\history{a}]\GG$ is $\GG'_{0x}=\GG_{0x}$.
\end{proof}

\begin{lem}\label{lem:time-elapse-prop}
  Let $\GG$ be a distance graph in normal form.
  Then, $\sem{\elapse{\GG}}=\elapse{\sem{\GG}}$,
  and $\elapse{\GG}$ can be computed in time $\mathcal{O}(|X|^{2})$.
\end{lem}

\begin{proof}
  The time elapse operation corresponds to (1) replacing the weight of $0\to\history{a}$ with
  $(<,+\infty)$ if it is not $(\leq,+\infty)$, and (2) replacing the weight of $0\to\prophecy{a}$
  with $(\leq,0)$ if it is not $(\leq,-\infty)$. Let $\GG'$ be the distance graph thus
  obtained from $\GG$.  Similar to timed automata it can be shown that 
  $\sem{\GG'}=\elapse{\sem{\GG}}$.

  Then, $\elapse{\GG}$ is obtained by normalization of $\GG'$.
  The transformation from $\GG$ to $\GG'$ does not decrease the weights of edges, 
  it may only increase the weights of edges from $0$ to $x\in X$.
  Therefore, no shorter paths may be obtained by this transformation.  
  We deduce that, for all edges $0\neq x\to y$ we have 
  $\elapse{\GG}_{xy}=\GG'_{xy}=\GG_{xy}$, and
  $\GG_{0y}\leq\elapse{\GG}_{0y}\leq\GG'_{0y}$ for all $y\in X$.
  
  Since $\GG$ is in normal form, the shortest path in $\GG'$ from $0$ to $y\in X$ is 
  either the edge $0\to y$ with weight $\GG'_{0y}$ or of the form $0\to x\to y$ which is 
  of weight $\GG'_{0x}+\GG_{xy}$. We show below that, if $\GG'_{0x}+\GG_{xy}<\GG'_{0y}$ 
  then $x\in X_{P}$ is a prophecy clock and $\GG'_{0x}=(\leq,0)$. We deduce that 
  $\elapse{\GG}_{0y}=\min(\{\GG'_{0y}\}\cup\{\GG_{xy}\mid x\in X_{P}\})
  =\min(\{\GG'_{xy}\mid x\in X_{P}\cup\{0\}\})$.
  
  Assume that $\GG'_{0x}+\GG_{xy}<\GG'_{0y}$.  Then,
  $\GG'_{0x}\neq(\leq,+\infty)\neq\GG_{xy}$ and $\GG'_{0y}\neq(\leq,-\infty)$.  Hence,
  $\GG_{0x}\neq(\leq,+\infty)$ and $\GG_{0y}\neq(\leq,-\infty)$.  Since $\GG$ is in
  normal form, we have $\GG_{0y}\leq\GG_{0x}+\GG_{xy}$ and we deduce that 
  $\GG_{0x}\neq(\leq,-\infty)\neq\GG_{xy}$.
  If $x\in X_{P}$, we get $\GG'_{0x}=(<,+\infty)=\GG'_{0x}+\GG_{xy}<\GG'_{0y}$. This 
  implies $\GG_{0y}=\GG'_{0y}=(\leq,+\infty)$, a contradiction with 
  $\GG_{xy}\neq(\leq,+\infty)$ since $\GG$ is in standard form.
  Hence, $x\in X_{H}$ is a history clock and $\GG_{0x}\neq(\leq,-\infty)$ implies 
  $\GG'_{0x}=(\leq,0)$.  
  
  Finally, from $\elapse{\GG}_{0y}=\min(\{\GG'_{0y}\}\cup\{\GG_{xy}\mid x\in X_{P}\})$ 
  and the definition of $\GG'_{0y}$, 
  we get
  $$
  \elapse{\GG}_{0y} = 
  \begin{cases}
    (\leq,+\infty) & \text{if } \GG_{0y} = (\leq,+\infty) \\
    (\leq,-\infty) & \text{if } \GG_{0y} = (\leq,-\infty) \\
    \min((<,+\infty),(\leqlt,c)) & \text{if $y$ is a history clock and $G_{0y}\neq(\leq,+\infty)$} \\
    \min((\leq,0),(\leqlt,c)) & \text{$y$ is a prophecy clock and $G_{0y}\neq(\leq,-\infty)$}
  \end{cases}
  $$
  where $(\leqlt,c)=\min\{\GG_{xy} \mid x\in X_{P}\}$.
\end{proof}

We conclude this section by showing some properties of reachable zones that will be 
useful in the sequel.

\begin{lem}\label{lem:reachable-properties}
  Let $Z$ be a non-empty reachable zone and let $\GG$ be its canonical distance graph.
  \begin{enumerate}
    \item  For all $x\in X_{H}$, we have $\GG_{x0}=(\leq,-\infty)$ or 
    $\GG_{0x}\leq(<,+\infty)$.
  
    \item  For all $x,y\in X$, if $\GG_{xy}=(\leq,-\infty)$ then 
    $\GG_{x0}=(\leq,-\infty)$ or $\GG_{0y}=(\leq,-\infty)$.
  \end{enumerate}
\end{lem}

\begin{proof}
  Let $\history{a}\in X_{H}$ be a history clock.  The weight of $\history{a}\to 0$ in the initial
  distance graph $Z_{0}$ is $(\leq,-\infty)$, which is the least possible weight.  It
  stays unchanged until we first apply the reset operation on $\history{a}$, resulting in the
  weighted edges $0 \xra{(\leq,0)} \history{a}$ and $\history{a} \xra{(\leq,0)} 0$. Then, the 
  weight of edge $0\to\history{a}$ may only be increased by the time elapse operation, which 
  sets it to $(<,+\infty)$. This proves the first property.
  
  For the second property, let $x,y\in X$ with $\GG_{xy}=(\leq,-\infty)$ and
  $\GG_{x0}\neq(\leq,-\infty)$. We have to show that $\GG_{0y}=(\leq,-\infty)$.
  If $x\in X_{H}$ then we get $\GG_{0x}\leq(<,+\infty)$ by the first property. 
  If $x\in X_{P}$ then we have $\GG_{0x}\leq(\leq,0)$.
  In both cases, since $\GG$ is normal, we obtain
  $\GG_{0y}\leq\GG_{0x}+\GG_{xy}=(\leq,-\infty)$ and we are done.
\end{proof}

We have so far seen the computation of the event-zone graph. As
discussed in Section~\ref{sec:event-zones}, we require a simulation to
truncate the event-zone graph to a finite prefix that is sound and complete for
reachability. We discuss a new simulation relation for ECAs in the
next section.

\section{A concrete simulation relation for ECAs}\label{sec:simulation}

\begin{figure}[!h]
    \centering	
    \scalebox{1}{
      \begin{tikzpicture}
        [state/.style={draw, thick, circle, inner sep=5pt}]
        \begin{scope}[xshift=4cm, yshift=0cm]
        \begin{scope}[every node/.style={state}]
          \node (q0) at (3, 3) {\scriptsize $q_0$}; 
          \node (q1) at (5, 3) {\scriptsize $q_1$}; 
          \node (q2) at (7, 3) {\scriptsize $q_2$}; 
          \node (q3) at (9, 3) {\scriptsize $q_3$}; 
        \end{scope}

        \begin{scope}[->, >=stealth, thick]
          \draw (2.25, 3) to (q0); 
          \draw (q0) to [out=120,in=60,looseness=8] (q0);
          \draw (q1) to [out=120,in=60,looseness=8] (q1);  
          \draw (q0) to (q1); 
          \draw (q1) to (q2); 
          \draw (q2) to (q3); 
          \draw (q3) to (9.75,3); 
        \end{scope}
        
        \node at (3, 5.1) {\footnotesize $a$}; 
        \node at (5, 4.6) {\footnotesize $b$}; 
        \node at (3, 4.6) {\scriptsize $\prophecy{b} \geq -1$}; 
        \node at (4, 3.25) {\footnotesize $a$}; 
        \node at (6, 2.75) {\scriptsize $\history{a} \geq 2$}; 
        \node at (8, 2.75) {\scriptsize $\history{a} \leq 5$}; 
  
        \node at (6, 3.25) {\footnotesize $b$}; 
        \node at (8, 3.25) {\footnotesize $b$};

        \node at (1, 2) {\footnotesize $\G(q_0) = P$}; 
        \node at (1, 1.5) {\footnotesize $\G(q_1) = P$}; 
        \node at (1, 1) {\footnotesize $\G(q_2) = P$}; 
        \node at (1, 0.5) {\footnotesize $\G(q_3) = P$}; 
  
        \node at (5, 2) {\footnotesize $\G(q_0) = P$}; 
        \node at (5, 1.5) {\footnotesize $\G(q_1) = P \cup \{\history{a} \geq 2\}$}; 
        \node at (5, 1) {\footnotesize $\G(q_2) = P \cup \{\history{a} \leq 5\}$};  
        \node at (5, 0.5) {\footnotesize $\G(q_3) = P$};  
  
        \node at (10, 2) {\footnotesize $\G(q_0) = P$}; 
        \node at (10, 1.5) {\footnotesize $\G(q_1) = P \cup \{\history{a} \geq 2,\history{a} \leq 5\}$}; 
        \node at (10, 1) {\footnotesize $\G(q_2) = P \cup \{\history{a} \leq 5\}$};  
        \node at (10, 0.5) {\footnotesize $\G(q_3) = P$};  
  
        \begin{scope}[->, >=stealth, thick]
          \draw (2, 1.25) to (3,1.25); 
          \draw (7, 1.25) to (8,1.25); 
        \end{scope}
    
      \end{scope}
  
    \end{tikzpicture} 
    } 
    \caption[Example of $\G$ mapping computation for an ECA]{An event-clock
    automaton $\Aa$ and illustration of the computation of the $\G$-map.
    Note that the set $P$ is used to denote the set of constraints
    involving prophecy clocks, i.e., $P = \{\prophecy{a} \leq 0, 0\leq\prophecy{a},
    \prophecy{b} \leq 0, 0\leq\prophecy{b}\}$ - this set is part of the $\G$-set of
    each state.  Further, observe that the constraints involving $\history{a}$ do not get
    propagated to state $q_0$ as the transition $q_0 \xra{} q_1$ is on the action $a$, and
    therefore resets $\history{a}$.}
    \label{fig:eca-g-map}
  \end{figure}
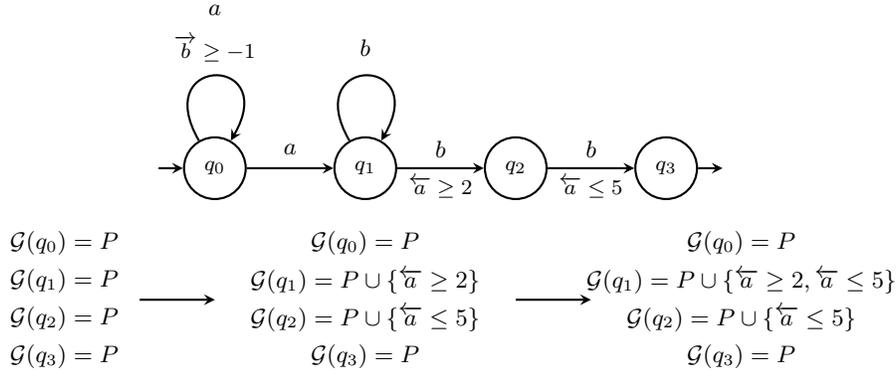

We fix an event-clock automaton $\A=(Q,\Sigma,X,T,q_0,F)$ for this section.  We will define a simulation relation $\preceq_{\A}$ on the configurations of the ECA. We first define a map $\G$ from $Q$ to sets of atomic constraints.  The map $\G$ is obtained as the least fixpoint of the set of equations:
$$
\G(q) = \{\prophecy{b}\leq0, 0\leq\prophecy{b} \mid b\in\Sigma\} \cup
\bigcup_{(q,a,g,q')\in T} \asplit(g) \cup \pre{a}{\G(q')}
$$
where $\asplit(g)$ is the set of atomic constraints on \emph{history} clocks occurring in $g$ and, for a set of atomic constraints $G$, $\pre{a}{G}$ is defined as the set of constraints on history clocks in $G$ except those on $\history{a}$.
Notice that constraints in $\G(q)$ use the constant $0$ and constants used in constraints
of $\A$.  As a consequence, it is easy to see that the least fixpoint computation terminates and results in a finite set.
Intuitively, $\G(q)$ contains an atomic constraint $\varphi$ of the form $\history{a} \leqlt c$ or $c \leqlt \history{a}$ if $\varphi$ appears in a transition out of some state $q'$, and there is a path $q \to \dots \to q'$ from $q$ to $q'$ over transitions that do not read $a$. This makes the value of $\history{a}$ at $q$ important for the verification of the constraint $\varphi$ at $q'$. Transitions on letter $a$ reset the history clock $\history{a}$ to $0$, and so, the actual value of $\history{a}$ at $q$ is irrelevant for constraints on $\history{a}$ that are witnessed after a ``reset''. This is why such constraints are ignored in $\G(q)$. 
Figure~\ref{fig:eca-g-map} illustrates the computation of the $\G$-sets associated with each state of a given event-clock automaton.

Let $G$ be a set of atomic constraints. The preorder $\preceq_{G}$ is 
defined on valuations by
$$
v\preceq_{G}v' \qquad \text { if }
\forall\varphi\in G,~\forall\delta\geq0,\qquad v+\delta\models\varphi
\implies v'+\delta\models\varphi \,.
$$

Notice that in the condition above, we \emph{do not} restrict $\delta$ to those such that $v+\delta$ is a valuation: we may have $v(\prophecy{a})+\delta>0$ for some $a\in\Sigma$\footnote{
  We highlight the distinction between time elapse operation in
  general, and the time elapse operation allowed by the semantics of ECA. In a general
  time elapse, there is no restriction on the amount of time that can be elapsed from a
  valuation.  But for the semantics of ECA, we restrict to time elapses such that
  prophecy clocks stay at most $0$.
  }.
In usual timed automata, this question does not arise, as elapsing any $\delta$ from any given valuation always results in a valuation.
But this is crucial for the proof of Theorem~\ref{thm:simulation} below.
We illustrate this using Example~\ref{eg:sim}.
It also allows us to get a clean characterization of the simulation
(Lemma~\ref{lem:simulation-properties}) which in turn is useful for deriving the
simulation test and in showing finiteness.

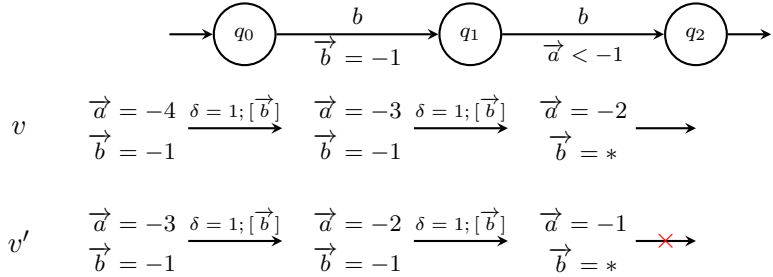
\begin{figure}[!h]
    \centering	
    \scalebox{1}{
      \begin{tikzpicture}
        [state/.style={draw, thick, circle, inner sep=5pt}]
        \begin{scope}[xshift=4cm, yshift=0cm]
        \begin{scope}[every node/.style={state}]
          \node (q0) at (2, 3) {\scriptsize $q_0$}; 
          \node (q1) at (5, 3) {\scriptsize $q_1$}; 
          \node (q2) at (8, 3) {\scriptsize $q_2$}; 
        \end{scope}

        \begin{scope}[->, >=stealth, thick]
          \draw (1, 3) to (q0); 
          \draw (q0) to (q1); 
          \draw (q1) to (q2); 
          \draw (q2) to (9,3);
        \end{scope}
        
        \node at (3.5, 3.25) {\footnotesize $b$}; 
        \node at (3.5, 2.75) {\footnotesize $\prophecy{b} = -1$}; 

        \node at (6.5, 3.25) {\footnotesize $b$}; 
        \node at (6.5, 2.75) {\scriptsize $\prophecy{a} < -1$};

        \node at (0.5, 2) {\footnotesize $\prophecy{a}=-4$}; 
        \node at (0.5, 1.5) {\footnotesize $\prophecy{b}=-1$
        }; 

        \node at (0.5, 0.5) {\footnotesize $\prophecy{a}=-3$}; 
        \node at (0.5, 0) {\footnotesize $\prophecy{b}=-1$}; 

        \node at (3.5, 2) {\footnotesize $\prophecy{a}=-3$}; 
        \node at (3.5, 1.5) {\footnotesize $\prophecy{b}=-1$
        }; 
        \node at (3.5, 0.5) {\footnotesize $\prophecy{a}=-2$}; 
        \node at (3.5, 0) {\footnotesize $\prophecy{b}=-1$}; 

        \node at (6.5, 2) {\footnotesize $\prophecy{a}=-2$}; 
        \node at (6.5, 1.5) {\footnotesize $\prophecy{b}=*$
        }; 
        \node at (6.5, 0.5) {\footnotesize $\prophecy{a}=-1$}; 
        \node at (6.5, 0) {\footnotesize $\prophecy{b}=*$}; 
  
        \begin{scope}[->, >=stealth, thick]
          \draw (1.25, 1.75) to (2.5,1.75); 
          \draw (1.25, 0.25) to (2.5,0.25); 
          
          \draw (4.25, 1.75) to (5.5,1.75); 
          \draw (4.25, 0.25) to (5.5,0.25); 
          
          \draw (7.2, 1.75) to (8,1.75); 
          \draw (7.2, 0.25) to (8,0.25); 
        \end{scope}
  
        \node at (-1, 1.75) {$v$}; 
        \node at (-1, 0.25) {$v'$};

        \node at (1.875, 2) {\tiny $\delta=1;[\prophecy{b}]$}; 
        \node at (1.875, 0.5) {\tiny $\delta=1;[\prophecy{b}]$}; 

        \node at (4.875, 2) {\tiny $\delta=1;[\prophecy{b}]$}; 
        \node at (4.875, 0.5) {\tiny $\delta=1;[\prophecy{b}]$}; 

        \node at (7.6, 0.25) { \textcolor{red}{$\times$}}; 

      \end{scope}
  
    \end{tikzpicture} 
    } 
    \caption[]{An example to explain why it is not sufficient to only look at time-elapse restricted to ECA valuations in the definition of $\Gg$-simulation. The $*$ symbol for the value of $\prophecy{b}$ denotes that $\prophecy{b}$ may take any legal value in this valuation.}
    \label{fig:sim-eg}
  \end{figure}

  \begin{exa}~\label{eg:sim}
    Consider two valuations $v$ and $v'$ defined as follows:
      $v(\prophecy{a}) = -4$, $v(\prophecy{b})=-1$ and
      $v'(\prophecy{a}) = -3$ and $v'(\prophecy{b}) = -1$. Let $G =
      \{\prophecy{a} \le 0, 0 \le \prophecy{a}, \prophecy{b} \le 0, 0 
      \le \prophecy{b}\}$.  Firstly, notice that $v
    \not \preceq_{G} v'$: if we take $\delta = 3.5$, we have $v +
    \delta \models \prophecy{a} \le 0$, whereas $v' + \delta \not
    \models \prophecy{a} \le 0$. Notice that in $v + \delta$ and $v' +
    \delta$, the value of $\prophecy{b}$ is $+2.5$, which is not
    a legal valuation reachable in an ECA. This means that if $v,
    v'$ were valuations at a certain state of an ECA, they
    cannot elapse $\delta$ units staying at that state.

    Now, suppose in the definition of $\preceq_{G}$, we restrict to
  $\delta$ such that $v + \delta$ of all prophecy clocks is at most
  $0$. Then, $v \preceq_{G} v'$: firstly $\delta \le 1$ due to the
  value of $\prophecy{b}$, and further, for all $\delta \le 1$,
  whenever $v + \delta$ satisfies a constraint in $G$, the valuation $v' +
  \delta$ satisfies the same constraint. However, this way of
  relating $v$ and $v'$ does not guarantee a
  simulation, especially after a transition that updates the value of
  $\prophecy{b}$ to a fresh value. In particular, we can come up with a
  sequence of transitions that is feasible from $v$, but not from
  $v'$. This is illustrated in Figure~\ref{fig:sim-eg}. 
\end{exa}
  
\begin{rem} \label{rmk:prophecyequality}
  Let $v,v'$ be valuations and $a\in\Sigma$.  If
  $v\preceq_{\{0\leq\prophecy{a},\prophecy{a}\leq0\}}v'$ then $v(\prophecy{a})=v'(\prophecy{a})$.
  
  The proof is easy.
  First, if $v(\prophecy{a})=-\infty$ then $v+\delta\models\prophecy{a}\leq 0$ for all $\delta\geq 0$.
  Since $v'$ simulates $v$ we deduce that $v'+\delta\models\prophecy{a}\leq 0$ for all $\delta\geq 0$.
  This implies $v'(\prophecy{a})=-\infty$.
  Next, if $-\infty<v(\prophecy{a})\neq v'(\prophecy{a})\leq 0$.  
  Then, for $\delta=-v(\prophecy{a})\geq 0$ we have
  $v(\prophecy{a})+\delta=0\neq v'(\prophecy{a})+\delta$, a contradiction with
  $v\preceq_{\{0\leq\prophecy{a},\prophecy{a}\leq0\}}v'$. 
  \qed
\end{rem}

Based on $\preceq_G$ and the $\Gg(q)$ computation, we can define a preorder $\preceq_{\A}$ between configurations of ECA $\A$.

\begin{defi}
  $(q,v)\preceq_{\A}(q',v')$ if $q=q'$ and $v\preceq_{\G(q)}v'$.
\end{defi}

With this definition in place, we can now show the following theorem.

\begin{thm}\label{thm:simulation}
  The relation $\preceq_{\A}$ is a simulation on the transition system $S_{\A}$ of ECA $\A$.
\end{thm}

\begin{proof}
  Assume that $(q,v_{1})\preceq_{\A}(q,v_{2})$, i.e., $v_{1}\preceq_{\G(q)}v_{2}$.
  \begin{description}
    \item[Delay transition]  Assume that $(q,v_{1})\xra{\delta}(q,v_{1}+\delta)$ is a 
    transition of $S_{\A}$. Then, $v_{1}+\delta\models\bigwedge_{x\in\Sigma}\prophecy{x}\leq0$.
    Since $\G(q)$ contains $\prophecy{x}\leq0$ for all $x\in\Sigma$ and 
    $v_{1}\preceq_{\G(q)}v_{2}$, we deduce that 
    $v_{2}+\delta\models\bigwedge_{x\in\Sigma}\prophecy{x}\leq0$.
    Therefore, $(q,v_{2})\xra{\delta}(q,v_{2}+\delta)$ is a transition in $S_{\A}$.
    It is easy to see that $v_{1}+\delta\preceq_{\G(q)}v_{2}+\delta$.
  
    \item[Action transition]  
    Let $t=(q,g,a,q')$ be a transition in $\A$ and assume that 
    $(q,v_{1})\xra{t}(q',v'_{1})$ is a transition in $S_{\A}$, i.e.,
    $v_{1}\models 0\leq\prophecy{a}$ and for some $v''_{1}\in[\prophecy{a}]v_{1}$ we have 
    $v''_{1}\models g$ and $v'_{1}=[\history{a}]v''_{1}$. 
    Since $\G(q)$ contains $0\leq\prophecy{a}$ and $v_{1}\preceq_{\G(q)}v_{2}$, we deduce that
    $v_{2}\models 0\leq\prophecy{a}$.
   
    We have $v''_{1}=v_{1}[\prophecy{a}\mapsto\alpha]$ for some $\alpha\in[-\infty,0]$
    and
    $v'_{1}=v''_{1}[\history{a}\mapsto 0]=v_{1}[\history{a}\mapsto 0,\prophecy{a}\mapsto\alpha]$.
    Define $v''_{2}=v_{2}[\prophecy{a}\mapsto\alpha]\in[\prophecy{a}]v_{2}$ and
    $v'_{2}=v''_{2}[\history{a}\mapsto 0]=v_{2}[\history{a}\mapsto 0,\prophecy{a}\mapsto\alpha]$.
    Notice that, $v'_{1}(x)=v''_{1}(x)=v''_{2}(x)=v'_{2}(x)$ for all prophecy clocks 
    $x\in X_{P}$ (follows from $v_{1}\preceq_{\G(q)}v_{2}$ and $\{x\leq0,0\leq 
    x\}\subseteq\G(q)$).
    From $v_{1}\preceq_{\G(q)}v_{2}$ and the definition of $v''_{2}$, we
    deduce that $v''_{1}\preceq_{\G(q)}v''_{2}$.
    Since $\G(q)$ contains $\asplit(g)$, and 
    $v''_{1}\preceq_{\G(q)}v''_{2}$, we deduce that $v''_{2}\models g$.

    Therefore, $(q,v_{2})\xra{t}(q',v'_{2})$ is a transition in $S_{\A}$.
    Since $\G(q)$ contains $\pre{a}{\G(q')}$, we easily get from 
    $v_{1}\preceq_{\G(q)}v_{2}$ and the definitions of $v'_{1},v'_{2}$ that
    $v'_{1}\preceq_{\G(q')}v'_{2}$.
    Note that, to get this, we crucially use the fact that in the definition of the
    simulation $\preceq_{G}$ we consider \emph{all} $\delta\geq0$ and not only those such
    that $v+\delta$ is a valuation. To see why, notice that the
      $\alpha$ that is picked 
      to make $v'_1(\prophecy{a}) = \alpha$ could be any value in
      $[-\infty, 0]$, in particular, it could be smaller than all
      values in $v_1$. This potentially allows $\delta$ such that $v'_1 +
      \delta$ is a valuation whereas $v_1 + \delta$ is not.
    \qedhere
  \end{description}
\end{proof}

When $G=\{\varphi\}$ is a singleton, we simply write $\preceq_{\varphi}$ for
$\preceq_{\{\varphi\}}$.  The definition of the $\preceq_{\A}$ simulation above in some
sense declares what is expected out of the simulation.  Below, we give a constructive
characterization of the simulation in terms of the constants used and the valuations.  For
example, if $v(\history{a}) = 3$ and $\history{a} \le 5$ is a constraint in $G$, point $2$ below
says that all $v'$ with $v'(\history{a}) \le 3$ simulate $v$.  The next lemma is a
generalization of \cite[Lemma 8]{Gastin0S20} to our setting containing prophecy clocks
and the undefined values $+\infty$ and $-\infty$.

\begin{lem}\label{lem:simulation-properties}
  Let $v,v'$ be valuations and $G$ a set of atomic constraints. We have
  \begin{enumerate}
    \item  $v\preceq_{G}v'$ iff $v\preceq_{\varphi}v'$ for all $\varphi\in G$.
  
    \item  $v\preceq_{x\leqlt c}v'$ iff
    $v(x)\not\leqlt c$ or 
    $v'(x)\leq v(x)$ or
    $(\leqlt,c)=(\leq,+\infty)$ or
    $(\leqlt,c)=(<,+\infty)\wedge v'(x)<+\infty$.
  
    \item $v\preceq_{c\leqlt x}v'$ iff
    $c\leqlt v'(x)$ or 
    $v(x)\leq v'(x)$ or 
    $(\leqlt,c)=(<,+\infty)$ or
    $(\leqlt,c)=(\leq, +\infty)\wedge v(x)<+\infty$.

  \end{enumerate}
\end{lem}
\begin{proof}\hfill
  \begin{enumerate}
    \item  is clear.
  
    \item  The right to left implication is easy. Notice for instance that if 
    $(\leqlt,c)=(<,+\infty)$ and $v'(x)<+\infty$ then for all $\delta\geq0$ we have 
    $v'+\delta\models x<+\infty$. 
    Conversely, assume that $v\preceq_{x\leqlt c}v'$ and $v(x)\leqlt c$ and
    $(\leqlt,c)\neq(\leq,+\infty)$.
    If $(\leqlt,c)=(<,+\infty)$ then, using $\delta=0$ and $v(x)\leqlt c$, we get $v'(x)<+\infty$.
    If $(\leqlt,c)=(\leq,-\infty)$ then, using $\delta=0$ and $v(x)\leqlt c$,
    we get $v'(x)\leq-\infty=v(x)$.
    Otherwise, $c\in\mathbb{Z}$ and we have to show that $v'(x)\leq v(x)$.
    Assume that $v(x)<v'(x)$. Then, we find $\delta\geq0$ such that $v(x)+\delta\leqlt c 
    <v'(x)+\delta$, a contradiction.
  
    \item Again, the implication from right to left is easy.  Notice that in the last two cases, $(\leqlt,c)=(<,+\infty)$ or $(\leqlt,c)=(\leq, +\infty)\wedge v(x)<+\infty$, then for all $\delta\geq0$ we have $v+\delta\not\models c\leqlt x$.
    Conversely, assume that $v\preceq_{c\leqlt x}v'$ and $c\not\leqlt v'(x)$ and $(\leqlt,c)\neq (<,+\infty)$.
    If $(\leqlt,c)=(\leq,+\infty)$ then, using $\delta=0$ and $c\not\leqlt v'(x)$, we get $c\not\leqlt v(x)$, i.e., $v(x)<+\infty$.
    If $(\leqlt,c)=(<,-\infty)$ then, using $\delta=0$ and $c\not\leqlt v'(x)$, we get $c\not\leqlt v(x)$, i.e., $v(x)=-\infty=v'(x)$.
    Otherwise, $c\in\mathbb{Z}$ and we have to show that $v(x)\leq v'(x)$.
    Assume that $v'(x)<v(x)$. Then, we find $\delta\geq0$ such that $c\leqlt v(x)+\delta$ but $c\not\leqlt v'(x)+\delta$, a contradiction.
    \qedhere
  \end{enumerate}
\end{proof}

We now state some useful properties that get derived from
Lemma~\ref{lem:simulation-properties}.
  
\begin{rem}\label{rem:simulation properties}
  Let $v,v'$ be valuations and $G$ a set of atomic constraints.
  \begin{enumerate}
    \item Let $x\leqlt_{1} c_{1}$ and $x\leqlt_{2} c_{2}$ be constraints with
    $(\leqlt_{1},c_{1})\leq(\leqlt_{2},c_{2})<(<,+\infty)$ 
    (we say that $x\leqlt_{1} c_{1}$ is subsumed by $x\leqlt_{2} c_{2}$).
    If $v\preceq_{x\leqlt_{2} c_{2}}v'$ then $v\preceq_{x\leqlt_{1} c_{1}}v'$.
    Indeed, from $(\leqlt_{2},c_{2})<(<,+\infty)$ and $v\preceq_{x\leqlt_{2} c_{2}}v'$ we get either (1) $v'(x)\leq v(x)$ or (2) $v(x)\not\leqlt_{2} c_{2}$ which implies $v(x)\not\leqlt_{1}c_{1}$ since $(\leqlt_{1},c_{1})\leq(\leqlt_{2},c_{2})$.
  
    \item Let $c_{1}\leqlt_{1} x$ and $c_{2}\leqlt_{2} x$ be constraints with
    $(c_{1},\leqlt_{1})\leq(c_{2},\leqlt_{2})<(+\infty,\leq)$
    (we say that $c_{1}\leqlt_{1} x$ is subsumed by $c_{2}\leqlt_{2} x$).
    If $v\preceq_{c_{2}\leqlt_{2} x}v'$ then $v\preceq_{c_{1}\leqlt_{1} x}v'$.
    Indeed, from $(c_{2},\leqlt_{2})<(+\infty,\leq)$ and $v\preceq_{c_{2}\leqlt_{2} x}v'$ we get either (1) $v(x)\leq v'(x)$ or (2) $c_{2}\leqlt_{2} v'(x)$ which implies
    $c_{1}\leqlt_{1} v'(x)$ since $(c_{1},\leqlt_{1})\leq(c_{2},\leqlt_{2})$.
    
    The ordering between \emph{lower weights} is defined by 
    $(c_{1},\leqlt_{1})<(c_{2},\leqlt_{2})$ if $c_{1}<c_{2}$ or $c_{1}=c_{2}$, 
    ${\leqlt}_{1}={\leq}$ and ${\leqlt}_{2}={<}$. We have 
    $(c_{1},\leqlt_{1})<(c_{2},\leqlt_{2})$ iff 
    $(\leqlt_{2},-c_{2})<(\leqlt_{1},-c_{1})$.
  \end{enumerate}
\end{rem}

Before lifting the simulation to event-zones, we present a central
technical object that will be used from time to time in the next set
of results.

\subsection{Distance graph for valuations that simulate a given valuation}~\label{sec:distance graph for up G v}

For a valuation $v$ and a set of constraints $G$, we let $\ua_{G}v=\{v'\in\V \mid v\preceq_{G} v'\}$,
i.e., the set of valuations $v'$ which simulate $v$. 
We will define a distance graph, denoted
$\graphv{G}{v}$, such that $\sem{\graphv{G}{v}}=\ua_{G}v$.  We remark that
$\sem{\graphv{G}{v}}$ is not really a zone since it may use constants that are not integers.

We assume that $G$ contains
$\{0\leq\prophecy{a},\prophecy{a}\leq0 \mid a\in\Sigma\}$ so
that $v\preceq_{G}v'$ implies $v(\prophecy{a})=v'(\prophecy{a})$ for all prophecy clocks $\prophecy{a}$ with $a\in\Sigma$ (see Remark~\ref{rmk:prophecyequality}).
We remove from $G$ constraints equivalent to true, such as $x\leq+\infty$, $-3<\history{a}$ or 
$0\leq\history{a}$, or equivalent to false, such as $\history{a}<0$ or $+\infty<x$.
Also, by Remark~\ref{rem:simulation properties}, we may remove from $G$ constraints that 
are subsumed by other constraints in $G$, while not
changing the simulation relation. 
Hence, for history clocks, we have at most one upper-bound constraint $\history{a}\leqlt c$ with
$(\leq,0)\leq(\leqlt,c)<(<,+\infty)$, and at most one lower-bound constraint $c\leqlt\history{a}$
with $(0,\leq)<(c,\leqlt)<(+\infty,\leq)$.
From now on, we always assume that the sets $G$ of atomic constraints that we
consider satisfy the above conditions.

\begin{figure}
	\centering
    \begin{tikzpicture}
        \begin{scope}[state/.style={draw, thick, circle, inner sep=1pt}]
          \node [state, fill=black] (p) at (0, 6) {}; 
          \node [state, fill=black] (q) at (3, 6) {}; 
          \node [state, fill=black] (r) at (5, 6) {}; 
          \node [state, fill=black] (s) at (-3, 6) {}; 
          \node [state, fill=black] (t) at (-5, 6) {}; 
        \end{scope}
        
        \begin{scope}[->, >=stealth, thick]
          \draw [bend left=25] (p) to (q); 
          \draw [bend left=25] (q) to (p); 
          \draw [bend left=60] (r) to (p); 
          \draw [bend left=25] (p) to (s); 
          \draw [bend left=25] (s) to (p); 
          \draw [bend left=60] (p) to (t); 
          \draw [bend left=60] (t) to (p); 
        \end{scope}
  
        \node at (0,5.6)  {$0$};
        \node at (3,5.6)  {$\history{a}$};
        \node at (5.2,5.6)  {$\history{b}$};
        \node at (-3,5.6)  {$\prophecy{b}$};
        \node at (-5.2,5.6)  {$\prophecy{a}$};
    
        \node at (-1.5,6.6)  {$(\leq, 5)$};
        \node at (-1.5,5.4)  {$(\leq, -5)$};
        \node at (-2.5,7.55)  {$(\leq, 7)$};
        \node at (-2.5,4.45)  {$(\leq, -7)$};

        \node at (1.5,6.6)  {$(\leq, 2)$};
        \node at (1.5,5.4)  {$(<, -1)$};
        \node at (2.5,4.45)  {$(\leq, -3)$};
    \end{tikzpicture}  
  \caption[]{
      The distance graph $\uparrow_G v$ when $v$ is such that $v(\history{a}) = 2,
  v(\prophecy{a}) = -7, v(\history{b}) = 3, v(\prophecy{b}) = -5$ and $\G = P \cup
  \{\history{a} \leq 3, \history{a} > 1, \history{b} \leq 2, \history{b} \geq 5\}$.  Here,
  the set $P = \{\prophecy{a} \leq 0, \prophecy{a} \geq 0, \prophecy{b} \leq 0,
  \prophecy{b} \geq 0\}$ denotes the set of constraints involving prophecy
  clocks.}
  \label{fig:dist-graph-up-G-v}
\end{figure}
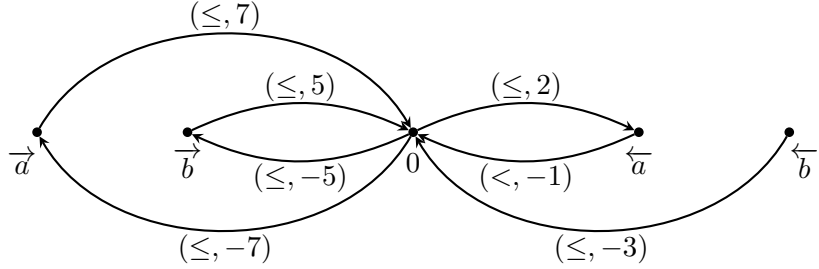

The definition of the distance graph $\graphv{G}{v}$ which
defines $\ua_{G}v$ is based on Lemma~\ref{lem:simulation-properties}.
\begin{itemize}
  \item For each prophecy clock $\prophecy{a}$, we have the edges
  $\prophecy{a}\xra{(\leq,-v(\prophecy{a}))}0$ and
  $0\xra{(\leq,v(\prophecy{a}))}\prophecy{a}$.

  \item  For each history clock $\history{a}$, we have the edge $0\to\history{a}$ with weight
  \begin{itemize}
    \item $(\leq,v(\history{a}))$ if $\history{a}\leqlt c\in G$ with $(\leqlt,c)<(<,+\infty)$ and
    $v(\history{a})\leqlt c$,

    \item  $(<,+\infty)$ if we are not in the case above and 
    $\history{a}<+\infty \in G$, $v(\history{a})<+\infty$,
  
    \item  $(\leq,+\infty)$ otherwise.
  \end{itemize}

  \item  For each history clock $\history{a}$, we have the edge $\history{a}\to 0$ with weight
  \begin{itemize}
    \item $(\leq,-\infty)$ if $+\infty\leq\history{a} \in G$ and $v(\history{a})= +\infty$,
    and if we are not in this case:
  
    \item $(\leqlt,-c)$ if $c\leqlt\history{a}\in G$ with $(c,\leqlt)<(+\infty,\leq)$ and
    $c\leqlt v(\history{a})$,

    \item $(\leq,-v(\history{a}))$ if $c\leqlt\history{a}\in G$ with $(c,\leqlt)<(+\infty,\leq)$
    and $c\not\leqlt v(\history{a})$,

    \item $(\leq,0)$ otherwise.
  \end{itemize}
\end{itemize}
An example of $\uparrow_G v$ is given in Figure~\ref{fig:dist-graph-up-G-v}.

With this definition, the distance graph $\graphv{G}{v}$ is in standard form, but not in
normal form.  Using Lemma~\ref{lem:simulation-properties}, we easily see that it has the
desired property:

\begin{lem}
  We have $v\preceq_{G} v'$ iff $v'$ satisfies all the constraints of $\graphv{G}{v}$.
\end{lem}

\subsection{Simulation for event-zones and an effective algorithmic check}

Let $Z,Z'$ be two event-zones and $G$ be a set of atomic constraints.
We say that $Z$ is $G$-simulated by $Z'$, denoted $Z\preceq_{G}Z'$, if for all $v\in Z$ 
there exists $v'\in Z'$ such that $v\preceq_{G}v'$. 
Finally, we define $(q,Z)\preceq_\A (q',Z')$ if $q=q'$ and $Z\preceq_{\G(q)} Z'$.  In the
rest of this section, we show how to check this relation efficiently.
We let $\da_{G}Z=\{v\in\V\mid v\preceq_{G}v' \text{ for some } v'\in Z\}$.
Notice that $Z\preceq_{G}Z'$ iff $Z\subseteq\da_{G}Z'$ iff $\da_{G}Z\subseteq\da_{G}Z'$.

To check $Z \not \preceq_G Z'$, we require a valuation $v \in Z$ with a witness that
$\ua_G{v} \cap Z'$ is empty.  Let $\GG'$ be a distance graph in standard form 
with $Z'=\sem{\GG'}$. Since $\graphv{G}{v}$ is also in standard form, so is the 
distance graph $\min(\graphv{G}{v},\GG')$ which defines $\ua_G{v} \cap Z'$. So we may apply 
Lemma~\ref{lem:empty-iff-negative-cycle} and the witness will be a
negative cycle in $\min(\graphv{G}{v},\GG')$.  
We show that if $\ua_G{v} \cap Z'$ is empty and $\GG'$ is in normal form, then there is a small witness, i.e., a
negative cycle in $\min(\graphv{G}{v},\GG')$ containing at most three edges, and belonging
to one of three specific forms.

\begin{lem}\label{lem:negative-cycles-v-Z}
  Let $v$ be a valuation, 
  $Z'$ a non-empty \emph{reachable} event-zone
  with canonical distance graph $\GG'$
  and $G$ a set of atomic constraints. Then, $\ua_G v \cap Z'$ is empty iff there is a
  negative cycle in one of the following forms:
  \begin{enumerate}
  \item $0 \to x \to 0$ with $0\to x$ from $\graphv{G}{v}$ and
    $x \to 0$ from $\GG'$,
  \item $0 \to y \to 0$ with $0 \to y$ from $\GG'$ and $y \to 0$
    from $\graphv{G}{v}$, and
  \item $0 \to x \to y \to 0$, with weight of $x \to y$ from
    $\GG'$ and the others from $\graphv{G}{v}$. Moreover, this
    negative cycle has finite weight.
  \end{enumerate}
\end{lem}
\begin{proof}
  Since the distance graph $\GG'$ is in normal form, it has no negative cycle. 
  Similarly, $\graphv{G}{v}$ has no negative cycle since $v\in\ua_{G}v\neq\emptyset$.  
  We know that $\ua_{G}v\cap Z'=\emptyset$ iff there is a (simple) negative cycle in $\min(\graphv{G}{v},\GG')$.  
  Since $\GG'$ is in normal form, we may restrict to negative cycles which do not use two consecutive edges from $\GG'$.  
  Now all edges of $\graphv{G}{v}$ are adjacent to node $0$.  
  Hence, if a simple cycle uses an edge from $\GG'$ which is adjacent to $0$, it consists of only two edges $0\to x\to 0$, one from $\GG'$ and one from $\graphv{G}{v}$.
  Otherwise, the simple cycle is of the form $0\to x\to y\to 0$ where the edge $x\to y$ is from $\GG'$ and the other two edges are from $\graphv{G}{v}$.  
  It remains to show that the two clock negative cycle $0 \to x \to y \to 0$ can be considered to have finite weight, i.e., weight is not $(\le, -\infty)$.

  For the cycle to have weight $(\le, -\infty)$, one of the edges
  should have weight $(\le, -\infty)$ and the others should have a
  weight different from $(\le, +\infty)$. We will show that for every
  such combination, there is a smaller negative cycle with a single
  clock and $0$. Hence we can ignore negative cycles of the form
  $0 \to x \to y \to 0$ with weight $(\le, -\infty)$.

  Suppose $\GG'_{xy}=(\le,-\infty)$. Since $Z'$ (the zone corresponding to the distance graph $\GG'$) is a reachable zone, using Lemma~\ref{lem:reachable-properties} we get $\GG'_{x0}=(\le,-\infty)$ or $\GG'_{0y}=(\le,-\infty)$.
  This gives a smaller negative cycle $0\to x \xra{(\leq,-\infty)} 0$ or $0 \xra{(\leq,-\infty)}
  y\to 0$ with the other edge $0 \to x$ or $y \to 0$ coming from $\graphv{G}{v}$,
  since by our hypothesis of a negative cycle, these edges have weight different from
  $(\le,+\infty)$.

  Suppose the weight of $0 \to x$ is $(\le, -\infty)$ in $\graphv{G}{v}$.  This can happen
  only when $x$ is a prophecy clock and $v(x)=-\infty$.
  Since $\GG'_{xy}\neq(\le,+\infty)$ and $\GG'$ is in standard form, we infer $\GG'_{x0}\neq(\le,+\infty)$.
  Hence $0 \xra{(\le, v(x))} x \xra{\GG'_{x0}} 0$ is also a negative cycle.

  Suppose $y\to 0$ has weight $(\le,-\infty)$ in $\graphv{G}{v}$.  This can happen only
  when $y$ is a history clock and $v(y)=+\infty$.  Since $\GG'_{xy}\neq(\le,+\infty)$ and
  $\GG'$ is in standard form, we obtain $\GG'_{0y}\neq(\le,+\infty)$.  
  Hence, $0 \xra{\GG'_{0y}} y \xra{(\le, -v(y))} 0$ is a negative cycle.
\end{proof}

We now have all the results required to state our inclusion test.  Using the above
lemma, and relying on a careful analysis 
we obtain the following theorem.

\begin{thm}\label{thm:simulation-check}
  Let $Z, Z'$ be non-empty \emph{reachable} zones, and $G$ a set of atomic constraints
  containing $\prophecy{a} \le 0$ and $0 \le \prophecy{a}$ for every prophecy
  clock $\prophecy{a}$. Then, $Z \not \preceq_G Z'$ iff one of the
  following conditions holds:\footnote{%
    Recall that for a non-empty zone $Z$, we simply write $Z_{xy}$ for the weight of the 
    edge $x\to y$ in the canonical distance graph of $Z$.}
  \begin{enumerate}
  \item $Z'_{x0} < Z_{x0}$ for some prophecy clock $x$, or for some
    history clock $x$ with
    \begin{itemize}
    \item $(x < +\infty) \in G$ and $Z'_{x0} = (\le, -\infty)$, or
    \item $(x \leqlt_1 c) \in G$ for $c \in \mathbb{N}$ and
      $(\le, 0) \le Z_{x0} + (\leqlt_1, c)$.
    \end{itemize}
  \item $Z'_{0y} < Z_{0y}$ for some prophecy clock $y$, or for some
    history clock $y$ with
    \begin{itemize}
    \item $(+\infty \le y) \in G$ and $Z_{0y} = (\le, +\infty)$, or
    \item $(d \leqlt_2 y) \in G$ for $d \in \Nat$ and
      $Z'_{0y} + (\leqlt_2, -d) < (\le, 0)$.
    \end{itemize}
    
    \item $Z'_{xy} < Z_{xy}$ and $Z'_{xy}$ is finite for two distinct (prophecy or
    history) clocks $x, y$ with $(x \leqlt_1 c), (d \leqlt_2 y) \in G$ for $c, d \in \Nat$
    and $(\le, 0) \le Z_{x0} + (\leqlt_1, c)$ and $Z'_{xy} + (\leqlt_2, -d) < Z_{x0}$.
  \end{enumerate}
\end{thm}

\begin{proof}
  By definition, $Z \not \preceq_G Z'$ iff
  there is a $v \in Z$ such that $\ua_G v \cap Z' =
  \emptyset$. Lemma~\ref{lem:negative-cycles-v-Z} gives three kinds of
  negative cycles that witness $\ua_G v \cap Z' = \emptyset$. We 
  show that the three conditions in the theorem respectively
  characterize the presence of the three kinds of negative cycles.
  
  \medskip\noindent
  \textbf{Case 1.} There is a negative cycle $0 \to x \to 0$ with
  $0 \to x$ from $\graphv{G}{v}$ and $x \to 0$ from $\graph{Z'}$ iff
  Item 1 above is true.

  ($\Rightarrow$).  Suppose there is such a negative cycle
  $0 \to x \to 0$. Weight of $0 \to x$ is either $(\le, v(x))$ or
  $(<, +\infty)$.

  \textit{When weight of $0\to x$ is $(\le,v(x))$}: We have
  $(\le, v(x)) + Z'_{x0} < (\le, 0)$. 
  Lemma~\ref{lem:weight-properties} implies $Z'_{x0}<(\leq,-v(x))$.
  Now, since $v \in Z$, it satisfies all constraints of $\graph{Z}$. 
  In particular, $(\leq,-v(x))\leq Z_{x0}$. We obtain $Z'_{x0}<Z_{x0}$.
  Notice that the weight of $0 \to x$ is $(\le, v(x))$ when either $x$ is a prophecy clock
  or $x$ is a history clock with $(x\leqlt_1 c)\in G$ for $c\in\Nat$ and $v(x)\leqlt_1 c$.
  We can rewrite $v(x) \leqlt_1 c$ as $(\le, v(x)) \le (\leqlt_1, c)$.
  Hence, we get $(\leq,0)\leq(\leqlt_{1},c)+(\leq,-v(x))\leq(\leqlt_{1},c)+Z_{x0}$.

  \textit{When weight of $0 \to x$ is $(<, +\infty)$}: This happens when $x$ is a history
  clock, $(x < +\infty) \in G$ and $v(x) < +\infty$ 
    (and we are not in the case above).
    For the cycle to be negative, we must have $Z'_{x0} = (\le, -\infty)$.  
  From $v \in Z$ and
  $v(x) < +\infty$, we get $Z_{x0} \neq (\le,-\infty)$. Hence
  $Z'_{x0} < Z_{x0}$.

  ($\Leftarrow$). Assume Item 1 is true.
 
  Suppose first that $x$ is a prophecy clock.  Since $Z'_{x0}<Z_{x0}$, using
  Lemma~\ref{lem:weight-properties} we find $\alpha$ satisfying both
  $Z'_{x0}<(\leq,-\alpha)\leq Z_{x0}$ and $(\leq,\alpha)\leq Z_{0x}$.  By
  Lemma~\ref{lem:fixing-values-distance-graph} we find $v\in Z$ with $v(x)=\alpha$.  This
  gives the negative cycle $0\xra{(\leq,v(x))}x\xra{Z'_{x0}}0$.

  Next, suppose that $x$ is a history clock with $x\leqlt_1 c$ in $G$ for $c\in\Nat$ and
  $(\le,0)\le Z_{x0} + (\leqlt_{1},c)$.  
  Since $Z'_{x0}<Z_{x0}$, using Lemma~\ref{lem:weight-properties}, we can find $\alpha$
  with $Z'_{x0}<(\leq,-\alpha)\leq Z_{x0}$ and $(\leq,\alpha)\leq\min(Z_{0x},(\leqlt_1,c))$.
  As above, by Lemma~\ref{lem:fixing-values-distance-graph} we find $v\in Z$ with
  $v(x)=\alpha$, resulting in the negative cycle $0\xra{(\leq,v(x))}x\xra{Z'_{x0}}0$.
  
  Now, suppose there is a history clock $x$ with $(x<+\infty)\in G$ and
  $(\leq,-\infty)=Z'_{x0}<Z_{x0}$.
  This implies $Z_{x0}\neq(\leq,-\infty)$ and by Lemma~\ref{lem:reachable-properties} we 
  obtain $Z_{0x}\leq(<+\infty)$. Let $v\in Z$. We have $v(x)<+\infty$.
  For this $v$, the weight of $0 \to x$ in $\ua_G v$ will be at most $(<, +\infty)$.  As
  $Z'_{x0} = (\le, -\infty)$, we get a negative cycle $0 \to x \to 0$ with $0 \to x$ from
  $\graphv{G}{v}$ and $x \to 0$ from $\graph{Z'}$.

  \medskip\noindent
  \textbf{Case 2.} There is a negative cycle $0 \to y \to 0$ with
  $0 \to y$ from $\graph{Z'}$ and $y \to 0$ from $\graphv{G}{v}$ iff
  Item 2 in the theorem statement is true.

  ($\Rightarrow$). Suppose there is such a negative cycle.

  \textit{When weight of $y \to 0$ is $(\le, -v(y))$.} This happens
  when $y$ is a prophecy clock or $y$ is a history clock with $(d
  \leqlt_2 y) \in G$ with $d \in \Nat$ and $d \not \leqlt_2 v(y)$. From
  the negative cycle, we have $Z'_{0y} + (\le, -v(y)) < (\le, 0)$, 
  i.e., $Z'_{0y}<(\leq,v(y))$.  Since $v\in Z$, we have $(\leq,v(y))\leq Z_{0y}$.
  We deduce that $Z'_{0y}<Z_{0y}$.
  In the case when $y$ is a history clock, we have $d \not \leqlt_2 v(y)$. Therefore, either
  $v(y) < d$ or $v(y) = d$ and ${\leqlt}_2 = {<}$. In both 
  cases, $(\leqlt_2, -d) \le (\le, -v(y))$. Hence $Z'_{0y} +
  (\leqlt_2, -d) < (\le, 0)$.
  
  \textit{When weight of $y \to 0$ is $(\le, -\infty)$}. This occurs
  when $y$ is a history clock, $+\infty \le y$ is in $G$ and $v(y) =
  +\infty$. As the cycle is negative we get
  $Z'_{0y} \neq (\le, +\infty)$. Since $v \in Z$, we get
  $Z_{0y} = (\le, +\infty)$. This gives $Z'_{0y} < Z_{0y}$.
  
  \textit{When weight of $y \to 0$ is $(\leqlt_2, -d)$}. This is
    when $y$ is a history clock, we are not in the subcase above and
  there is $d \leqlt_2 y$ in $G$ with $d \in \Nat$ and $d
  \leqlt_2 v(y)$. The negative cycle gives $Z'_{0y} + (\leqlt_2, -d)
  < (\le, 0)$. From $d \leqlt_2 v(y)$, we can infer that $(\le,
  -v(y)) \le (\leqlt_2, -d)$. Therefore we also have $Z'_{0y} +
  (\le, -v(y)) < (\le, 0)$. As in the first subcase above, we get
  $Z'_{0y} < Z_{0y}$.
  
  The remaining case is when $y$ is a history clock and 
  the weight of $y \to 0$ is $(\le, 0)$.  Since $(\le, 0) \le Z'_{0y}$, the cycle
  $0 \to y \to 0$ cannot be negative and we can ignore this case.

  ($\Leftarrow$). Assume Item 2 is true.
    
  Let us start with the case when $y$ is a prophecy clock.  We have $Z'_{0y}<Z_{0y}$.
  By Lemma~\ref{lem:weight-properties}, we can find $\alpha$ with both
  $Z'_{0y}<(\leq,\alpha)\leq Z_{0y}$ and $(\leq,-\alpha)\leq Z_{y0}$.
  By Lemma~\ref{lem:fixing-values-distance-graph} we find $v\in Z$ with
  $v(y)=\alpha$.  This gives the negative cycle $0\xra{Z'_{0y}}y\xra{(\leq,-v(y))}0$.  

  Suppose $y$ is a history clock with $+\infty \le y$ in $G$, and $Z_{0y} = (\le, +\infty)$.
  For reachable zones, this implies that $Z_{y0}=(\leq,-\infty)$ 
  (Lemma~\ref{lem:reachable-properties}).
  Hence every valuation in $Z$ has $y$-value to be $+\infty$.  Pick an arbitrary
  $v \in Z$.  We have $v(y) = +\infty$.  For this $v$, the value of $y \to 0$ in
  $\graphv{G}{v}$ is $(\le, -\infty)$.  Now, since $Z'_{0y} < Z_{0y}$, the cycle $Z'_{0y}
  + (\le, -\infty)$ is negative.

  Finally, let $y$ be a history clock with $d \leqlt_2 y$ in $G$ for $d \in \Nat$ and
  $Z'_{0y} + (\leqlt_2, -d) < (\le, 0)$. 
  Since $Z'_{0y}<Z_{0y}$, using again Lemma~\ref{lem:weight-properties}, we find $\alpha$
  with both $Z'_{0y}<(\leq,\alpha)\leq Z_{0y}$ and $(\leq,-\alpha)\leq Z_{y0}$.  By
  Lemma~\ref{lem:fixing-values-distance-graph} we find $v\in Z$ with $v(y)=\alpha$.
  If $v(y) \not \leqlt_2 d$, then the weight of $y \to 0$ is at most $(\le, -v(y))$.
  Using $Z'_{0y}<(\leq,\alpha)$ gives $0 \to y \to 0$ to be negative cycle.  Suppose
  $d\leqlt_2 v(y)$.  Then weight of $y \to 0$ is at most $(\leqlt_2, -d)$.  But we already
  have $Z'_{0y} + (\leqlt_2, -d) < (\le, 0)$ in our hypothesis, which gives the required
  negative cycle.
  
  \medskip\noindent
  \textbf{Case 3.} There is a \emph{finite weight} negative cycle
  $0 \to x \to y \to 0$ with $0 \to x$ and $y \to 0$ from
  $\graphv{G}{v}$ and $x \to y$ from $\graph{Z'}$ iff the third
  condition of the theorem is true.

  ($\Rightarrow$).  Suppose there is such a negative cycle.  
  Since the weight of the cycle is finite, the weight of each edge is also finite.
  Hence, the weight of the edge $0\to x$ is $(\leq,v(x))$.  We find $x\leqlt_{1} c$ in $G$
  with $c\in\Nat$ and $v(x)\leqlt_1 c$ (if $x$ is a prophecy clock then
  $(\leqlt_{1},c)=(\leq,0)$).  As in case 1 above, from $v\in Z$ we deduce that
  $(\leq,0)\leq Z_{x0}+(\leq,v(x))$ and from $v(x)\leqlt_{1}c$ we get
  $(\leq,v(x))\leq(\leqlt_{1},c)$.
  We obtain $(\le,0) \le Z_{x0}+(\leqlt_1,c)$.
  
  Now, since the weight of $y\to 0$ is finite, we find $d\leqlt_{2}y$ in $G$ (if 
  $y$ is a prophecy clock we take $(d,\leqlt_{2})=(0,\leq)$).
  The weight of $y \to 0$ is
  $\max( (\le, -v(y)), (\leqlt_2, -d) )$. Therefore:
  $Z'_{xy} + (\le, v(x) - v(y)) < (\le, 0)$ and
  $(\le, v(x)) + Z'_{xy} + (\leqlt_2, -d) < (\le, 0)$. 
  Since $v \in Z$, we have $(\leq,v(y)-v(x))\leq Z_{xy}$.
  Using $Z'_{xy}<(\leq,-(v(x)-v(y)))=(\leq,v(y)-v(x))$ (recall that $v(x)$ and $v(y)$ are 
  finite in the present case),
  we get $Z'_{xy} < Z_{xy}$. Again, as $v \in Z$, we have
  $(\leq,-v(x))\leq Z_{x0}$. Together with
  $Z'_{xy}+(\leqlt_2,-d)<(\leq,-v(x))$, we get
  $Z'_{xy} + (\leqlt_2, -d) < Z_{x0}$.

  ($\Leftarrow$). Suppose the third condition is true. Instead of
  explicitly constructing a $v$ as in the previous two cases, we will simply prove that
  there exists a valuation that forms the required negative
  cycle.  We start by defining some new weights and observe some
  properties that will be used later. Define 
  $w_1 = (<, -e')$ if $Z'_{xy} = (\le, e')$ and $w_1 = (\le, -e')$ if
  $Z'_{xy} = (<, e')$, and $w_2 = (\le, -e' + d)$ if either $Z'_{xy}$
  or $(\leqlt_2,-d)$ has a strict inequality, and $w_2 = (<, -e' + d)$
  otherwise. Notice that $w_1 + Z'_{xy} = (<, 0)$ and 
  $w_2 + Z'_{xy} + (\leqlt_2, -d) = (<, 0)$ are negative.
    Using $Z'_{xy} + (\leqlt_2, -d) < Z_{x0}$, we get 
  $(<,0)=w_{2}+Z'_{xy}+(\leqlt_{2},-d)<w_{2}+Z_{x0}$ and we obtain
  $(\leq,0)\leq w_{2}+Z_{x0}$. Similarly, using $Z'_{xy}<Z_{xy}$ we get
  $(<,0)=w_{1}+Z'_{xy}<w_{1}+Z_{xy}$ and $(\leq,0)\leq w_{1}+Z_{xy}$.

  Consider a distance graph $\GG$ formed by taking $\graph{Z}$
  and modifying two of its edges as follows: change $0 \to x$ to
  $\min(Z_{0x}, (\leqlt_1, c), w_2)$; change $y \to x$ to
  $\min(Z_{yx}, w_1)$. Notice first that $\GG$ is in standard form.
  Next, we claim that $\GG$ has only
  non-negative cycles. It is sufficient to show that $0 \to x \to 0$
  and $y \to x \to y$ are non-negative. From
  $(\le, 0) \le Z_{x0} + (\leqlt_1, c)$, $(\le, 0) \le w_2 + Z_{x0}$
  and the fact that $\graph{Z}$ does not have negative cycles, we
  infer that $0 \to x \to 0$ is non-negative in $\GG$.  The
  weight of $y \to x \to y$ is either $Z_{yx} + Z_{xy}$ or
  $w_1 + Z_{xy}$. The former is non-negative as $Z$ is non-empty. We
  have shown above that $w_1 + Z_{xy}$ is non-negative. Hence
  $\GG$ is standard with no negative cycles, and its solution set is
  non-empty.

  Now, pick a $v$ that satisfies constraints of
  $\GG$. Valuation $v$ is in $Z$, and additionally satisfies
  $v(x) \leqlt_1 c$, $(\le, v(x)) \le w_2$ and
  $(\le, v(x) - v(y)) \le w_1$. Recall that $w_1 + Z'_{xy}=(<,0)$. 
  Using $(\le, v(x) - v(y)) \le w_1$, we obtain
  $(\le, v(x) - v(y)) + Z'_{xy} < (\le, 0)$. If $d \not \leqlt_2 v(y)$,
  then this corresponds to the weight of the cycle
  $0 \to x \to y \to 0$ which we have now shown to be
  negative. Otherwise, we have $d \leqlt_2 v(y)$ and the weight of
  edge $y \to 0$ is $(\leqlt_2, -d)$. The weight of the cycle would
  be: $(\le, v(x)) + Z'_{xy} + (\leqlt_2, -d)$. But, $(\le, v(x)) \le
  w_2$ and 
  $w_2 + Z'_{xy} + (\leqlt_2, -d) = (<,0)$. Hence the required cycle is negative.
\end{proof}

From Theorem~\ref{thm:simulation-check}, we can see that the inclusion test requires
iteration over clocks $x, y$ and checking if the conditions are satisfied by the
respective weights.

\begin{cor}
  Checking if $(q,Z) \preceq_\A (q',Z')$ can be done in time $\mathcal{O}(|X|^2)$ =
  $\mathcal{O}(|\Sigma|^2)$.
\end{cor}

\section{Finiteness of the simulation relation}\label{sec:termination}

In this section, we will show that the simulation relation $\preceq_{\A}$ defined in
Section~\ref{sec:simulation} is finite, which implies that the reachability algorithm of
Definition~\ref{def:reach-algo} terminates.  Recall that given an event-clock automaton
$\A$, we have an associated map $\G$ from states of $\A$ to sets of atomic constraints.
Let $M=\max\{|c| \mid c\in\mathbb{Z} \text{ is used in some constraint of } \A\}$, the
maximal constant of $\A$.  We have $M\in\mathbb{N}$ and constraints in the sets $\G(q)$
use constants in $\{-\infty,+\infty\}\cup\{c\in\mathbb{Z}\mid|c|\leq M\}$.

Recall that the simulation relation $\preceq_{\A}$ was
defined on nodes of the event-zone graph
$\ezg(\A)$  by $(q,Z)\preceq_{\A}(q',Z')$ if $q=q'$ and
$Z\preceq_{\G(q)}Z'$.
This simulation relation $\preceq_{\A}$
is \emph{finite} if for any
infinite sequence $(q,Z_{0}),(q,Z_{1}),(q,Z_{2}),\ldots$ of
\emph{reachable} nodes in $\ezg(\A)$ we find $i<j$ with
$(q,Z_{j})\preceq_{\A}(q,Z_{i})$, i.e., $Z_{j}\preceq_{\G(q)}Z_{i}$.
Notice that we restrict to \emph{reachable} zones in the definition
above.  Our goal now is to prove that the relation $\preceq_{\A}$ is
finite.  The structure of the proof is as follows.  
\begin{enumerate}
  \item We prove in Lemma~\ref{lem:dagger} of Section~\ref{sec:dagger} that for any
  \emph{reachable} node $(q,Z)$ of $\ezg(\A)$, the canonical distance graph $\graph{Z}$
  satisfies a set of conditions, that we call $(\dagger)$ conditions below, which
  depend only on the maximal constant $M$ of $\A$.

  \item We introduce an equivalence relation $\sim_{M}$ of \emph{finite index} on
  valuations (depending on $M$ only) and show in Lemma~\ref{lem:sim_M main property} of
  Section~\ref{sec:eqreln} that, if $G$ is a set of atomic constraints using constants in
  $\{c\in\mathbb{Z}\mid|c|\leq M\}\cup \{-\infty,+\infty\}$ and if $Z$ is a zone such that
  its canonical distance graph $\graph{Z}$ satisfies $(\dagger)$ conditions, then
  $\da_{G}Z$ is a union of $\sim_{M}$ equivalence classes.
\end{enumerate}

\subsection{Prophecy clocks and distance graphs in reachable event-zones}
\label{sec:dagger}

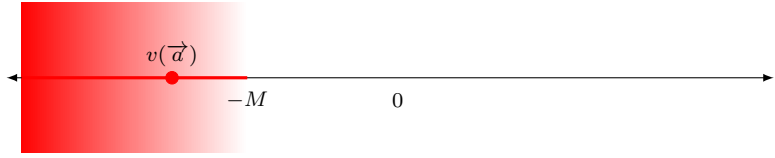
\begin{figure}[!h]
  \centering	
      \centering	
      \begin{tikzpicture}

        \shade[left color=red, right color=white] (-2,-1) rectangle (-5,1);

        \draw[latex-latex] (-5.2,0) -- (5,0) ; 
      
        \node at (0, -0.3) {\scriptsize$0$}; 
        \node at (-2, -0.3) {\scriptsize$-M$}; 
        \node at (-3, 0.3) {\scriptsize $v(\prophecy{a})$}; 
        \draw[red,fill=red] (-3,0) circle (.5ex);					
        \draw[color=red,very thick] (-2,0) -- (-5,0);

      \end{tikzpicture}
  \caption[]{A pictorial depiction of
  Lemma~\ref{lem:reachable-zones-prophecy-clocks}, where the shaded region depicts the zone $Z$.
  }
  \label{fig:lem-invariant}
\end{figure}

We start with a lemma which highlights an important and surprising property of
\emph{prophecy} clocks in reachable event-zones.  Namely, we show that if an $M$-reachable
event-zone contains a valuation $v$ in which a prophecy clock has a finite value that is
below $-M$, then it must contain all valuations obtained from $v$ by setting that prophecy
clock to any finite value smaller than $-M$.  
The implication of this property is illustrated in Figure~\ref{fig:lem-invariant}.
This property is essential for the proof of
the $(\dagger)$ conditions.  The proof follows from the observation that the property is
true in the initial zone, and is invariant under the zone operations, namely, guard
intersection, reset, release and time elapse.

\begin{lem}\label{lem:reachable-zones-prophecy-clocks}
  Let $Z$ be an $M$-reachable event-zone.  For every valuation $v \in Z$,
  and for every prophecy clock $\prophecy{x}$, if
  $-\infty < v(\prophecy{x}) < -M$, then $v[\prophecy{x}\mapsto\alpha] \in Z$
  for every $-\infty < \alpha < -M$.
\end{lem}

\begin{proof}
  The initial zone is
  $(\bigwedge_{\history{x}} \history{x} = +\infty) \land (\bigwedge_{\prophecy{x}}
  -\infty \le \prophecy{x} \le 0)$. The property is true in this zone. We
  now show that the property is invariant under guard intersection (with maximal constant 
  $M$), reset, release and time elapse.
  Assume $Z$ is a zone that satisfies the property.  Let $v$ be an
  arbitrary valuation (not necessarily in $Z$) with $-\infty < v(\prophecy{x}) < -M$. For
  $-\infty<\alpha<-M$, we simply write $v_\alpha=v[\prophecy{x}\mapsto\alpha]$.
  \begin{description}
  \item[Guard intersection] Let $g$ be a guard, which is in general a
    conjunction of atomic constraints. Suppose $g$ contains an atomic
    constraint on $\prophecy{x}$, either $c \leqlt \prophecy{x}$ or
    $\prophecy{x} \leqlt c$. The constant $c$ is either $-\infty$, or
    $-M \le c \le 0$. Therefore if $v$ satisfies the constraint, every
    $v_\alpha$ will satisfy the atomic constraint. Now, suppose $g$
    contains $d \leqlt y$ or $y \leqlt d$ for some $y \neq
    \prophecy{x}$. Once again, notice that if $v$ satisfies this guard,
    every $v_\alpha$ will satisfy the guard since
    $v(y) = v_\alpha(y)$. Hence, if $v$ satisfies the guard,
    $v_\alpha$ satisfies the guard. The property is true in the zone
    $Z \land g$.
    
  \item[Release]  The release operation on $\prophecy{y}$ takes a
    valuation $u$ and adds a valuation $u[\prophecy{y}\mapsto\beta]$ for
    each $-\infty \le \beta \le 0$.
    Let $v \in [\prophecy{y}]Z$. Then, there is some valuation $u \in Z$
    such that $v=u[\prophecy{y}\mapsto\beta]$ for some $-\infty \le \beta \le 0$.
    When $\prophecy{y} = \prophecy{x}$, associating $\alpha$ to $\prophecy{x}$ in the
    release operation starting from $u$ gives $v_\alpha$.  When
    $\prophecy{y} \neq \prophecy{x}$, we use the induction hypothesis: valuation
    $u_\alpha \in Z$.
    Setting $\prophecy{y}$ to $\beta$ in the release operation starting
    from $u_\alpha$ gives $v_\alpha$.

  \item[Reset] The reset operation takes every valuation in $Z$ and
    makes some history clock to $0$. This does not perturb the
    required property.

  \item[Time elapse]  Suppose $v + \delta \in \elapse{Z}$ for some
    $v \in Z$ and $\delta\geq0$.  This means
    $(v+\delta)(\prophecy{y}) \le 0$ for every prophecy clock $\prophecy{y}$.
    We have to show that if $-\infty < (v + \delta)(\prophecy{x}) < -M$,
    then $(v + \delta)_\alpha \in \elapse{Z}$ for all
    $-\infty < \alpha < -M$. Since
    $-\infty < (v + \delta)(\prophecy{x}) < -M$, we also have
    $-\infty < v(\prophecy{x}) < -M$. For some given $\alpha$, let
    $\beta = \alpha - \delta$. We have $v_\beta \in Z$ by
    hypothesis. Notice that $(v+\delta)_{\alpha}=v_\beta + \delta $.
    Moreover, $v_\beta + \delta \in \elapse{Z}$: for each clock
    $\prophecy{y} \neq \prophecy{x}$,
    $(v_\beta + \delta)(\prophecy{y}) = (v + \delta)(\prophecy{y}) \le 0$ and
    $(v_\beta + \delta)(\prophecy{x}) = \alpha < -M \le 0$.  \qedhere
  \end{description}
\end{proof}

\begin{rem}
  We now make three significant observations about the above lemma.
  \begin{enumerate}
    \item The lemma holds for $M$-reachable event-zones but not for all event-zones.

    \item There is no similar version of the above lemma for history clocks.  A reset of a
    history clock makes its value exactly equal to $0$ in every valuation and creates
    non-trivial diagonal constraints with other clocks.  Moreover, repeated resets can
    generate arbitrarily large diagonal constraints, for e.g., a loop with guard $x = 1$
    and reset $x$.  This is why simulations are particularly needed to control history
    clocks.

    \item In our simulation $v \preceq_G v'$, we have $v(\prophecy{a}) = v'(\prophecy{a})$
    (Remark~\ref{rmk:prophecyequality}): there is no abstraction of the value
    of prophecy clocks and the simulation relation by itself does not have any means to
    show finiteness.  However, as we show below, due to the above property, the
    reachable zones themselves take care of finiteness with respect to prophecy clocks.
  \end{enumerate}
  The challenge is then to combine this observation on prophecy clocks along with the
  non-trivial simulation happening for history clocks to prove that we still get a finite
  simulation.  This is what we do in Section~\ref{sec:eqreln}.
\end{rem}

Now, we give the $(\dagger)$ conditions and prove that they are satisfied by canonical distance
graphs of reachable zones.

\begin{lem}\label{lem:dagger}
  Let $Z\neq\emptyset$ be an $M$-\emph{reachable} zone.
  Then, the canonical distance graph $\graph{Z}$ satisfies the $(\dagger)$ conditions:
  \begin{enumerate}
    \item[$\dagger_{1}$] If $Z_{\prophecy{x}0}=(<,+\infty)$ then
    for all $y\neq\prophecy{x}$, either $y$ is a prophecy clock which is undefined in $Z$ and
    $Z_{\prophecy{x}y}=Z_{0y}=(\leq,-\infty)$ or $Z_{\prophecy{x}y}\in\{(<,+\infty),(\leq,+\infty)\}$.
    
    \item[$\dagger_{2}$] If $Z_{\prophecy{x}0}<(<,+\infty)$ then
    $(\leq,0)\leq Z_{\prophecy{x}0}\leq(\leq,M)$.
    
    \item[$\dagger_{3}$] If $Z_{\prophecy{x}\history{y}}<(<,+\infty)$ then
    $(\leq,0)\leq Z_{\prophecy{x}0}\leq(\leq,M)$.
    
    \item[$\dagger_{4}$] Either $Z_{0\prophecy{x}}=(\leq,-\infty)$ or
    $(<,-M)\leq Z_{0\prophecy{x}}\leq(\leq,0)$.
    
    \item[$\dagger_{5}$] Either $Z_{0\prophecy{y}}=(\leq,-\infty)$ 
    or $Z_{x0} + (<,-M) \leq Z_{x\prophecy{y}}$ for all $x\in X$ with $x\neq\prophecy{y}$.
    
    \item[$\dagger_{6}$] Either
    $Z_{\prophecy{x}\prophecy{y}}\in\{(\leq,-\infty),(<,+\infty),(\leq,+\infty)\}$
    or $(<,-M) \leq Z_{\prophecy{x}\prophecy{y}} \leq (\leq,M)$.
  \end{enumerate}
\end{lem}

\begin{proof}\hfill
  \begin{enumerate}
    \item[$\dagger_{1}$] Assume that $Z_{\prophecy{x}0}=(<,+\infty)$ and let $y\neq\prophecy{x}$.
    Consider first the case $Z_{0y}=(\leq,-\infty)$, i.e., $y$ is a prophecy clock which
    is undefined in $Z$.  Then, since $\graph{Z}$ is in normal form, we have
    $Z_{\prophecy{x}y}\leq Z_{\prophecy{x}0}+Z_{0y}=(<,+\infty)+(\leq,-\infty)=(\leq,-\infty)$.
    
    The second case is when $Z_{0y}\neq(\leq,-\infty)$.  
    This implies $Z_{\prophecy{x}y}\neq(\leq,-\infty)$ since otherwise we would get $Z_{0y}\leq
    Z_{0\prophecy{x}}+Z_{\prophecy{x}y}=(\leq,-\infty)$.
    We claim that there is a valuation $v\in Z$ with $-\infty<v(y)$ and
    $-\infty<v(\prophecy{x})<-M$.
     
    Consider the distance graph $\GG'$ obtained from $\graph{Z}$ by setting the weight of
    edge $y\to 0$ to $\min(Z_{y0},(<,+\infty))$ and of edge $0\to\prophecy{x}$ to
    $\min(Z_{0\prophecy{x}},(<,-M))$.  Notice that $\GG'$ is in standard form.
    We show that $\GG'$ has no negative cycles.
    Since $\graph{Z}$ is in normal form,
    the candidates for being negative must use the new weight 
    $(<,-M)$ of $0\to\prophecy{x}$ or the new weight $(<,+\infty)$ of $y\to 0$ or both. This 
    gives the cycle $0\to\prophecy{x}\to 0$ with weight $(<,-M)+Z_{\prophecy{x}0}=(<,+\infty)$,
    the cycle $0 \to y \to 0$ with weight $Z_{0y}+(<,+\infty)$ which is not negative since 
    $Z_{0y}\neq(\leq,-\infty)$, and the cycle $y \to 0 \to \prophecy{x} \to y$ with weight
    $(<,+\infty)+(<,-M)+Z_{\prophecy{x}y}$
    which is not negative since $Z_{\prophecy{x}y}\neq(\le,-\infty)$.
    Since $\GG'$ is standard and has no negative cycle,
    Lemma~\ref{lem:empty-iff-negative-cycle} implies $\sem{\GG'}\neq\emptyset$.  Note that
    $\sem{\GG'}\subseteq\sem{\graph{Z}}=Z$.  Finally, $-\infty<v(y)$ and
    $-\infty<v(\prophecy{x})<-M$ for all $v\in\GG'$, which proves the claim.
    
    By Lemma~\ref{lem:reachable-zones-prophecy-clocks},
    $v_{\alpha}=v[\prophecy{x}\mapsto\alpha]\in Z$ for all $-\infty<\alpha<-M$.  Now,
    $v_{\alpha}(y)-v_{\alpha}(\prophecy{x})=v(y)-\alpha$ satisfies the constraint
    $Z_{\prophecy{x}y}$.  We deduce that $Z_{\prophecy{x}y}$ is either $(<,+\infty)$ or
    $(\le,+\infty)$.
    
    \item[$\dagger_{2}$]  Suppose $(\leq,M)<Z_{\prophecy{x}0}<(<,+\infty)$.  By
    Lemma~\ref{lem:weight-properties}, we find $\alpha$ such that $(\leq,\alpha)\leq
    Z_{\prophecy{x}0}$, $(\leq,-\alpha)\leq Z_{0\prophecy{x}}$ and $\alpha\not\leq M$.  Now, by
    Lemma~\ref{lem:fixing-values-distance-graph}, we find $v\in Z$ with
    $-\infty<v(\prophecy{x})=-\alpha<-M$.    
    By Lemma~\ref{lem:reachable-zones-prophecy-clocks},
    $v[\prophecy{x}\mapsto\alpha]\in Z$ for all $-\infty < \alpha < -M$, a contradiction with 
    $Z_{\prophecy{x}0}$ finite.    
    
    \item[$\dagger_{3}$] Assume $Z_{\prophecy{x}\history{y}}<(<,+\infty)$ is finite. Since 
    $\graph{Z}$ is standard, we get $Z_{\prophecy{x}0}\neq(\leq,+\infty)$. Now, $\dagger_{1}$ 
    implies $Z_{\prophecy{x}0}<(<,+\infty)$. We conclude with $\dagger_{2}$.
    
    \item[$\dagger_{4}$] 
    The proof is similar to $\dagger_{2}$. Suppose $(\leq,-\infty)<Z_{0\prophecy{x}}<(<,-M)$.  By
    Lemma~\ref{lem:weight-properties}, we find $\alpha$ such that $(\leq,\alpha)\leq
    Z_{0\prophecy{x}}$, $(\leq,-\alpha)\leq Z_{\prophecy{x}0}$ and $\alpha\neq-\infty$.  Now, by
    Lemma~\ref{lem:fixing-values-distance-graph}, we find $v\in Z$ with
    $-\infty<v(\prophecy{x})=-\alpha<-M$. Since $Z_{0\prophecy{x}}<(<,-M)$, we can 
    find $-\infty\neq\beta<-M$ such that $Z_{0\prophecy{x}}<(\leq,\beta)$. Now, 
    Lemma~\ref{lem:reachable-zones-prophecy-clocks} implies that 
    $v[\prophecy{x}\mapsto\beta]\in Z$, which is a contradiction with 
    $Z_{0\prophecy{x}}<(\leq,\beta)$.
    
    \item[$\dagger_5$]  Suppose $Z_{0\prophecy{y}}\neq(\leq,-\infty)$.
    If $Z_{x0}=(\leq,-\infty)$ or $Z_{x\prophecy{y}}=(\leq,+\infty)$, the condition is 
    trivially true. So we also assume in the following that $Z_{x0}\neq(\leq,-\infty)$ 
    and $Z_{x\prophecy{y}}\neq(\leq,+\infty)$. Since $\graph{Z}$ is in standard form, we get 
    $Z_{x0}\neq(\leq,+\infty)$. By Lemma~\ref{lem:reachable-properties}, we also have 
    $Z_{x\prophecy{y}}\neq(\leq,-\infty)$.
    
    Consider the distance graph $\GG'$ obtained from the canonical graph $\graph{Z}$ by
    setting $\GG'_{0x}=\min(Z_{0x},(<,+\infty))$,
    $\GG'_{\prophecy{y}0}=\min(Z_{\prophecy{y}0},(<,+\infty))$, and keeping other weights unchanged.
    The graph $\GG'$ is in standard form. Moreover, it has no negative cycles. Indeed, 
    since $\graph{Z}$ is in normal form, the new cycles that we have to check in $\GG'$ 
    are $0\to x\to 0$, $0\to\prophecy{y}\to 0$ and $\prophecy{y}\to 0\to x\to \prophecy{y}$.
    Since $Z_{x0}\neq(\leq,-\infty)$, $Z_{0\prophecy{y}}\neq(\leq,-\infty)$ and 
    $Z_{x\prophecy{y}}\neq(\leq,-\infty)$, these cycles are non-negative.
    Let $\GG''$ be the normalization of $\GG'$.  Since $Z_{x0}\neq(\leq,+\infty)$, we have
    $\GG''_{x0}=\GG'_{x0}=Z_{x0}$.  Notice that $\GG''\leq\GG'\leq\graph{Z}$, hence
    $\sem{\GG''}\subseteq\sem{\GG'}\subseteq Z$.

    We claim that for all $\varepsilon>0$ and all $c\in\mathbb{R}$ with $(\leq,c)<Z_{x0}$
    we have $(\leq,c-M-\varepsilon)<Z_{x\prophecy{y}}$.  Indeed, if
    $(\leq,c)<Z_{x0}=\GG''_{x0}$ then by Lemma~\ref{lem:weight-properties} we find
    $\alpha$ such that $(\leq,\alpha)\leq\GG''_{x0}$, $(\leq,-\alpha)\leq\GG''_{0x}$ and
    $c<\alpha$.  Next, by Lemma~\ref{lem:fixing-values-distance-graph}, we find
    $v\in\sem{\GG''}\subseteq Z$ such that $v(x)=-\alpha$.  If $v(\prophecy{y})<-M$ then by
    Lemma~\ref{lem:reachable-zones-prophecy-clocks} we get
    $v'=v[\prophecy{y}\mapsto-M-\varepsilon]\in Z$. Otherwise, we let $v'=v$. We have 
    $c-M-\varepsilon<\alpha-M-\varepsilon\leq v'(\prophecy{y})-v'(x)$. We obtain
    $(\leq,c-M-\varepsilon)<(\leq,v'(\prophecy{y})-v'(x))\leq Z_{x\prophecy{y}}$ and the claim is 
    proved.
    
    Finally, if $Z_{x0}=(<,+\infty)$ then we may take $c$ arbitrarily large and the claim
    implies $(<,+\infty)\leq Z_{x\prophecy{y}}$.  Otherwise, $Z_{x0}=(\leqlt,d)$ is finite and
    taking $c=d-\varepsilon$ we obtain $(\leq,d-M-2\varepsilon)<Z_{x\prophecy{y}}$ for all
    $\varepsilon>0$. This implies $Z_{x0}+(<,-M)=(<,d-M)\leq Z_{x\prophecy{y}}$.
    
    \item[$\dagger_{6}$] Assume that
    $Z_{\prophecy{x}\prophecy{y}}\notin\{(\leq,-\infty),(<,+\infty),(\leq,+\infty)\}$.
    Since $\graph{Z}$ is in standard form, we have $Z_{\prophecy{x}0}\neq(\leq,+\infty)$.
    Now, $(\leq,-\infty)\neq Z_{\prophecy{x}\prophecy{y}}\leq
    Z_{\prophecy{x}0}+Z_{0\prophecy{y}}$ implies that
    $Z_{0\prophecy{y}}\neq(\leq,-\infty)$.
    Next, since $Z_{0 \prophecy{y}}\neq(\leq,-\infty)$ and
    $Z_{\prophecy{x}\prophecy{y}}\notin\{(<,+\infty),(\leq,+\infty)\}$, by
    $\dagger_{1}$ we deduce that $Z_{\prophecy{x}0}\neq(<,+\infty)$.  

    Applying $\dagger_5$, we deduce that
    $(<,-M)\leq Z_{\prophecy{x}0}+(<,-M)\leq Z_{\prophecy{x}\prophecy{y}}$.
    
    Finally, $Z_{\prophecy{x}\prophecy{y}}\leq Z_{\prophecy{x}0}+Z_{0\prophecy{y}}\leq
    Z_{\prophecy{x}0}\leq(\leq,M)$ by $\dagger_{2}$.  
    \qedhere
  \end{enumerate}
\end{proof}

\begin{figure}[!h]
  \centering	
      \centering	
      \begin{tikzpicture}
      [state/.style={draw, thick, circle, inner sep=4pt}]
      \begin{scope}[xshift=1cm, yshift=0cm]
      \begin{scope}[every node/.style={state}]
        \node (q0) at (4, 3) {\scriptsize $q_0$}; 
        \node (q1) at (6, 3) {\scriptsize $q_1$}; 
        \node (q2) at (8, 3) {\scriptsize $q_2$}; 
      \end{scope}	
      
      \begin{scope}[->, >=stealth, thick]
        \draw (3.25, 3) to (q0); 
        \draw (q0) to [out=120,in=60,looseness=8] (q0);
        \draw (q2) to [out=120,in=60,looseness=8] (q2);  
        \draw (q0) to (q1); 
        \draw (q1) to (q2); 
        \draw (q2) to (8.75,3); 
      \end{scope}
      
      \node at (6, 2) {\footnotesize $\Aa_2$}; 

      \node at (4, 4.6) {\footnotesize $b$}; 
      \node at (4, 4.25) {\scriptsize $\prophecy{b}=-1 \wedge \prophecy{a}<-1$}; 
      \node at (5, 3.25) {\footnotesize $b$}; 
      \node at (5, 2.75) {\scriptsize $\prophecy{a}=-1$}; 

      \node at (7, 3.25) {\footnotesize $a$}; 

      \node at (8, 4.25) {\footnotesize $d$}; 
    \end{scope}

      \begin{scope}[xshift=0cm, yshift=0cm, zone/.style={draw,rectangle,minimum width=2.3cm, minimum height=2cm, rounded corners}]
      \node [zone] (z0) at (1, 0) {}; 
      \node [zone] (z1) at (1, 3) {};
      \node [zone] (z2) at (4.5, 0) {}; 
      \node [zone] (z3) at (8, 0) {};
      \end{scope}
      
      \begin{scope}[->, >=stealth, thick]
      \draw (-1,0) to (z0); 
      \draw (z0) to (z1); 
      \draw (z0) to (z2);
      \draw (z1) to (z2);
      \draw (z2) to (z3);
      \draw (z1) to [out=120,in=180,looseness=4] (z1);
      \draw (z3) to [out=55,in=30,looseness=4] (z3);
      \draw (z3) to (10,0);
      \end{scope}

      \node at (0.25, 0) {\footnotesize $q_0$}; 
      \node at (1.2, 0.5) {\scriptsize $\prophecy{a} \leq 0$}; 
      \node at (1.2, 0) {\scriptsize $\prophecy{b} \leq 0$}; 
      \node at (1.2, -0.5) {\scriptsize $\prophecy{d} \leq 0$};

      \node at (0.2, 3) {\footnotesize $q_0$}; 
      \node at (1.2, 3.7) {\scriptsize $-1 \leq \prophecy{b} \leq 0$}; 
      \node at (1.2, 3.2) {\scriptsize $\prophecy{a} - \prophecy{b} < 0$}; 
      \node at (1.2, 2.7) {\scriptsize $\prophecy{d} - \prophecy{b} \leq 1$}; 
      \node at (1.2, 2.2) {\scriptsize $\prophecy{a} < 0, ~ \prophecy{d} \leq 0 $};       
    
      \node at (3.6, 0) {\footnotesize $q_1$}; 
      \node at (4.7, 0.75) {\scriptsize $-1 \leq \prophecy{a} \leq 0$}; 
      \node at (4.7, 0.25) {\scriptsize $\prophecy{b} - \prophecy{a} \leq 1$}; 
      \node at (4.7, -0.25) {\scriptsize $\prophecy{d} - \prophecy{a} \leq 1$}; 
      \node at (4.7, -0.75) {\scriptsize $\prophecy{b} \leq 0, ~ \prophecy{d} \leq 0 $}; 

      \node at (7.3, 0) {\footnotesize $q_2$}; 
      \node at (8.2, 0.5) {\scriptsize $\prophecy{a} \leq 0$}; 
      \node at (8.2, 0) {\scriptsize $\prophecy{b} \leq 0$}; 
      \node at (8.2, -0.5) {\scriptsize $\prophecy{d} \leq 0$}; 

      \node at (1.25, 1.5) {\footnotesize $b$}; 
      \node at (0.2, 1.65) {\scriptsize $\prophecy{b}=-1~\wedge$}; 
      \node at (0.2, 1.35) {\scriptsize $\prophecy{a}<-1$}; 
      
      \node at (-1, 4.25) {\footnotesize $b$}; 
      \node at (-1, 4.75) {\scriptsize $\prophecy{b}=-1 \wedge \prophecy{a}<-1$}; 

      \node at (2.75, 0.25) {\footnotesize $b$}; 
      \node at (2.75, -0.25) {\scriptsize $\prophecy{a}=-1$}; 

      \node at (2.5, 1.5) {\footnotesize $b$}; 
      \node at (3.2, 1.75) {\scriptsize $\prophecy{a}=-1$}; 

      \node at (6.25, 0.25) {\footnotesize $a$}; 
      \node at (9.6, 1.5) {\footnotesize $d$}; 
    \end{tikzpicture} 
  \caption[]{Event-\emph{predicting} automaton $\A_2$, for which there exists no finite
  time-abstract bisimulation on valuations, and its (finite) event-zone graph $\ezg(\A_{2})$.}
  \label{fig:eca-zone-graph-ex-2}
\end{figure}
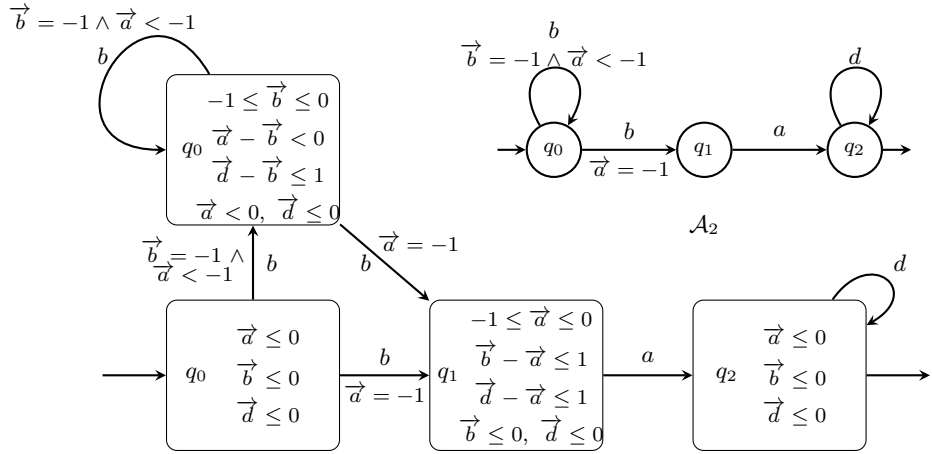

We now see that the $(\dagger)$ conditions imply that the weights of edges of the form $0 \to \prophecy{x}$, $\prophecy{x} \to 0$ and $\prophecy{x} \to \prophecy{y}$ belong to the finite set
$$
\{(\leq,-\infty),(<,+\infty),(\leq,+\infty)\}\cup \{(\leqlt,c)\mid
c\in\mathbb{Z} \wedge -M\leq c\leq M\} \,.
$$  
For an example, see Figure~\ref{fig:eca-zone-graph-ex-2}.
Note that the event-zone graph is sound and complete for reachability. 
Thus, we obtain as a corollary that, for event-predicting automata (EPA), we do not even need simulation to obtain finiteness.

\begin{cor}
  Let $\A$ be an EPA. The zone graph $\ezg(\A)$ is finite.
\end{cor}

Thus, for EPA, one could simply construct the event-zone graph
  without simulations to solve the reachability problem: more
  precisely, the reachability algorithm of
  Definition~\ref{def:reach-algo} can be modified by taking equality
  $=$ instead of simulation $\preceq$. At the same time, notice that
  in the definition of $v \preceq_G v'$, we have $v(\prophecy{a}) =
  v'(\prophecy{a})$ for prophecy clocks. Therefore, for EPA, the check
  $v \preceq_G v'$ boils down to $v = v'$ and the check $Z \preceq_G
  Z'$ amounts to checking $Z \incl Z'$, an inclusion between
  zones.

  We can also see that Theorem~\ref{thm:simulation-check}
  when restricted to prophecy clocks boils down to mere inclusion. We
  provide a short explanation for this fact. The first two conditions
  of Theorem~\ref{thm:simulation-check} simply
  check $Z'_{x0} < Z_{x0}$ and $Z'_{0y} <
  Z_{0y}$ when there are only prophecy clocks. In the third 
  condition, it remains to show that the extra
  tests (1) $(\le, 0) \le Z_{x0} + (\leqlt_1, c)$ and (2) $Z'_{xy} +
  (\leqlt_2, -d) < Z_{x0}$ are true whenever $Z'_{xy} < Z_{xy}$, and
  $x, y$ are prophecy clocks. So
  there is no need to check the extra conditions separately.

  When $x$ and $y$ are prophecy clocks, we have $x \le 0$
  and $0 \le y$ in $G$. By definition of standard form, we have $(\le,
  0) \le Z_{x0}$. This automatically gives (1). For (2), since
  $Z'_{xy} < Z_{xy}$ and $Z_{xy} \le Z_{x0} + Z_{0y}$, we have
  $Z'_{xy} < Z_{x0} + Z_{0y}$. But $Z_{0y} \le (\le, 0)$ since $\GG(Z)$ is
  in standard form. Therefore, we deduce $Z_{xy} \le Z'_{x0}$. 

For general ECA, i.e., when we have both history and prophecy clocks, we need to make sure
that the finiteness of the simulation that we do for handling history clocks does not get
affected in the presence of prophecy clocks.  We turn to this issue in the next section.

\subsection{An equivalence relation of finite index on valuations}
\label{sec:eqreln}

We define an equivalence relation of finite index
$\sim_{M}$ on valuations.
First, we define $\sim_{M}$ on 
$\alpha,\beta\in\overline{\mathbb{R}}=\mathbb{R}\cup\{-\infty,+\infty\}$
by $\alpha\sim_{M}\beta$ if
$(\alpha\leqlt c \Longleftrightarrow \beta\leqlt c)$ for all
$(\leqlt,c)$ with ${\leqlt}\in\{<,\leq\}$ and
$c\in\{-\infty,+\infty\}\cup\{d\in\mathbb{Z}\mid|d|\leq M\}$.  In
particular, if $\alpha\sim_{M}\beta$ then
$(\alpha=-\infty \Longleftrightarrow \beta=-\infty)$ and
$(\alpha= +\infty \Longleftrightarrow \beta= +\infty)$.

Next, for valuations $v_{1},v_{2}\in\V$, we define
$v_{1}\sim_{M}v_{2}$ by two conditions: $v_{1}(x)\sim_{M}v_{2}(x)$ and
$v_{1}(x)-v_{1}(y)\sim_{2M}v_{2}(x)-v_{2}(y)$ for all clocks
$x,y\in X$. Notice that we use $2M$ for differences of values.
Clearly, $\sim_{M}$ is an equivalence relation of finite index on
valuations.

The next result relates the equivalence relation $\sim_{M}$ and the simulation relation
$\preceq_{G}$ when the finite constants used in the constraints are bounded by $M$.
Recall from Section~\ref{sec:distance graph for up G v} the definition of the distance
graph $\graphv{G}{v}$ for the set of valuations $\ua_{G}v$.

\begin{lem}\label{lem:sim_M and graphv}
  Let $v_{1},v_{2}\in\V$ be valuations with $v_{1}\sim_{M}v_{2}$ and
  let $G$ be a set of atomic constraints using constants in
  $\{-\infty,+\infty\}\cup\{c\in\mathbb{Z}\mid|c|\leq M\}$.  By
  replacing the weights $(\leq,v_{1}(x))$ (resp.\ $(\leq,-v_{1}(x))$)
  by $(\leq,v_{2}(x))$ (resp.\ $(\leq,-v_{2}(x))$) in the graph
  $\graphv{G}{v_{1}}$ we obtain the graph $\graphv{G}{v_{2}}$.
\end{lem}
    
\begin{proof}
  This is clear by definition for edges $\prophecy{x}\to 0$ and
  $0\to\prophecy{x}$ adjacent to a prophecy clock. We consider now edges
  adjacent to history clocks $\history{x}$.
  \begin{itemize}
  \item Consider the edge $0\to\history{x}$.
    \begin{itemize}
    \item If its weight is $(\leq,v_{1}(\history{x}))$ in
      $\graphv{G}{v_{1}}$ then there is some $\history{x}\leqlt c\in G$
      with $(\leqlt,c)<(<,+\infty)$ and $v_{1}(\history{x})\leqlt c$.
      Since $v_{1}\sim_{M}v_{2}$ we deduce that
      $v_{2}(\history{x})\leqlt c$ and the edge $0\to\history{x}$ has weight
      $(\leq,v_{2}(\history{x}))$ in $\graphv{G}{v_{2}}$.
      
    \item If its weight is $(<,+\infty)$ in $\graphv{G}{v_{1}}$ then we
      are not in the case above and $\history{x}<+\infty \in G$,
      $v_{1}(\history{x})<+\infty$.  Since $v_{1}\sim_{M}v_{2}$ we deduce
      that $v_{2}(\history{x})<+\infty$ and the edge $0\to\history{x}$ has
      weight $(<,+\infty)$ in $\graphv{G}{v_{2}}$.
      
    \item Otherwise, the weight is $(\leq,+\infty)$ in both
      $\graphv{G}{v_{1}}$ and $\graphv{G}{v_{2}}$.
    \end{itemize}
    
  \item Consider the edge $\history{x}\to 0$.
    \begin{itemize}
    \item If its weight is $(\leq,-\infty)$ in $\graphv{G}{v_{1}}$
      then $+\infty\leq\history{x} \in G$ and $v_{1}(\history{x})= +\infty$.
      Since $v_{1}\sim_{M}v_{2}$ we deduce that
      $v_{2}(\history{x})= +\infty$ and the edge $\history{x}\to0$ has weight
      $(\leq,-\infty)$ in $\graphv{G}{v_{2}}$.
            
    \item If its weight is $(\leqlt,-c)$ in $\graphv{G}{v_{1}}$ then
      we are not in the case above and there is some
      $c\leqlt\history{x}\in G$ with $(c,\leqlt)<(+\infty,\leq)$ and
      $c\leqlt v_{1}(\history{x})$.  Since $v_{1}\sim_{M}v_{2}$ we deduce
      that $c\leqlt v_{2}(\history{x})$ and the edge $\history{x}\to0$ has
      weight $(\leqlt,-c)$ in $\graphv{G}{v_{2}}$.
      
    \item If its weight is $(\leq,-v_{1}(\history{x}))$ in
      $\graphv{G}{v_{1}}$ then we are not in the cases above and there
      is some $c\leqlt\history{x}\in G$ with $(c,\leqlt)<(+\infty,\leq)$
      and $c\not\leqlt v(\history{x})$, Since $v_{1}\sim_{M}v_{2}$ we
      deduce that $c\not\leqlt v_{2}(\history{x})$ and the edge
      $\history{x}\to0$ has weight $(\leq,-v_{2}(\history{x}))$ in
      $\graphv{G}{v_{2}}$.
      
    \item Otherwise, the weight is $(\leq,0)$ in both
      $\graphv{G}{v_{1}}$ and $\graphv{G}{v_{2}}$.  \qedhere
    \end{itemize}
  \end{itemize}
\end{proof}

Next we state the central lemma that says that $\da_{G}Z$ is a union of $\sim_{M}$
equivalence classes.

\begin{lem}\label{lem:sim_M main property}
  Let $v_{1},v_{2}\in\V$ be valuations with $v_{1}\sim_{M}v_{2}$ and
  let $G$ be a set of atomic constraints using constants in
  $\{-\infty,+\infty\}\cup\{c\in\mathbb{Z}\mid|c|\leq M\}$.
  Let $Z$ be a zone with a canonical distance graph $\graph{Z}$ satisfying
  $(\dagger)$.  Then, 
  $v_{1}\in\da_{G}Z$ iff $v_{2}\in\da_{G}Z$.
\end{lem}

\begin{proof}
  Notice that $v\in\da_{G}Z$ iff $\ua_{G}v\cap Z\neq\emptyset$.
  The proof is by contradiction.
  We assume that $\ua_{G}v_{1}\cap Z\neq\emptyset$ and
  $\ua_{G}v_{2}\cap Z=\emptyset$.  By Lemma~\ref{lem:negative-cycles-v-Z}
  we find a negative cycle $C_{2}$ using one edge from
  $\graph{Z}$ and one or two edges from $\graphv{G}{v_{2}}$.  By
  Lemma~\ref{lem:sim_M and graphv}, we have a corresponding cycle
  $C_{1}$ using the same edge from $\graph{Z}$ and the same one or two
  edges from $\graphv{G}{v_{1}}$.  The cycle $C_{1}$ is not negative
  since $\ua_{G}v_{1}\cap Z\neq\emptyset$.  Therefore, the negative
  cycle $C_{2}$ should use at least one edge labelled
  $(\leq,v_{2}(x))$ or $(\leq,-v_{2}(x))$.  We consider the different
  cases.
  \begin{enumerate}
  \item Cycle
    $C_{2}=0 \xra{(\leq,v_{2}(\history{y}))} \history{y} \xra{Z_{\history{y}0}}
    0$.  We have
    $C_{1}=0 \xra{(\leq,v_{1}(\history{y}))} \history{y} \xra{Z_{\history{y}0}}
    0$.

    Since we have the edge $0 \xra{(\leq,v_{1}(\history{y}))} \history{y}$ in
    $\graphv{G}{v_{1}}$, there is a constraint $\history{y}\leqlt' c'$ in
    $G$ with $(\leqlt',c')<(<,+\infty)$ and $v_{1}(\history{y})\leqlt' c'$.
    We deduce that $0\leq v_{1}(\history{y})\leq M$.

    Let $Z_{\history{y}0}=(\leqlt,c)$.
    Since $C_{1}$ is not a negative cycle, we get
    $(\leq,0)\leq(\leqlt,c+v_{1}(\history{y}))$, which is equivalent to
    $-c\leq v_{1}(\history{y})$. Using $0\leq v_{1}(\history{y})\leq M$ and
    $v_{1}\sim_{M}v_{2}$ we deduce that $-c\leq v_{2}(\history{y})$.  This
    is equivalent to $(\leq,0)\leq(\leqlt,c+v_{2}(\history{y}))$, a
    contradiction with $C_{2}$ being a negative cycle.
    
  \item Cycle
    $C_{2}=0 \xra{Z_{0\history{y}}} \history{y} \xra{(\leq,-v_{2}(\history{y}))}
    0$.  We have
    $C_{1}=0 \xra{Z_{0\history{y}}} \history{y} \xra{(\leq,-v_{1}(\history{y}))}
    0$.
    
    Since we have the edge $\history{y} \xra{(\leq,-v_{1}(\history{y}))} 0$ in
    $\graphv{G}{v_{1}}$, there is a constraint $c'\leqlt'\history{y}$ in
    $G$ with $(c',\leqlt')<(+\infty,\leq)$ and
    $c'\not\leqlt' v_{1}(\history{y})$.  We deduce that
    $0\leq v_{1}(\history{y})\leq M$.

    Let $Z_{0\history{y}}=(\leqlt,c)$.
    Since $C_{1}$ is not a negative cycle, we get
    $(\leq,0)\leq(\leqlt,c-v_{1}(\history{y}))$, which is equivalent to
    $v_{1}(\history{y})\leq c$.  Using $v_{1}\sim_{M}v_{2}$ and
    $0\leq v_{1}(\history{y})\leq M$, we deduce that
    $v_{2}(\history{y})\leq c$.  This is equivalent to
    $(\leq,0)\leq(\leqlt,c-v_{2}(\history{y}))$, a contradiction with
    $C_{2}$ being a negative cycle.
    
  \item Cycle
    $C_{2}=0 \xra{(\leq,v_{2}(\prophecy{x}))} \prophecy{x} \xra{Z_{\prophecy{x}0}}
    0$.  We have
    $C_{1}=0 \xra{(\leq,v_{1}(\prophecy{x}))} \prophecy{x} \xra{Z_{\prophecy{x}0}}
    0$.
    
    Since $C_{2}$ is negative, we have $Z_{\prophecy{x}0}\neq(\leq,+\infty)$.  Also, if
    $Z_{\prophecy{x}0}=(<,+\infty)$ then we must have $v_{2}(\prophecy{x})=-\infty$, which implies
    $v_{1}(\prophecy{x})=-\infty$ since $v_{1}\sim_{M}v_{2}$, a contradiction with $C_{1}$
    being non-negative.  Hence, $Z_{\prophecy{x}0}=(\leqlt,c)<(<,+\infty)$ and by
    $(\dagger_{2})$, we infer $0\leq c\leq M$.
    
    Since $C_{1}$ is not negative, we get
    $(\leq,0)\leq(\leqlt,c+v_{1}(\prophecy{x}))$, which is equivalent to
    $-c\leq v_{1}(\prophecy{x})$.  Using $v_{1}\sim_{M}v_{2}$ and
    $0\leq c\leq M$ we deduce that $-c\leq v_{2}(\prophecy{x})$.  This is
    equivalent to $(\leq,0)\leq(\leqlt,c+v_{2}(\prophecy{x}))$, a
    contradiction with $C_{2}$ being a negative cycle.
    
  \item Cycle
    $C_{2}=0 \xra{Z_{0\prophecy{x}}} \prophecy{x} \xra{(\leq,-v_{2}(\prophecy{x}))}
    0$.  We have
    $C_{1}=0 \xra{Z_{0\prophecy{x}}} \prophecy{x} \xra{(\leq,-v_{1}(\prophecy{x}))}
    0$.
    
    Let $Z_{0\prophecy{x}}=(\leqlt,c)$.  Since $C_{2}$ is negative, we
    deduce that $v_{2}(\prophecy{x})\neq-\infty$.  Using
    $v_{1}\sim_{M}v_{2}$, we infer $v_{1}(\prophecy{x})\neq-\infty$.  Since
    $C_{1}$ is not negative, we get $Z_{0\prophecy{x}}\neq(\leq,-\infty)$.
    From
    $(\dagger_{4})$, we infer
    $(<,-M)\leq Z_{0\prophecy{x}}$ and $-M\leq c\leq 0$.
    
    Since $C_{1}$ is not a negative cycle, we get
    $(\leq,0)\leq(\leqlt,c-v_{1}(\prophecy{x}))$, which is equivalent to
    $v_{1}(\prophecy{x})\leq c$.
    Using $v_{1}\sim_{M}v_{2}$ and $-M\leq c\leq 0$, we deduce that
    $v_{2}(\prophecy{x})\leq c$.  This is equivalent to
    $(\leq,0)\leq(\leqlt,c-v_{2}(\prophecy{x}))$, a contradiction with
    $C_{2}$ being a negative cycle.
  
  \item Cycle
    $C_{2}=0 \xra{(\leq,v_{2}(\history{y}))} \history{y}
    \xra{Z_{\history{y}\prophecy{x}}} \prophecy{x} \xra{(\leq,-v_{2}(\prophecy{x}))} 0$.
        
    We have
    $C_{1}=0 \xra{(\leq,v_{1}(\history{y}))} \history{y}
    \xra{Z_{\history{y}\prophecy{x}}} \prophecy{x} \xra{(\leq,-v_{1}(\prophecy{x}))} 0$.

    Let $Z_{\history{y}\prophecy{x}}=(\leqlt,c)$.  As in case 1 above, we get
    $0 \leq v_{1}(\history{y}) \leq M$.  From the fact that the cycle
    $0 \xra{(\leq,v_{1}(\history{y}))} \history{y} \xra{Z_{\history{y}0}} 0$ is
    not negative, we get $(\leq,-M)\leq Z_{\history{y}0}$.  Since $C_{2}$
    is negative, we get $v_{2}(\prophecy{x})\neq-\infty$.  Using
    $v_{1}\sim_{M}v_{2}$, we infer $v_{1}(\prophecy{x})\neq-\infty$.  From
    the fact that the cycle
    $0 \xra{Z_{0\prophecy{x}}} \prophecy{x} \xra{(\leq,-v_{1}(\prophecy{x}))} 0$ is
    not negative, we deduce $Z_{0\prophecy{x}}\neq(\leq,-\infty)$.  Using
    $(\dagger_{5})$
    we obtain
    $$
    (\leq,-M)+(<,-M)\leq Z_{\history{y}0}+(<,-M) \leq Z_{\history{y}\prophecy{x}}
    =(\leqlt,c)
    $$
    and we deduce that $-2M\leq c\leq 0$.
    
    Since $C_{1}$ is not a negative cycle, we get
    $(\leq,0)\leq(\leqlt,c+v_{1}(\history{y})-v_{1}(\prophecy{x}))$, which is
    equivalent to $v_{1}(\prophecy{x})-v_{1}(\history{y})\leq c$.  Using
    $v_{1}\sim_{M}v_{2}$ and $-2M\leq c\leq 0$ we deduce that
    $v_{2}(\prophecy{x})-v_{2}(\history{y})\leq c$.  We conclude as in the
    previous cases.
    
  \item Cycle
    $C_{2}=0 \xra{(\leq,v_{2}(\prophecy{x}))} \prophecy{x}
    \xra{Z_{\prophecy{x}\history{y}}} \history{y} \xra{(\leq,-v_{2}(\history{y}))} 0$.

    We have
    $C_{1}=0 \xra{(\leq,v_{1}(\prophecy{x}))} \prophecy{x}
    \xra{Z_{\prophecy{x}\history{y}}} \history{y} \xra{(\leq,-v_{1}(\history{y}))} 0$.

    Since $C_{2}$ is negative but not $C_{1}$, we get first 
    $Z_{\prophecy{x}\history{y}}\neq(\leq,+\infty)$ and then $v_{1}(\prophecy{x})\neq-\infty$.
    As in case 2 above, we get $0\leq v_{1}(\history{y})\leq M$. 
    We deduce that $Z_{\prophecy{x}\history{y}}=(\leqlt,c)<(<,+\infty)$ and $c\neq+\infty$.
    From $(\dagger_{3})$ we obtain $Z_{\prophecy{x}0}\leq(\leq,M)$.
    Since $0 \xra{(\leq,v_{1}(\prophecy{x}))} \prophecy{x} \xra{Z_{\prophecy{x}0}} 0$ is
    not a negative cycle, we get $-M\leq v_{1}(\prophecy{x})\leq 0$.
    Finally, we obtain $0\leq v_{1}(\history{y})-v_{1}(\prophecy{x})\leq 2M$.

    Since $C_{1}$ is not a negative cycle, we get
    $(\leq,0)\leq(\leqlt,c+v_{1}(\prophecy{x})-v_{1}(\history{y}))$, which is
    equivalent to $v_{1}(\history{y})-v_{1}(\prophecy{x})\leq c$.  Using
    $v_{1}\sim_{M}v_{2}$ and
    $0\leq v_{1}(\history{y})-v_{1}(\prophecy{x})\leq 2M$, we deduce that
    $v_{2}(\history{y})-v_{2}(\prophecy{x})\leq c$.  We conclude as in the
    previous cases.
            
  \item Cycle
    $C_{2}=0 \xra{(\leq,v_{2}(\prophecy{x}))} \prophecy{x}
    \xra{Z_{\prophecy{x}\history{y}}} \history{y} \xra{(\leqlt',-c')} 0$.

    We have
    $C_{1}=0 \xra{(\leq,v_{1}(\prophecy{x}))} \prophecy{x}
    \xra{Z_{\prophecy{x}\history{y}}} \history{y} \xra{(\leqlt',-c')} 0$.

    Let $Z_{\prophecy{x}\history{y}}=(\leqlt,c)$.  
    As in case 6 above, we show that $c\neq+\infty$ and $-M\leq v_{1}(\prophecy{x})\leq 0$.

    Since $C_{1}$ is not a negative cycle, we get
    $0\leq c+v_{1}(\prophecy{x})-c'$, which is equivalent to
    $c'-c\leq v_{1}(\prophecy{x})$.  Using $v_{1}\sim_{M}v_{2}$ and
    $-M\leq v_{1}(x)\leq 0$, we deduce that $c'-c\leq v_{2}(\prophecy{x})$,
    which is equivalent to $0\leq c+v_{2}(\prophecy{x})-c'$, a
    contradiction with $C_{2}$ being a negative cycle.
    
  \item Cycle
    $C_{2}=0 \xra{(\leq,v_{2}(\prophecy{x}))} \prophecy{x}
    \xra{Z_{\prophecy{x}\prophecy{y}}} \prophecy{y} \xra{(\leq,-v_{2}(\prophecy{y}))} 0$
    with $x\neq y$.

    We have
    $C_{1}=0 \xra{(\leq,v_{1}(\prophecy{x}))} \prophecy{x}
    \xra{Z_{\prophecy{x}\prophecy{y}}} \prophecy{y} \xra{(\leq,-v_{1}(\prophecy{y}))} 0$.

    Since $C_{2}$ is negative but not $C_{1}$, using $v_{1}\sim_{M}v_{2}$ we get
    successively $Z_{\prophecy{x}\prophecy{y}}\neq(\leq,+\infty)$, $v_{2}(\prophecy{y})\neq-\infty\neq
    v_{1}(\prophecy{y})$, $v_{1}(\prophecy{x})\neq-\infty\neq v_{2}(\prophecy{x})$, and finally
    $(\leq,-\infty)< Z_{\prophecy{x}\prophecy{y}}<(<,+\infty)$.
    
    Let $Z_{\prophecy{x}\prophecy{y}}=(\leqlt,c)$. 
    From $(\dagger_{6})$, we deduce that $-M\leq c \leq M$.

    Since $C_{1}$ is not a negative cycle, we get
    $(\leq,0)\leq(\leqlt,c+v_{1}(\prophecy{x})-v_{1}(\prophecy{y}))$, which is
    equivalent to $v_{1}(\prophecy{y})-v_{1}(\prophecy{x})\leq c$.  Using
    $v_{1}\sim_{M}v_{2}$ and $-M\leq c \leq M$, we deduce that
    $v_{2}(\prophecy{y})-v_{2}(\prophecy{x})\leq c$.  We conclude as in the
    previous cases.
    
  \item Cycle
    $C_{2}=0 \xra{(\leq,v_{2}(\history{x}))} \history{x}
    \xra{Z_{\history{x}\history{y}}} \history{y} \xra{(\leq,-v_{2}(\history{y}))} 0$
    with $x\neq y$.

    We have
    $C_{1}=0 \xra{(\leq,v_{1}(\history{x}))} \history{x}
    \xra{Z_{\history{x}\history{y}}} \history{y} \xra{(\leq,-v_{1}(\history{y}))} 0$.
    
    As in case 1 above, we get $0\leq v_{1}(\history{x})\leq M$.  As in
    case 2 above, we get $0\leq v_{1}(\history{y})\leq M$.  We obtain
    $-M\leq v_{1}(\history{y})-v_{1}(\history{x}) \leq M$.

    Let $Z_{\history{x}\history{y}}=(\leqlt,c)$.  Since $C_{1}$ is not
    negative, we get
    $(\leq,0)\leq(\leqlt,c+v_{1}(\history{x})-v_{1}(\history{y}))$, which is
    equivalent to $v_{1}(\history{y})-v_{1}(\history{x})\leq c$.  Using
    $v_{1}\sim_{M}v_{2}$ and
    $-M\leq v_{1}(\history{y})-v_{1}(\history{x}) \leq M$, we deduce that
    $v_{2}(\history{y})-v_{2}(\history{x})\leq c$.  We conclude as in the
    previous cases.
    
  \item Cycle
    $C_{2}=0 \xra{(\leq,v_{2}(\history{x}))} \history{x}
    \xra{Z_{\history{x}\history{y}}} \history{y} \xra{(\leqlt',-c')} 0$ with
    $x\neq y$.

    We have
    $C_{1}=0 \xra{(\leq,v_{1}(\history{x}))} \history{x}
    \xra{Z_{\history{x}\history{y}}} \history{y} \xra{(\leqlt',-c')} 0$.
    
    As in case 1 above, we get $0\leq v_{1}(\history{x})\leq M$.

    Let $Z_{\history{x}\history{y}}=(\leqlt,c)$.  Since $C_{1}$ is not
    negative, we get $0\leq c+v_{1}(\history{x})-c'$, which is equivalent
    to $c'-c\leq v_{1}(\history{x})$.  Using $v_{1}\sim_{M}v_{2}$ and
    $0\leq v_{1}(\history{x})\leq M$, we deduce that
    $c'-c\leq v_{2}(\history{x})$, which is equivalent to
    $0\leq c+v_{2}(\history{x})-c'$, a contradiction with $C_{2}$ being a
    negative cycle. 
  \end{enumerate}
   Notice that we have crucially used the ``$2M$'' occurring in  the
     definition of $v_1 \sim_M v_2$ (as $v_1(x) - v_1(y)
     \sim_{2M} v_2(x) - v_2(y)$) in the cases where we deal with
     cycles containing one prophecy clock and one history clock (Cases
     5 and 6). In the rest of the cases, it was sufficient to use
     $v_1(x) \sim_M v_2(x)$.\qedhere
\end{proof}

Finally, from Lemmas~\ref{lem:dagger} and \ref{lem:sim_M main property}, we obtain our
main theorem of the section.

\begin{thm}\label{thm:simulation finite}
  The simulation relation $\preceq_{\A}$ is finite.
\end{thm}

\begin{proof}
  Let $(q,Z_{0}),(q,Z_{1}),(q,Z_{2}),\ldots$ be an infinite sequence
  of \emph{reachable} nodes in $\ezg(\A)$.  By Lemma~\ref{lem:dagger},
  for all $i$, the distance graph $\graph{Z_{i}}$ in canonical form
  satisfies conditions $(\dagger)$.

  The atomic constraints in $G=\G(q)$ use constants in
  $\{-\infty,+\infty\}\cup\{c\in\mathbb{Z}\mid|c|\leq M\}$.
  From Lemma~\ref{lem:sim_M main property} we deduce that for all $i$, 
  $\da_{G}Z_{i}$ is a union of $\sim_{M}$-classes.
  Since $\sim_{M}$ is of finite index, there are only finitely many
  unions of $\sim_{M}$-classes.  Therefore, we find $i<j$ with
  $\da_{G}Z_{i}=\da_{G}Z_{j}$, which implies $Z_{j}\preceq_{G}Z_{i}$.
\end{proof}

Note that the number of enumerated zones is bounded by $2^r$
  where $r$ is the number of equivalence classes of $\sim_M$. This is
  similar to the known upper bound in classical timed automata, where
  instead of $r$, we can use the number of regions as defined
  in~\cite{Alur:TCS:1994}.
Indeed, despite this blow up the interest in zone algorithms is that, at least
in the timed setting, they work significantly better in practice.  We hope the above
zone-based approach for ECA will also pave the way for fast implementations for ECA.

\section{Conclusion}
In this paper, we propose a simulation-based approach for reachability
in ECAs. The main difficulty and difference from timed automata is the
use of prophecy clocks and undefined values. We believe that the
crux of our work has been in identifying the new representation for
prophecy clocks and undefined values. With this as the starting point,
we have been able to adapt the zone graph computation and the
$\Gg$-simulation technique to the ECA setting. This process required
us to closely study the mechanics of prophecy clocks in the zone
computations and we discovered this surprising property that prophecy
clocks by themselves do not create a problem for finiteness.

  The final reachability algorithm looks almost identical to the timed
automata counterpart and hence provides a mechanism to transfer timed automata technology to the ECA setting. 
As a follow-up to this work, we have extended our reachability algorithm to a more general model, called \emph{Generalized timed automaton (GTA)}~\cite{AkshayGGJS23} (which subsumes event-clock automata). 
In~\cite{AkshayGGJS23}, we also developed a prototype implementation of our algorithm in \textsc{Tchecker}, an open-source platform for timed automata analysis.
To the best of our knowledge, this is the first tool that can handle event-clock automata, and it provides a way to efficiently check event-clock specifications on timed automata models. 

\bibliographystyle{alphaurl}
\bibliography{m.bib}

\end{document}